\RequirePackage[hyphens]{url}
\documentclass[9pt,review]{livecoms}

\usepackage{lipsum} 
\usepackage[version=4]{mhchem}
\usepackage{siunitx}
\usepackage[italic]{mathastext}
\graphicspath{{figures/}}
\newcommand{\versionnumber}{1.0}  

\newcommand{\githubrepository}{\url{https://github.com/jhenin/Methods-for-enhanced-sampling-and-free-energy-calculations}}  

\title{Enhanced sampling methods for molecular dynamics simulations [Article v\versionnumber]}

\author[1,2*]{J\'er\^ome H\'enin}
\author[3*]{Tony Leli\`evre}
\author[4*]{Michael R.  Shirts}
\author[5,6*]{Omar Valsson}
\author[7*]{Lucie Delemotte}
\affil[1]{Laboratoire de Biochimie Th\'eorique UPR 9080, CNRS, Paris, France}
\affil[2]{Institut de Biologie Physico-Chimique--Fondation Edmond de Rothschild, Paris, France}
\affil[3]{CERMICS, Ecole des Ponts, INRIA, Marne-la-Vall\'ee, France}
\affil[4]{Department of Chemical and Biological Engineering, University of Colorado Boulder, Boulder, CO, USA, 80309}
\affil[5]{University of North Texas, Department of Chemistry, Denton, TX, USA}
\affil[6]{Max Planck Institute for Polymer Research, Mainz, Germany}
\affil[7]{KTH Royal Institute of Technology, Science for Life Laboratory, Stockholm, Sweden}

\corr{jerome.henin@ibpc.fr}{JH}
\corr{tony.lelievre@enpc.fr}{TL}
\corr{michael.shirts@colorado.edu}{MRS}
\corr{omar.valsson@unt.edu}{OV}
\corr{lucied@kth.se}{LD}

\orcid{J\'er\^ome H\'enin}{0000-0003-2540-4098}
\orcid{Tony Leli\`evre}{0000-0002-3412-113X}
\orcid{Michael R. Shirts}{0000-0003-3249-1097}
\orcid{Omar Valsson}{0000-0001-7971-4767}
\orcid{Lucie Delemotte}{0000-0002-0828-3899}

\blurb{This LiveCoMS document is maintained online on GitHub at \githubrepository; to provide feedback, suggestions, or help improve it, please visit the GitHub repository and participate via the issue tracker.}

\pubDOI{10.XXXX/YYYYYYY}
\pubvolume{<volume>}
\pubissue{<issue>}
\pubyear{<year>}
\articlenum{<number>}
\datereceived{Day Month Year}
\dateaccepted{Day Month Year}

\usepackage{graphicx}
\usepackage{amsmath}
\usepackage{amssymb}
\usepackage{bbm} 
\usepackage{color}
\usepackage[utf8]{inputenc}
\usepackage{hyperref}
\usepackage{caption}
\usepackage{subcaption}

\newcommand{\vx}{\mathbf{x}}
\newcommand{\vp}{\mathbf{p}}
\newcommand{\vz}{\mathbf{z}}
\newcommand{\vF}{\mathbf{F}}
\newcommand{\vb}{\mathbf{b}}
\newcommand{\A}{\mathrm{A}}
\newcommand{\B}{\mathrm{B}}
\usepackage{tabularx}

\begin{document}

\begin{frontmatter}

\maketitle
\begin{abstract}
Enhanced sampling algorithms have emerged as powerful methods to extend the utility of molecular dynamics simulations and allow the sampling of larger portions of the configuration space of complex systems in a given amount of simulation time. This review aims to present the unifying principles and differences of many of the  computational methods currenly used for enhanced sampling in molecular simulations of biomolecules, soft matter and molecular crystals. Indeed, despite the apparent abundance and divergence of such methods, the principles at their core can be boiled down to a relatively limited number of statistical and physical principles.
To enable comparisons, the various methods are introduced using similar terminology and notation. We then illustrate in which ways many different methods combine principles from a smaller class of enhanced sampling concepts.
This review is intended for scientists with an understanding of the basics of molecular dynamics simulations and statistical physics who want a deeper understanding of the ideas that underlie various enhanced sampling methods and the relationships between them.
This living review is intended to be updated to continue to reflect the wealth of sampling methods as they continue to emerge in the literature.
\end{abstract}

\end{frontmatter}

\clearpage
\tableofcontents

\section{Introduction}

Molecular dynamics (MD) simulations are nowadays routinely employed to gain insights into the atomistic-level behavior of molecular systems. They are often used in combination with experiments, usually to provide the atomistic counterpart to a more macroscopic description afforded by other techniques.
MD simulations rely on the numerical and iterative solution of the equations of motion, using small timesteps for integration, on the order of femtoseconds. While they are useful to monitor the time evolution of a system, for instance, in response to a perturbation, they are also very often used as an efficient sampling tool to recover statistical \hyperlink{ref:Ensemble} {ensembles}, much in the same way \hyperlink{ref:MetropolisMonteCarlo} {Monte Carlo} (MC) based methods of configurational sampling are.

In this review, we assume that we have a MD simulation algorithm that samples a single specified \hyperlink{ref:Ensemble} {ensemble} (constant number of particles, constant volume or pressure, constant temperature---NVT or NPT, respectively)~\footnote{Many of the methods work for constant chemical potential, but as such simulations cannot be carried out in standard MD simulations because of changing particle numbers, and dedicated simulations are challenging and rare, we will not explicitly address the application of the methods to these systems.}. A large number of algorithms have been proposed to achieve this type of configurational sampling and the algorithm choice does not impact what is covered in this review. Two important criteria remain: that this algorithm samples the \hyperlink{ref:Distribution} {distribution} of choice, and that the sampling is \hyperlink{ref:ergodic} {ergodic}, i.e. it will eventually cover the entire configuration space. However, the samples are allowed to be, and almost always are, correlated, and the time needed to approach this \hyperlink{ref:ergodic} {ergodic} behavior could effectively be infinite, or at least beyond the time scale of any reasonable computer simulation. We note that many methods described here can also be applied if \hyperlink{ref:MetropolisMonteCarlo} {Monte Carlo} algorithms are used instead of MD simulations to sample conformation space.

MD simulations are generally considered to suffer from three main limitations:
\begin{itemize}
    \item the accuracy of the interaction model or force field (MM, QM/MM, semi-empirical, \emph{ab initio}...) may not enable the desired insights.
    \item the simulation output (the trajectory) is high-dimensional, noisy and can be difficult to interpret and describe using a meaningful and relevant lower-dimension level of description.
    \item given the limitation on the timestep, which needs to be small enough for integration to be stable and accurate, the timescales that can be sampled are often shorter than the process of interest to the researcher.
\end{itemize}

This review focuses on describing the numerous methods that have been put forth to address the third issue, broadly referred to as ``enhanced sampling MD simulation'' techniques (see also~\cite{10.1016/j.bbagen.2014.10.019,10.1063/1.5109531,10.1039/d1cp04809k} for earlier reviews surveying the field). 

We do note note that the second issue, of finding a low-dimensional projection for the interpretation of MD simulations, is directly related to several of the algorithms presented in this review, grouped under the umbrella term ``collective variable (\hyperlink{ref:CV} {CV})-based methods''. Indeed, the configurational ensemble of systems of interest is generally very peaked, featuring several \hyperlink{ref:metastab} {metastable} and well-defined states with high probability, while the regions between these states have probabilities close to zero in the full high-dimensional Cartesian space. This explains why schemes that reduce the \hyperlink{ref:DimRed} {dimensionality} of the space sampled by projecting it onto a lower-dimensional surface can be successful if they accelerate sampling along the CV.
If the process of interest is a transition between two states, then the ideal CV accelerating transitions between these states is the committor function, which describes the progress of the transition between the two states, and the transition state between the two endpoints is the region where the committor value is around $\frac{1}{2}$~\cite{doi:10.1146/annurev.physchem.53.082301.113146,LiMa_2014,BaronPeters_RC_ARPC_2015,BanushkinaKrivov_2016,Tuckerman2010}. In such cases, CV is synonymous with reaction coordinate. However, finding CVs that approximate the committor function is a very challenging task and rarely done in practice, though several recent methods show promise in achieving this~\cite{LiLinRen_2019,MoriMatubayasi_2020,PalacioRodriguez_Pietrucci_2022,JungHummer_2021,FrassekBolhuis_2021,WuMa_2022}. Instead, CVs are generally chosen using chemical and physical intuition, or by using methods that aim to automatically extract CVs from simulations. While finding these ``good'' \hyperlink{ref:CV} {CV}s is crucial for the success of CV-based methods, we will only briefly touch on the various methods that exist for identifying such CVs. The reader is referred to~\cite{WANG2020139, doi:10.1080/00268976.2020.1737742, doi:10.1021/acs.jctc.0c00355} for a more extensive discussion of this issue.

\subsection{Scope}
The concept of accelerated sampling is so broad that we must make some decisions as to what scope of approaches to cover in a coherent review. 
Some enhanced sampling schemes are purely exploratory, i.e. they enable to discover uncharted regions of the configuration space efficiently but only allow semi-quantitative estimates of probability distributions. Many schemes, in addition, enable the estimation of probability \hyperlink{ref:Distribution} {distribution}s and free energies from the sampled space. Thus, all methods fall onto a exploration / free energy estimation tradeoff continuum. Some methods, such as metadynamics, were initially proposed as simply exploration schemes for configuration space, but were later refined and shown to be useful to calculate probability densities as well.  Others started off from their derivation being grounded in the estimation of free energies. An entire class of methods remains exclusively useful to broadly survey the configuration space, and learn to do that in optimal ways using \hyperlink{ref:Adaptive} {adaptive} schemes~\cite{ChiavazzoE5494}. In this review, we focus instead on algorithms that recover original statistical \hyperlink{ref:Ensemble} {ensembles} and free energy \hyperlink{ref:FES} {landscape}s. Note that exploratory methods can nevertheless be useful to some of these schemes, as they offer initial configurations to start simulations from.

Another type of information that can be valuable to the scientist is the evolution of the system along time, which gives access to the rates of transition between states. It is therefore important to note that many enhanced sampling schemes do not preserve the kinetics of the system, and are therefore primarily useful to recover equilibrium probability \hyperlink{ref:Distribution} {distribution}s. In fact, the most efficient sampling methods arguably alter the dynamics as much as possible while preserving thermodynamics. Some methods do preserve kinetic information, or at least allow kinetic information to be recovered, and we will generally note which methods which have this property.

New algorithms are constantly proposed to increase exploration, and/or reduce the variance of the estimates on the conformational landscape. These usually combine several of the original strategies in advantageous ways. However, it is often difficult to compare these schemes, or even to simply understand their similarities and differences, due to the proliferation of acronyms, and the use of inhomogeneous and diverse notation schemes. One of the main aims of this review is to thus list and describe the basic methods that are based on leveraging a single statistical or physical principle, using a unified framework to orient the MD simulation practitioner in the forest of available schemes.

With this in mind, we find useful to summarize the scope of this review.

\begin{itemize}
    \item We focus on methods of interest to chemical and biological systems, soft matter systems, and other molecular systems amenable to molecular dynamics simulations.
    \item We describe methods for accelerated sampling of a given probability \hyperlink{ref:Distribution} {distribution} at equilibrium.
    We do not review purely exploratory methods that cannot be used to recover equilibrium statistics.
    \item We do not describe methods that are mainly used to characterize kinetic rates. See instead Refs~\cite{BRUCE20181,doi:10.1146/annurev-physchem-042018-052340,Dickson:2017:1568-0266:2626,10.1021/acs.biochem.8b00977,https://doi.org/10.1002/wcms.1455,Kieninger2020} for reviews of methods to estimate kinetics from enhanced sampling schemes.
    \item We do not exhaustively review path sampling algorithms, though some of the pathway-based methods that allow to recover statistical ensembles are mentioned. See instead Refs~\cite{DellagoBolhuis,CHONG201788,Peters2017,Elber2020}
    for reviews of methods for finding and sampling pathways linking states.
    \item We do not review methods that aim to extract \hyperlink{ref:CV} {collective variable}s from simulations. See instead Refs~\cite{WANG2020139,doi:10.1080/00268976.2020.1737742,doi:10.1021/acs.jctc.0c00355} for recent reviews of the topic. When relevant, we assume that any necessary collective variables are already known and specified.
\end{itemize}

Given the large scope of this project, the current version of this review is necessarily incomplete. Because of LiveCoMS' unique updating process, we look forward to including reader suggestions and contributions in later versions of this perpetually updated review. Such updates will include major methods that were unintentionally missed, better ways to explain methods, combinations of methods that escaped our notice, or additional ways or organizing and structuring the classification of theoretical ideas presented here. We encourage contributors to post their suggestions for improvements as issues on the GitHub repository at \url{https://github.com/jhenin/Methods-for-enhanced-sampling-and-free-energy-calculations}.

\subsection{An attempt at classification}

Several mathematical and physical principles have been recognized as useful to enhance the sampling and converge the probability distribution with a low variance.
From probability theory, several strategies have been borrowed: \hyperlink{ref:IS}{importance sampling} (carrying out a biased simulation to reduce the variance of the estimated property), localization (restricting the system to the sampling of a specific region of space), and conditioning (using conditional probabilities), for example. Originating more from the physics community, several levers have been proposed: sampling at higher temperatures, adding external forces or potentials, driving an \hyperlink{ref:AdiabaticDyn} {adiabatically} decoupled degree of freedom or expanding the ensemble considered, with or without exchanges between systems sampling a different but related configuration space.

\begin{figure*}[!htb]
\includegraphics[scale=0.85]{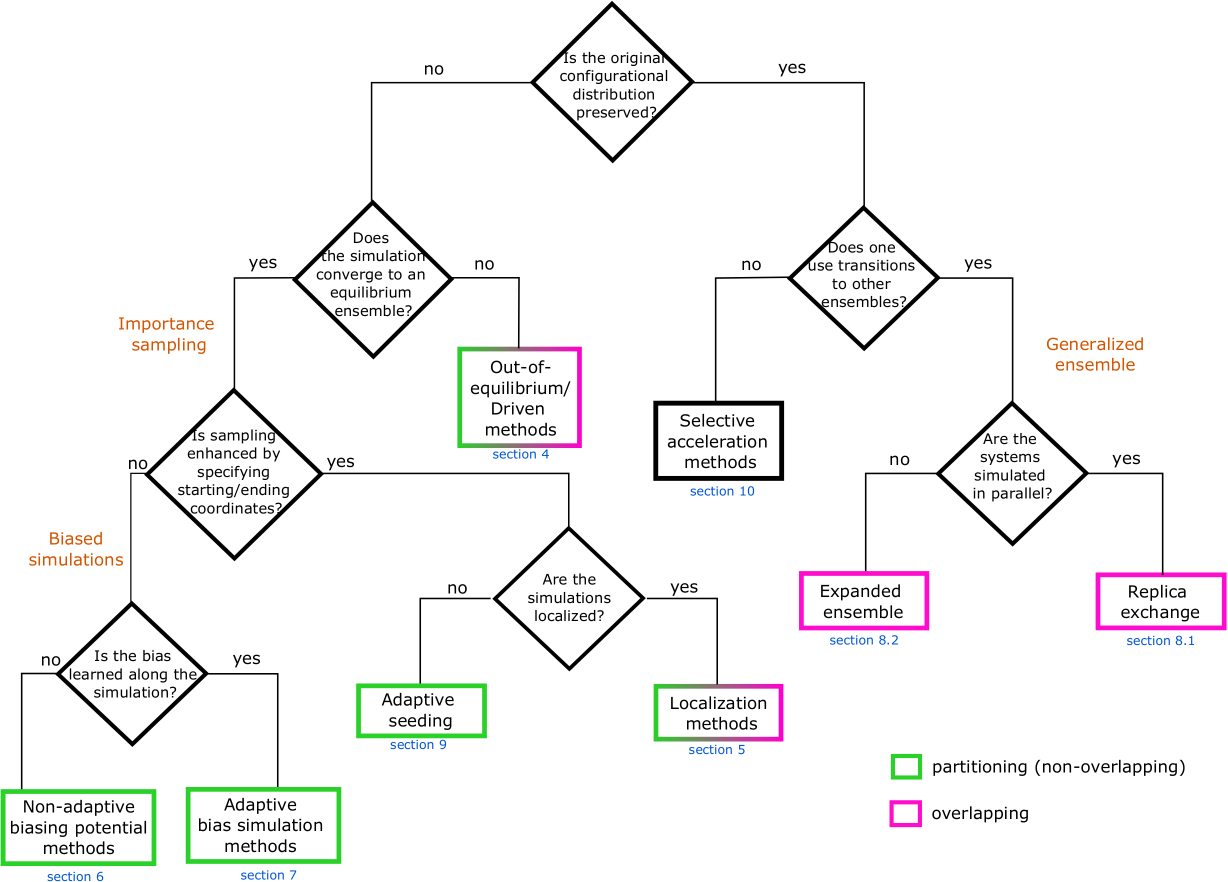}
  \caption{An attempt at classifying enhanced sampling schemes, answering yes/no questions that delineate various strategies based on physical or statistical principles (black diamonds). The sorting algorithm results in eight different classes of methods (boxes). These methods can further be sorted according to other classification schemes. An example is given by the division of methods into overlapping and partitioning schemes, highlighted by the coloring of the boxes. The section of the review describing the family of methods is shown in blue below the corresponding box. Labels in orange refer to families of methods that can be grouped under an umbrella term.}
  \label{fig:scheme}
\end{figure*}

\begin{figure}[!htb]
  \includegraphics[width=0.99\columnwidth] {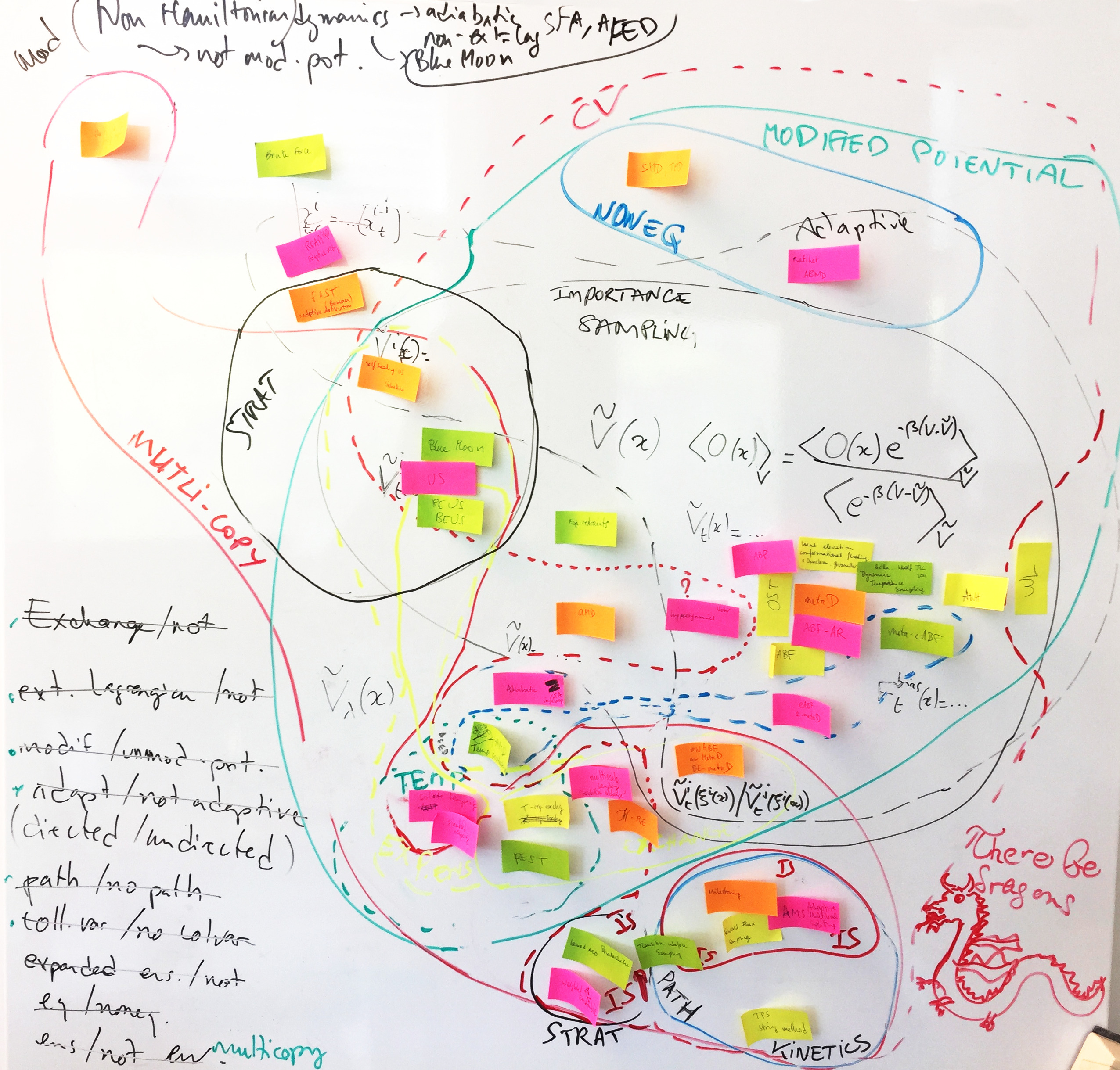}
  \caption{Early attempt at listing and classifying existing enhanced sampling schemes.}
  \label{fig:VennD}
\end{figure}

Crucially, most methods hinge on a relatively small number of statistical or physical ideas or principles. Here, we offer one possible way to organize these methods. We  start by explicitly listing these methodological ingredients, and classify the methods according to dichotomies, asking which of the different ingredients is leveraged (diamonds in Figure~\ref{fig:scheme}). The decision tree in the figure presents one of the possible alternatives by answering these different questions in a semi-arbitrary order, but attempting to list the questions from the most fundamental to the most specialized ones. The ordering of the questions inherently contains a level of subjectivity and we recognize that ordering these dichotomies in different ways can be equally reasonable, and that other representations may be useful (See Figure~\ref{fig:VennD} for an early attempt at a Venn Diagram).

There are certainly other binary methods of classification that could be used which are not immediately obvious in the decisions tree in Figure~\ref{fig:scheme}. For example, there are two main ways to impart additional structure to a system in a way that can that can aid in enhanced sampling: 
\begin{itemize}
\item One of these ways is to create partitions of configuration space that are completely non-overlapping; this can be done by creating ''level sets'' of a \hyperlink{ref:CV} {collective variable}, which all have the same value of some function of the coordinates, such as all the configurations with the same energy, or same distance between two specified particles, or same dihedral torsion between four particles. Collective variables generally should be defined on a contiguous region so that all values of interest can be visited. 
\item The second main way to add structure to a system is to create new ensembles that all share the same configurations, but for which their probabilities in each ensemble are different. For example, each separate ensemble could have a different temperature, meaning that configurations of different energies will have different probabilities in each ensemble. Alternatively, different ensembles could have a harmonic bias centered around a different point in \hyperlink{ref:CV} {CV} space for each ensemble, creating ensembles that are centered around different values of the CV. Generally, to be useful, the different ensembles must be interconnected by sharing at least some configurations that have non-negligible probabilities in multiple ensembles.  Each ensemble need not overlap with \emph{all} other ensembles, but there must be a interconnected network such that one can move stepwise between all the ensembles. 
\end{itemize}
This division into non-overlapping (partitioning) and overlapping structure is a fundamental one, because the algorithms used to calculate free energies and perform sampling, are different in the two cases. This is because different algorithms are needed depending on whether or not \hyperlink{ref:Microstate} {microstates} have defined probabilities in multiple ensembles or each microstate only belongs to one ensemble~\cite{Escobedo_unified_2005,abreu_framework_2006}. For example, when moving between ensemble members in the overlapping case, one generally uses Monte Carlo methods to generate moves to neighboring states. In the non-overlapping case, one usually uses standard dynamics to change between values of collective variables, but with the effect of any biases as a function of collective variables back-calculated to determine the resulting forces on the system. This particular division of methods is one that is not directly used in our classification, but shows up repeatedly in the descriptions of the algorithm. We have thus indicated this division using a green (partitioning) and purple (overlapping) coloring scheme in Figure~\ref{fig:scheme}.

Another complementary classification might to be consider the probability \hyperlink{ref:Distribution} {distribution} sampled at convergence by the enhanced sampling methods (sometimes called \hyperlink{ref:targetdist}{target distribution}), instead of the specifics of how the enhanced sampling methods achieve this in practice~\cite{invernizzi2020unified}.




The next sections provide the background information needed to understand the scheme presented in Figure~\ref{fig:scheme}, starting with the notation used throughout the paper and a glossary in Section~\ref{sec:Notion_Notation}, a description of the free energy estimators that are used in various enhanced sampling schemes and will be referred to later in the text in Section~\ref{sec:fe_estimators}, and followed by the more detailed description of the various families of enhanced sampling methods that emerge from our classification. We then list a selection of hybrid schemes that combine different principles. Finally, we summarize the software packages (MD simulation codes and libraries) that are publicly and openly available to run these different types of enhanced sampling simulations.

\subsection{Can enhanced sampling methods be compared critically?}
\label{sec:critical_comparison}

The efficiency of an enhanced sampling method depends on:
\begin{itemize}
 \item the application;
 \item the choice of parameters---and the optimal parameters are themselves application dependent;
 \item the expertise of the user.
\end{itemize}

Comparing methods requires defining a benchmark for which the results are known to a high accuracy and precision, and then attempting to study this benchmark by using each method ``fairly'', that is, either using the optimal parameters for the particular application, or the best parameters that a typical user will be able to set in practice. Such studies~\cite{Rizzi:JCAMD:2020a,Hrustak:JCP:2018} are challenging to carry about, and usually can only cover a few methods at a time.  A fundamental issue is that the ground truth is seldom known beyond simple systems such as alanine dipeptide or a few fast-folding miniproteins, and there is no guarantee that methods that work well for those systems will work well on more complex systems. In fact, the features of the conformational landscape might be very different and make methods that work well on model systems perform particularly badly on "real" systems. In addition, sometimes, a method may be less efficient than another when used optimally, but more robust to non-optimal circumstances.
A method may lead quickly to an approximate or qualitative result and go no further, whereas another may guarantee a precise result but require much more resources or take an uncertain time to converge.

For the current time, we have found that it would not be feasible to identify optimal methods for all applications within this review, as optimality indeed depends on the type of problem. Instead, we hope that presenting methods in a unified way can help guide the practitioner in their choice of enhanced sampling scheme when tackling their problem of interest.

\section{Useful notions and notations}
\label{sec:Notion_Notation}

One of the aims of this review is to use consistent notations to enable the reader to compare different methods and find similarities and differences across enhanced sampling
schemes. We introduce here the notation we will use throughout the paper. Note that given the existence of different notations in the literature, we have chosen the following one, while recognizing the validity of others. When especially crucial to understand the cited literature, we sometimes explicitly mention an alternative notation in following sections.

\subsection{Basic notations}
\label{sec:Notation}
\begin{itemize}
\item Cartesian coordinates of atoms (or coarse-grain particles) denoted by $\vx \in \mathbb{R}^{3N}$ and momenta denoted by $\vp \in \mathbb{R}^{3N}$, where $N$ is the number of particles.
\item Hamiltonian $H(\vx,\vp)=U(\vx)+K(\vp)$ that is a sum of the potential energy function denoted by $U(\vx)$, defined by a classical molecular mechanics force field, various levels of electronic structure theory in an \emph{ab initio} dynamics framework, or a QM/MM hybrid and the kinetic energy $K(\vp)$.
\item Force on particles $\vF(\vx) = -\nabla_\vx U(\vx)$, where $\nabla_\vx$ is the gradient with respect to $\vx$.
\item Absolute temperature $T$, inverse temperature $\beta = (k_\mathrm{B} T)^{-1}$.
\item \hypertarget{ref:AuxVar} {Extended (auxiliary) variable}: $\lambda$. This auxiliary variable can be a thermodynamic parameter, such as the temperature $T$ or the pressure $P$, or a parameter of the energy function $U_\lambda(\vx)$. The auxiliary variable can have a fixed value per simulation, can follow a pre-determined schedule, or can obey some dynamical equation of motion. It can have multiple dimensions, which we represent by the bold-faced vector $\boldsymbol{\lambda}$.

\end{itemize}

\subsection{Glossary of essential notions}
\label{sec:glossary}

\hypertarget{ref:Microstate} {\paragraph{Configuration or microstate}}
\index{configuration}
\index{microstate}
A single spatial arrangement of particles, represented by coordinates $\vx$. The set of possible configurations defines the configuration space $\Gamma$.
Configuration space augmented with the momenta variables is called the phase space, and therefore has twice the dimensionality. In many methods and calculations, we can take advantage of the fact that the momentum distribution obeys the analytical Maxwell-Boltzmann distribution at equilibrium to compute this quantity analytical rather than using sampling methods.  In many cases this contribution will be the same at both endpoints, and thus cancels out of the overall calculation.
In some cases, the  definition of configuration $\vx$ also includes the cell vectors of a periodic system (for example, in an isobaric \hyperlink{ref:Ensemble} {ensemble}).  This is used below where applicable.

We are using the original statistical mechanical definition of a microstate; we note that in the Markov State Modeling (MSM) literature, a microstate can also refer to an ensemble of configurations grouped together according to one or a set of order parameters, which is not intended here.

\hypertarget{ref:MolecularDynamics}{Molecular dynamics simulation}
\index{molecular dynamics}
A process that generates a trajectory, or sequence of configurations $\vx_t$.
The best-known classical dynamics is Hamiltonian dynamics:
\begin{equation}
\left\{
\begin{array}{ll}
    d\vx &= M^{-1} \vp \,  dt \\
    d\vp &= -\nabla_\vx U(\vx) \, dt
\end{array}
\right.
    \label{eq:md}
\end{equation}
where $M$ is the diagonal mass matrix.
A simplistic, discrete-time version of the above with time step $\delta t$ is:
\begin{equation}
\left\{
\begin{array}{ll}
    \delta \vx &= M^{-1} \vp \, \delta t \\
    \delta \vp &= -\nabla_\vx U(\vx) \, \delta t
    \label{eq:md_discrete}
\end{array}
\right.
\end{equation}
In practice, however, trajectories are often generated numerically using Verlet-style integrators.

\hypertarget{ref:Ensemble} {\paragraph{Statistical ensembles from molecular dynamics}}
\index{ensemble}
Hamiltonian dynamics conserves mechanical energy, and can be used to sample from the microcanonical (constant number of particles $N$, constant volume $V$ and constant energy $E$, NVE) ensemble under an \hyperlink{ref:ergodic} {ergodicity} assumption.

Isothermal dynamics includes modifications that function as a thermostat, simulating the equilibrium with an external temperature bath at a target temperature T.
As a result, as long as the simulation is at or near equilibrium, its temperature fluctuates around T, and it samples the canonical ensemble (NVT).
An example of isothermal dynamics is \hypertarget{ref:Langevin} {Langevin dynamics}:

\begin{equation}
\left\{
\begin{array}{ll}
    d\vx &= M^{-1} \vp \,  dt \\
    d\vp &= \left(-\nabla_\vx U(\vx) - \gamma \vp \right) dt
    + \sqrt{ \frac{2 \gamma M}{\beta}} \; d\mathbf{W}_t
\end{array}
\right.
\end{equation}
where $\gamma$ is a friction coefficient and ${W}_t$ is a Brownian motion in dimension $3N$.

To make the difference between Hamiltonian dynamics and Langevin dynamics more intuitive, consider this discrete-time approximation of Langevin dynamics:
\begin{equation}
\left\{
\begin{array}{ll}
    \delta\vx &= M^{-1} \vp \, \delta t \\
    \delta\vp &= \left(-\nabla_\vx U(\vx) - \gamma \vp + \sqrt{ \frac{2 \gamma M}{ \beta \delta t}} \, \mathbf{G} \right) \delta t
    \label{eq:Langevin_discrete}
\end{array}
\right.
\end{equation}
where $\mathbf{G}$ is a Gaussian-distributed stochastic 3N-vector of zero mean and variance 1.
Note that in practice, more sophisticated discrete Langevin integration schemes are used, which bring much better accuracy, stability, and performance~\cite{Skeel2002, Leimkuhler2012}.
Still, comparing the relatively simple Equations~\ref{eq:md_discrete} and \ref{eq:Langevin_discrete} shows that Langevin dynamics can be interpreted intuitively as similar to Hamiltonian dynamics, but including a modified force with added friction and stochastic collision terms. When $\gamma$ is zero, it reduces exactly to Hamiltonian dynamics.
Langevin dynamics will be used as a basic example in later sections of this review.

\hypertarget{ref:Distribution}{\paragraph{Distribution}}

The Boltzmann distribution (which characterizes the canonical ensemble) in phase space has the following probability density:
\begin{equation}
\mu(\vx, \vp) = \frac{1}{Q} e^{-\beta (U(\vx) + K(\vp))}
\label{eq:BoltzmannDistr}
\end{equation}
where $ \displaystyle Q = \int e^{-\beta (U(\vx) + K(\vp))} d\vx d\vp$ is the normalization factor, known as partition function.

The physical meaning of $\mu$ is a probability per unit volume of $(\vx, \vp)$ space. The probability of a region of phase space $\Sigma$ is:
\begin{equation}
    P(\Sigma) = \int_\Sigma \mu(\vx, \vp) \, d\vx d\vp
\end{equation}

In most cases, the energy is a sum of a potential term that depends only on positions and a kinetic term that depends only on momenta, as written in Eq.~\ref{eq:BoltzmannDistr}.
Then the momenta are statistically independent from the system configuration, hence their distribution is that of the ideal gas and does not bear significant information on any specific system.
This leads to a simple expression for the configurational distribution, where the momenta and kinetic energy do not appear:
\begin{equation}
\nu(\vx) = \int \mu(\vx, \vp) \, d\vp = \frac{1}{Z} \, e^{-\beta U(\vx)}
\end{equation}
where $Z$ is the configurational partition function, $ Z = \int e^{-\beta U(\vx)} \, d\vx$ . Sometimes, it may also be useful to define an unnormalized version of the configurational distribution, $q(\vx)$, such that $\nu(\vx) = \frac{1}{Z}q(\vx)$. There exist equivalent definitions of distributions for the isothermal-isobaric ensemble (NPT), which can be found in most statistical mechanics books~\cite{Zuckerman2010, Tuckerman2010}.

Note that there are other notation conventions: in some texts and papers, $Q$ denotes the configurational partition function and $Z$ denotes the configurational and momenta partition function.

\hypertarget{ref:Macrostate} {\paragraph{Macrostate}}

Macrostates are experimentally distinguishable or measurable states of a system.
They can be described formally either in terms of the thermodynamic state variables ($E$, $T$, $P$, $V$, or parameters of the Hamiltonian) or by specifying specific regions of configuration space (that is, disjoint sets of \hyperlink{ref:Microstate} {microstates}). A macrostate, besides being just a collection of microstates, also specifies a probability associated with each microstate that is contained in the microstate. The term ``thermodynamic state" is often used synonymously with macrostate, as the macrostates that we are most generally interested in studying with molecular simulation are macrostates that are completely defined by the specification of the macroscopic thermodynamic variables.

\hypertarget{ref:density_of_states}{\paragraph{Density of states}}
The density of states $\Omega(E)$ is defined as the number of states in a system that have a specific total energy $E$.  It can be mathematically expressed as $\Omega(E) = \int \delta(E-H(\vx,\vp))\, d\vx d\vp$, where $\delta$ is a Dirac delta function that is zero everywhere except where $E=H(\vx,\vp)$. The entropy $S$ as a function of $E$ is then simply $S(E)=k_\mathrm{B} \ln \Omega(E)$. Additionally, the configurational density of states is defined as the number of states that have a specific potential energy $U$ and is computed as $\Omega(U) = \int \delta(U-U(\vx))\, d\vx$, and the configurational entropy $S$ as a function of $U$ is $S(U)=k_\mathrm{B} \ln \Omega(U)$.

\hypertarget{ref:FE} {\paragraph{Free energy}}
In the canonical \hyperlink{ref:Ensemble} {ensemble}, the Helmholtz free energy $F$ is a property of a \hyperlink{ref:Macrostate} {macrostate} of a system, and is proportional to the logarithm of its partition function, which measures its statistical weight compared to other \hyperlink{ref:Macrostate} {macrostate}s:
\begin{equation}
F \propto -\beta^{-1} \ln Z_{\Sigma} = -\beta^{-1} \ln \int_\Sigma e^{-\beta U(\vx)} \, d\vx,
\end{equation}
where the integration is done over a subset $\Sigma$ of configuration space corresponding to the \hyperlink{ref:Macrostate} {macrostate}. $F$ thus depends on $\Sigma$, $U(x)$, and $\beta$, although this dependency is often not stated explicitly, but implied by the context.

When using a classical energy function, $F$ is only defined up to an arbitrary additive constant.
In practice, this is not a limitation, as quantities of measurable physical interest involve only free energy \textit{differences} between two \hyperlink{ref:Macrostate} {macrostate}s, rather than absolute free energies.
If two macrostates $\A$ and $\B$ can be distinguished experimentally, the ratio of the time they spend in each system ($P_A = Z_A/Z$ and $P_B = Z_B/Z$) is an experimental observable, and is the free energy difference between the two states:
\begin{align}
  \Delta F_{\A,\B} &= F_{\A} - F_{\B} \nonumber\\
    & = -\beta^{-1} \ln \frac{Z_{\A}}{Z_{\B}} \nonumber\\
  & = -\beta^{-1} \ln \frac{P_{\A}}{P_{\B}}
\end{align}

The Helmholtz free energy is sometimes notated $A$ in the literature. In this review, we will use $F$ for Helmholtz free energy, and the the symbol $A$ will be used for the \hyperlink{ref:FES} {free energy surface}. Gibbs free energy $G$ is the equivalent quantity in the isothermal-isobaric \hyperlink{ref:Ensemble} {ensemble}.

\hypertarget{ref:FEestimator} {\paragraph{Free energy estimator}}
An expression or algorithm that takes simulation data and estimates a numerical value for free energies or their differences. See Section~\ref{sec:fe_estimators} for a list and description of useful free energy estimators.

\hypertarget{ref:reduced} {\paragraph{Reduced quantities for homogeneous treatment of different ensembles}}

We define the reduced energy function $u_i(\vx)$
for state $i$ to be
\begin{eqnarray}
u_i(\vx) &=& \beta_i ( U_i(\vx) \;
+ p_i V(\vx)) \label{equation:reduced-energy}
\end{eqnarray}
where the pressure-volume term $p_i V(\vx)$ is only included in the case of a constant pressure \hyperlink{ref:Ensemble} {ensemble}.
Other terms, such as chemical potentials, may be added to generalize to other ensembles.
For each state $i$, $\beta_i$ is the inverse temperature, $U_i(\vx)$ the potential energy function (which may include external \hyperlink{ref:biasingE} {biasing} potentials), $p_i$ the external pressure.
This formalism allows a very large number of different
situations to be described by the same mathematics.

The reduced free energy $f$ is defined as $f = \beta F$ for the canonical ensemble,
or $f = \beta G$ for the isothermal-isobaric ensemble.
Then all \hyperlink{ref:Distribution} {Boltzmann-like distributions} for the thermodynamic ensembles mentioned above are given in their un-normalized form as $q(\vx) = e^{-u(\vx)}$ and normalized form as
  $\nu(\vx) = e^{f-u(\vx)}$.

\hypertarget{ref:CV}{\paragraph{Collective variable (CV)}} A function $\xi$ mapping the full $n$ dimensions configurations $\vx$ to a lower-dimensional representation $\vz$ (sometimes denoted as $\mathbf{s}$ in the literature):
\begin{equation}
\vz = \xi(\vx)
\end{equation}
In the literature, the letter $\xi$ is sometimes used for both the function and the variable. The same goes for $\vz$ (and $\mathbf{s}$).

The multi-dimensional case $\xi= (\xi_1,\xi_2,\ldots,\xi_d)$ can be described either as a single vector CV or a family of scalar CVs:
\begin{equation}
\vz = (z_1, z_2, \ldots, z_d) = \xi(\vx),
\end{equation}
where $d$ is the number of scalar collective variables (i.e., the dimension of the CV space), with $d\ll 3N$.

\hypertarget{ref:DimRed} {\paragraph{Dimensionality reduction}}
The process of finding functions $\xi$ mapping high-dimensional $\vx$ to low-dimensional $\vz$.

\hypertarget{ref:Alchemical} {\paragraph{Alchemical transformation}}
Non-physical parameters are characterized as ``alchemical'', and describe transition between Hamiltonians representing different molecular systems. Typically the change consists in either transforming a molecule into another one, or decoupling a molecule from its environment. Alchemical transformations are often used to estimate free energies of binding or solvation, as they create a continuous pathway of intermediate ensembles between two end states of physical interest.

\hypertarget{ref:FES} {\paragraph{Free energy profile / landscape / surface (FES)}}
\label{sec:FES}
While \hyperlink{ref:FE} {free energy} can be expressed as the logarithm of a partition function, a free energy surface (FES) is the logarithm of a \textit{partially integrated partition function}.
Given a chosen set of collective variables $\vz = \xi(\vx)$, this partially integrated partition function is, up to a normalization factor, the marginal probability density $\rho(\vz)$, and is obtained by integrating the Boltzmann density over all variables except $\vz$ (at constant $\vz$):
\begin{equation}
\label{eq:fes_definition}
    A(\vz) = -\beta^{-1} \ln \int
    \delta\left(\vz-\xi(\vx)\right) \, \nu(\vx)\, d\vx
    = -\beta^{-1} \ln \rho(\vz) ,
\end{equation}
where $\delta$ indicates the multivariate Dirac delta distribution defined by $\delta\left(\vz-\xi(\vx)\right)= \prod_{i=1}^{d} \delta\left(z_i-\xi_i(\vx)\right)$ with the single variable Dirac delta distribution defined by $\int f(x) \, \delta(x-y) \,  dx = f(y)$, for any function $f$. Intuitively, $\delta$ behaves like a function that is zero everywhere but in 0, and whose integral is 1.

In practice, most free energy surface calculations concern a Helmholtz free energy, but Gibbs free energy surfaces could be computed as well. The difference is small, unless the transformation of interest entails a measurable change in overall density.

If two \hyperlink{ref:Macrostate} {macrostate}s $\A$ and $\B$ correspond to domains of \hyperlink{ref:CV} {collective variable} space, then the probability ratio can be obtained from a free energy surface $A(\vz)$ by integrating its exponential (the associated density) over the corresponding domains:
\begin{align}
  \Delta F_{\A,\B} &=
  -\beta^{-1} \ln \frac{Z_\A}{Z_\B}\\
  & =  -\beta^{-1} \ln
  \frac{\int_\A e^{-\beta A(\vz)} \, d\vz}
  {\int_\B e^{-\beta A(\vz)} \, d\vz}.
\end{align}

Similarly to other \hyperlink{ref:reduced} {reduced} quantities, one may define the reduced free energy surface
\begin{equation}
    a(\vz) = \beta A(\vz)
\end{equation}

Note that in \hyperlink{ref:ExpEns} {expanded ensemble} approaches, a free energy as a function of the extended variable $\lambda$ can be defined as:
\begin{align}
    A(\lambda) &= -\beta^{-1} \ln Z_\lambda
    \nonumber \\
    &= -\beta^{-1} \ln \int
    \nu(\vx, \lambda)\, d\vx
\end{align}
Note the difference between this definition and that of a free energy surface as a function of a collective variable (Equation~\ref{eq:fes_definition}).
Here there is no Dirac distribution $\delta$ because $\vx$ and $\lambda$ are separate variables, so we obtain the marginal distribution by integrating over $\vx$, which preserves the dependence on $\lambda$. In other words, a slice of configuration space at constant $\lambda$ is the complete space of $\vx$ coordinates, whereas a slice of configuration space at constant $\vz = \xi(\vx)$ is only  a slice or subset of this coordinate space.

There is a very general relation between thermodynamic quantities (with the dimension of an energy) and probabilistic quantities.
The latter can be expressed as minus thermal energy ($-k_\mathrm{B} T = -\beta^{-1}$) times the natural logarithm of the former: this process is called Boltzmann inversion.
The main quantities discussed above are summarized in Table~\ref{tab:quantities}.

\begin{table}[]
\small
    \centering
\begin{tabular}{c|c}
probabilistic quantity & thermodynamic quantity  \\
$\bullet$ &   $-\beta^{-1} \ln(\bullet)$ \\
\hline
probability density $\nu(\vx)$  & potential energy  $U(\vx)$\\
$\downarrow$ &  \\
\textit{integrate over $\vx$ at constant $\vz$} & \\
$\downarrow$ &  \\
marginal probability density $\rho(\vz)$ & free energy surface $A(\vz)$ \\
$\downarrow$ &  \\
\textit{integrate over $\vz$ in $\Sigma$} & \\
$\downarrow$ &  \\
probability (measure) $P(\Sigma)$  & free energy  $F(\Sigma)$\\
\end{tabular}
    \caption{Relations between key statistical and thermodynamic quantities.
    The right column is $-\beta^{-1}$ times the logarithm of the left column.
    Probabilistic quantities are related by successive integration over larger slices of configuration space.}
    \label{tab:quantities}
\end{table}

The free energy surface can be interpreted as an effective potential energy surface defined on collective variables.
The free energy surface is related to a probability density in the same way a free energy is related to the probability of a state (Table~\ref{tab:quantities}).
Beware, however, that a probability density is not a probability measure: a value of the density is not the probability of any particular event.
Probability values are unitless and between 0 and 1; in contrast a probability density has units inverse volume, and can take values greater than 1. The probability of a state is obtained from a probability density by integrating over the relevant region of configuration space. In a continuous configuration space, the probability of each individual configuration is zero.
A density is normalized: its integral over the whole space is 1, which is the probability of the whole space.
Similarly, a free energy surface is not directly interpretable as a macroscopic free energy. Unlike a free energy, it is not an experimental observable.
Very importantly, free energy surfaces do not generally have a simple interpretation in terms of dynamics (e.g. free energy maxima may not be kinetic barriers), because of distortions due to the nonlinear geometry of the variables~\cite{10.1063/5.0020240,Dietschreit2022}.

\hypertarget{ref:PMF} {\paragraph{Potential of mean force (PMF)}}
Beware that this phrase is used in the literature in two different, incompatible meanings. The colloquial meaning today is the \hyperlink{ref:FES} {free energy surface} as defined above.  However, this can sometimes be misleading. The historic notion of PMF is used to describe the structure of simple liquids.
The potential of mean force $W(r)$ between two particles was defined based on its radial distribution function $g(r)$. The RDF is calculated from the probability density for the interparticle distance $r$, $\rho(r)$, divided by a normalization term (proportional to $r^2$) that makes it constant, conventionally equal to 1, at large distances in a homogeneous fluid.
\begin{align}
    W(r) &= -\beta^{-1} \ln g(r) \nonumber \\
    &= -\beta^{-1} \ln \frac{\rho(r)}{r^2} + C  \nonumber \\
    &= A(r) + 2 \beta^{-1} \ln(r) + C,
    \label{eq:pmf_fes}
\end{align}
where $A(r)$ is the \hyperlink{ref:FES} {FES} along the interparticle distance and $C$ is an arbitrary constant.
The historic PMF and FES are therefore related, but distinct quantities.

The phrase ``potential of mean force'' describes very literally that $W(r)$ is a potential arising from an average force that would act on a particle at that location. Crucially, the potential of mean force is zero in non-interacting systems: if $U(\vx) = 0$ for all $\vx$ then $W(r) = 0$ for all $r$, which is not generally the case for \hyperlink{ref:FES} {free energy surfaces} of nonlinear CVs, due to a \textit{Jacobian} term describing a purely geometric entropy.
For details, refer to Section~\ref{sec:fe_estimators:TI}.

Here again, one may define the \hyperlink{ref:reduced} {reduced} version of $W(r)$:
\begin{equation}
    w(r) = \beta W(r)
\end{equation}

\hypertarget{ref:multimodal} {\paragraph{Multimodal distribution}}

A probability \hyperlink{ref:Distribution} {distribution} featuring many disconnected regions of high probability (each local maximum is a \textit{mode}) separated by regions of low probability. Typically, if the dimension is high, most of the volume of configuration space has a very small probability, but there are often  a large number of significantly separated high probability regions.

\hypertarget{ref:ensemble_average} {\paragraph{Ensemble average}}

The ensemble average of an observable $O(\vx,\vp)$, which is a function of the phase space, is defined as:
\begin{align}
\langle O(\vx,\vp) \rangle &= \int O(\vx, \vp) \mu(\vx, \vp) \, d\vx d\vp
\end{align}

Or if $O(\vx)$ is only a function of configurations, then
\begin{align}
\langle O(\vx) \rangle &= \int O(\vx) \nu(\vx) \, d\vx
\end{align}

All measurable thermodynamic quantities of interest are equal to ensemble averages of some observable. For example, the total energy $E$ of a system is the expectation value of the Hamiltonian $\langle H \rangle$. Furthermore, the marginal probability density $\rho(\vz)$ is an ensemble average of a Dirac $\delta$ ``function'', $\rho(\vz)=\langle \delta\left(\vz-\xi(\vx)\right) \rangle$ (see Equation~\ref{eq:fes_definition}).  With finite sampling, expectation values have a measurable uncertainty. However, in the thermodynamic limit (when the number of samples approaches Avogadro's number), these can be assumed to be exact, as any uncertainties are on the order of $10^{-10}$ or smaller.  

\hypertarget{ref:ergodic} {\paragraph{Ergodic dynamics}}
The dynamics of a system is said to be ergodic if samples taken from any single, infinitely long trajectory describe the complete statistical properties of the dynamics.
This allows the estimation of \hyperlink{ref:ensemble_average} {ensemble averages} by the time average or (\textit{ergodic average}).
If we want to compute the average of observable $O(\vx, \vp)$ according to \hyperlink{ref:Distribution} {distribution} $\mu(\vx, \vp)$, and we can generate a discrete dynamics $(\vx_t, \vp_t)$ that is ergodic with respect to distribution $\mu$, then the \hyperlink{ref:ensemble_average} {ensemble average} $\langle O \rangle$ can be estimated as a time average:
\begin{align}
\langle O \rangle &= \int O(\vx, \vp) \mu(\vx, \vp) \, d\vx d\vp
\nonumber \\
&\approx \frac{1}{M} \sum_{t=1}^M O(\vx_t, \vp_t) \text{ for sufficiently large $M$,}
\end{align}
where $M$ is the number of samples. More precisely:
\begin{equation}
   \langle O \rangle = \lim_{M \to \infty} \frac{1}{M} \sum_{t=1}^M O(\vx_t, \vp_t)
    \label{eq:ergodic}
\end{equation}

Solution-phase molecular dynamics of small molecules is nearly always ergodic in practice (i.e in simulations of more than 10 ns), and many biological and soft materials problems are ergodic in the limit of sufficient samples: from here on, we generally assume the existence of ergodic dynamics in the systems to which the accelerated methods are applied, as long as enough sampling is performed.
However, ergodicity is a theoretical notion that characterizes asymptotic behavior over infinitely long times (the limit in Equation~\ref{eq:ergodic}).
In practice, molecular dynamics simulations are often very short compared to the longest relaxations times, so that trajectories do not explore the full configuration space.
This situation is described as ``quasi-nonergodicity''.
Enhanced sampling methods target precisely this case, and aim to recover the statistical properties of ergodic dynamics from trajectories of limited duration.

\hypertarget{ref:metastab} {\paragraph{Metastability and metastable states}}
When a system resides in some some regions of configuration space for long times, with rare transitions between those regions, those regions are called metastable regions or metastable states. Metastable regions are typically modes of a \hyperlink{ref:multimodal} {multimodal distributions}.

\hypertarget{ref:biasingE} {\paragraph{Biasing and biased energy}}
A biasing energy, also called bias energy, is an extra energetic term $U^\mathrm{bias}$ added to obtain a potential energy $\tilde{U}(\vx)$ biased to behave a certain way:
\begin{equation}
\tilde{U}(\vx) =  U(\vx) + U^\mathrm{bias}(\vx).
\end{equation}
In molecular dynamics, a bias energy
$U^\mathrm{bias}(\vx)$ gives rise to a bias force $\vF^\mathrm{bias}(\vx) = -\nabla_{\vx} U^\mathrm{bias}(\vx)$, so that the total biased force $\tilde{\vF}(\vx)$ is
\begin{equation}
\tilde{\vF}(\vx) = -\nabla_{\vx} \tilde U(\vx) = \vF(\vx) + \vF^\mathrm{bias}(\vx)
\end{equation}
The bias $U^\mathrm{bias}(\vx)$ is often a function of low-dimension \hyperlink{ref:CV} {collective variable}s $\vz = \xi(\vx)$, and can be written $U^\mathrm{bias}(\xi(\vx))$ so that in this case, the biased potential energy function is:
\begin{equation}
\tilde{U}(\vx) =  U(\vx) + U^\mathrm{bias}(\xi(\vx)).
\end{equation}
The resulting biasing force on atoms is calculated using the chain rule:
\begin{align}
\tilde{\vF}(\vx) &= -\nabla_{\vx} \tilde U(\vx) \\
&= \vF(\vx) -\nabla_{\vx} [ U^\mathrm{bias}(\xi(\vx))]\\
\tilde{\vF}(\vx)&= \vF(\vx) - \left . \frac{dU^\mathrm{bias}}{dz}\right |_{z=\xi(\vx)} \; \nabla_{\vx}\xi(\vx)
\end{align}
This requires the computation of the gradient $\nabla_{\vx}\xi(\vx)$ of the collective variable with respect to atomic Cartesian coordinates.

\hypertarget{ref:BiasedDist} {\paragraph{Biased configurational distribution}}
Under the influence of a biased potential energy $\tilde{U}(\vx) =  U(\vx) + U^\mathrm{bias}(\vx)$, the simulations will sample a biased configurational \hyperlink{ref:Distribution} {distribution} $\tilde{\nu}(\vx)$ given by
\begin{equation}
\label{eq:biased_configurational distribution}
\tilde{\nu}(\vx) = \frac{1}{\tilde{Z}} \, e^{-\beta \tilde{U}(\vx)}
\end{equation}
where $\tilde{Z} = \int e^{-\beta \tilde{U}(\vx)} \, d\vx = \int e^{-\beta\left [ U(\vx) + U^\mathrm{bias}(\vx) \right]} \, d\vx$ is the biased partition function. 
This can be re-written as 
\begin{align}
\label{eq:biased_x_distribtion_rewritten}
\tilde{\nu}(\vx) & = 
\frac{1}{\tilde{Z}} \, e^{-\beta\left [ U(\vx) + U^\mathrm{bias}(\vx) \right]} 
\nonumber \\ 
& =
\frac{1}{Z} \, e^{-\beta U(\vx)} \, \frac{Z}{\tilde{Z}} \, e^{-\beta U^\mathrm{bias}(\vx)} =
\nu(\vx) \, \frac{Z}{\tilde{Z}} \, e^{-\beta U^\mathrm{bias}(\vx)}, 
\end{align}
where 
\begin{align}
\label{eq:fraction_of_partition_functions}
\frac{Z}{\tilde{Z}} & = 
\frac{\int e^{-\beta U(\vx)} \, d\vx} {\int e^{-\beta \tilde{U}(\vx)} \, d\vx} = 
\frac{\int  e^{\beta U^\mathrm{bias}(\vx)} \, e^{-\beta \tilde{U}(\vx)} \, d\vx} {\int e^{-\beta \tilde{U}(\vx)} \, d\vx} = 
\nonumber \\ 
& =
\int  e^{\beta U^\mathrm{bias}(\vx)} \, \tilde{\nu}(\vx) \, d\vx =  
\langle e^{\beta U^\mathrm{bias}(\vx)} \rangle_{\tilde U},
\end{align}
where $\langle \cdots \rangle_{\tilde U}$ indicates an \hyperlink{ref:ensemble_average} {ensemble average} under the biased \hyperlink{ref:Distribution} {distribution} $\tilde{\nu}(\vx)$.

\hypertarget{ref:IS} {\paragraph{Importance sampling}}
\label{sec:importance_sampling} A family of methods where a separate, distinct probability \hyperlink{ref:Distribution} {distribution} $\tilde\nu(\vx)$ from the target one is sampled, but in such a way that the ratio of the two distributions is known or estimated numerically. Therefore, the \hyperlink{ref:targetdist} {target} probability and averages using the \hyperlink{ref:targetdist} {target} probability can be calculated.  Usually this is done to focus sampling on configurations that contribute more to any averages of interest.


The name refers to the idea of favoring sampling of the regions of importance, or if they are unknown, to flatten sampling towards a more uniform distribution. Frequently, the difference in probability is expressed in terms of some sort of \hyperlink{ref:biasingE} {biasing potential} $\tilde U(\vx)$. Importance sampling methods include the biasing potential and biasing force methods (\hyperlink{ref:Adaptive} {adaptive} or not), localization methods and adaptive \hyperlink{ref:Seeding} {seeding} methods described in Sections~\ref{sec:localization}, \ref{sec:biasing_potential}, \ref{sec:AdaptiveBiasSimulations} and \ref{sec:seeding}.

As explained below, unbiased properties of the original distribution may be obtained by \hyperlink{ref:Reweighting} {reweighting} from the modified, sampled distribution to the desired distribution.
In practice, the  sampled distribution is chosen to emphasize samples that contribute strongly to the averages of interest, making the reweighted average over samples from the modified distribution a low-variance estimator.

\hypertarget{ref:Reweighting} {\paragraph{Reweighting}}
Reweighting involves calculating averages and free energies of one \hyperlink{ref:Distribution} {distribution} using samples from another one, as occurs in \hyperlink{ref:IS} {importance sampling}, though it can be used in other situations as well. The distribution that one samples from  may be one explicitly generated by a biased simulation or performed at a different temperature, or one that is a mixture of several sampled distributions~\cite{reweighting_mixture_distribution}. 

If the sampled distribution is $\tilde \nu(x)$ and the original, unmodified distribution is $\nu(\vx)$, then the average of some observable $O(\vx)$ is calculated as 
\begin{eqnarray}
\langle O(\vx) \rangle &=& \int  O(\vx) \nu(\vx) d\vx \nonumber \\
&=& \int O(\vx) \frac{\nu(\vx)}{\tilde \nu(\vx)} \tilde \nu(\vx) d\vx\nonumber \\
&=& \left \langle O(\vx)  \frac{\nu(\vx)}{\tilde{\nu}(\vx)} \right \rangle_{\tilde \nu}
\end{eqnarray}
where the final average is taken from samples obtained from the modified distribution $\tilde \nu$, but the expectation is in the original distribution $\nu$.  This reweighting can be done effectively whenever the ratio $\frac{\nu(\vx)}{\tilde \nu(\vx)}$ doesn't vary much over the $\vx$ sampled. 

In the specific example of where $\tilde \nu(\vx)$ corresponds to a simulation with an added bias potential $U^{bias}$ (see Equation~\ref{eq:biased_x_distribtion_rewritten}), then the average of an observable $O(\vx)$ over the distribution $\nu(\vx)$ can be estimated by:
\begin{eqnarray}
\label{eq:reweighting_bias_potential}
\langle O(\vx) \rangle &=& \left \langle O(\vx) \frac{\tilde{Z}}{Z}e^{\beta U^\mathrm{bias}(\vx)} \right \rangle_{\tilde U} \nonumber  \\
 &=& \frac{\langle O(\vx) \, e^{\beta U^\mathrm{bias}(\vx)} \rangle_{\tilde U}}
{\langle e^{\beta U^\mathrm{bias}(\vx)} \rangle_{\tilde U}},
\end{eqnarray}
where the \hyperlink{ref:ensemble_average} {ensemble average}s are taken according to the \hyperlink{ref:BiasedDist} {biased distribution} arising from the biased potential energy $\tilde U(\vx)$. The term in the denominator is the exponential averaging estimate of $\frac{Z}{\tilde Z}$ (see Equation~\ref{eq:fraction_of_partition_functions} and Section~\ref{sec:fe_estimators:EXP} for more detail).

\hypertarget{ref:BiasedCVDist} {\paragraph{Biased CV distribution}}
Under the influence of a bias potential acting in \hyperlink{ref:CV} {CV} space $U^\mathrm{bias}(\vz)=U^\mathrm{bias}(\xi(\vx))$, the \hyperlink{ref:CV} {CV} will follow a biased \hyperlink{ref:CV} {CV} \hyperlink{ref:Distribution} {distribution} given by
\begin{equation}
\label{eq:BiasedCVDistrib}
\tilde \rho(\vz) = \frac{1}{\tilde{Z}} 
e^{-\beta \left[
A(\vz) + U^\mathrm{bias}(\vz)
\right]},
\end{equation}
where $\tilde{Z} = \int e^{-\beta\left [ U(\vx) + U^\mathrm{bias}(\xi(\vx)) \right]} \, d\vx = \int e^{-\beta\left [ A(\vz) + U^\mathrm{bias}(\vz) \right]} \, d\vz$ is the biased partition function.

\hypertarget{ref:targetdist}{\paragraph{Target distribution}}
Common in some enhanced sampling methods is the concept of a target distribution. This represents a desired probability \hyperlink{ref:Distribution} {distribution} that an enhanced sampling simulation is trying to achieve. The target distribution can be in the space of some collective variables (normally, the ones being biased) or another space such as the state space in $T$ or along an auxiliary variable $\lambda$. In some methods, the target distribution is set by the user. In others, the target distribution can be inferred from experimental observations. A common choice is to set the target distribution to be uniform over the variables or states of interest, with all values sampled equally. However, most methods theoretically support the usage of any given non-uniform target distributions if the user desires.

\hypertarget{ref:Replica} {\paragraph{Replicas}}
A replica is a copy of a molecular system. Replicas might simply be independent copies started from different random number seeds for velocities or different initial configurations, or they might each have a different value of some thermodynamic parameter like temperature or $\lambda$, or have a different biasing distribution. In many types of accelerated simulations, replicas can exchange information with each other, and this exchange is key to the success of the method.  In most methods, however, they do have the sane chemical composition and number of atoms. 

\hypertarget{ref:GenEns} {\paragraph{Generalized ensemble}}
Generalized ensemble methods encompass \hyperlink{ref:ExpEns} {expanded ensemble methods} and \hyperlink{ref:ReplEx} {replica exchange} methods (Section~\ref{sec:generalized-ensemble}).

\hypertarget{ref:ExpEns} {\paragraph{Expanded ensemble}}
While a usual statistical-mechanical ensemble describes one set of macroscopic conditions, an expanded (or extended) ensemble allows for additional degrees of freedom (physical or non-physical) to vary.  An expanded ensemble may be sampled mainly in two ways. The first option is to run a collection of simulations (replicas, $i$) among which an \hyperlink{ref:AuxVar} {auxiliary parameter} $\lambda_i$ takes a discrete set of values. In replica $i$, the dynamics of coordinates $\vx_i(t)$ are then propagated under a potential energy $U_{\lambda_i}(\vx_i)$ or at inverse temperature $\beta_i$.
The second option is to propagate $\lambda_i$ as an additional dynamic variable (see \hyperlink{ref:ExtL} {Extended Lagrangian}).

\hypertarget{ref:ExtL}{\paragraph{Extended Lagrangian dynamics, $\lambda$ dynamics}}
A special case of \hyperlink{ref:ExpEns} {expanded ensemble} simulation, whose equations of motion include \hyperlink{ref:AuxVar} {``fictitious'' (auxiliary) dynamical degrees of freedom} $\lambda$ that are not the spatial coordinates of physical objects or the associated momenta.
To sample the canonical \hyperlink{ref:Distribution} {distribution} of $(\vx, \lambda)$, they can be propagated following e.g. \hyperlink{ref:Langevin} {Langevin dynamics}:

\begin{equation}
\left\{
\begin{array}{ll}
d\vx &= M^{-1} \vp \, dt\\
d\vp &= \left(-\nabla_\vx U^\text{ext}(\vx,  \lambda) - \gamma \vp \right) dt
    + \sqrt{ \frac{2 \gamma M}{\beta}} \; d\mathbf{W}_t\\
d\lambda &= m_\lambda^{-1} p_\lambda \, dt \\
dp_\lambda &= \left(-\frac{\partial U^\text{ext}(\vx, \lambda)}{\partial \lambda} - \gamma_\lambda p_\lambda \right) dt
    + \sqrt{ \frac{2 \gamma_\lambda m_\lambda}{ \beta }} dW_t
\end{array}
\right.
\end{equation}

where $m_\lambda$ is a fictitious mass associated to $\lambda$.
While the two phrases are essentially synonymous, the term $\lambda$-dynamics is mostly used when a continuous dynamic variable $\lambda$ connects physically meaningful, and sometimes discrete, states, such as in \hyperlink{ref:Alchemical} {alchemical} transformations.
On the other hand, the term extended Lagrangian (or extended Hamiltonian) is more often used when the fictitious coordinate $\lambda$ follows a \hyperlink{ref:CV} {collective variable} $\xi(\vx)$, typically coupled through a harmonic potential: $U^\text{ext}(\vx, \lambda) = U(\vx) + \frac{k^\mathrm{ext}}{2}|\xi(\vx)-\lambda|^2$.
These methods are also referred to as ``extended-system dynamics''.  Note that much of this additional baggage can be avoided by making moves in $\lambda$ space using \hyperlink{ref:MetropolisMonteCarlo}{Monte Carlo} approaches.

\hypertarget{ref:MetropolisMonteCarlo} {\paragraph{Metropolis Monte Carlo}}
A Metropolis Monte Carlo step makes some change in the system in a way that preserves the underlying probability \hyperlink{ref:Distribution} {distribution} (usually the Boltzmann distribution).  Taking the canonical \hyperlink{ref:Ensemble} {ensemble} for specificity, the simplest rule is to propose some new coordinate $\vx^\prime$, and change from $\vx$ to $\vx^\prime$ with probability:
\begin{eqnarray}
P(\vx\rightarrow \vx^\prime) = \min\left\{1, e^{-\left[\beta U(\vx^\prime)-\beta U(\vx)\right]}\right\}
\label{eq:bal}
\end{eqnarray}
This means that the move always occurs if the energy is lowered (probability is increased), and sometimes occurs if the energy is higher, with the probability given in Equation~\ref{eq:bal}.
In order for this to preserve the underlying Boltzmann distribution, proposals must be made in a symmetric way, such that the probability of being at $\vx^\prime$ and proposing $\vx$ is the same as being in $\vx$ and proposing $\vx^\prime$. This is satisfied by simple rules like translating or rotating particles a symmetric amount from the current position, but care must be taken for more complex coordinate transformations like torsional displacement volume changes, or polymer chain moves~\cite{Siepmann_mp_1992}.  It is in fact possible to choose new states with asymmetric or biased probabilities if those biases and asymmetries are properly accounted for; this generalization is called a \emph{Metropolis-Hastings} step~\cite{Hastings_biometrika_1970}.

In addition, the Metropolis (or Metropolis-Hastings algorithms), can also be used to propose new parameters, such as a change in the temperature of the system, or a change in the potential energy of the system governed by an auxiliary paramter $\lambda$:
\begin{equation}
    P(i\rightarrow i^{\prime}) = \min\left\{1,e^{-\left[\beta U(\lambda_i,\vx)-\beta U(\lambda_{i^{\prime}},\vx)\right]}\right\}
\end{equation}

\hypertarget{ref:GibbsSampler} {\paragraph{Gibbs sampler}}
\hyperlink{ref:MetropolisMonteCarlo} {Metropolis Monte Carlo} steps are ways to move from one state of the system to another, considering one trial step at a time.  The Gibbs sampler~\cite{shirts_gibbssamp} is a  way to move to another state considering many trial states simultaneously. It is generally used when there are two (or more) different types of variables defined in the system that could be changed, $x$ and $y$, so the probability of the system is defined by $\nu(x,y)$. The Gibbs sampler is defined by taking steps in $x$ first and then $y$, by the following algorithm:
\begin{enumerate}
\item Start at $x_i$,$y_i$.
\item Pick a variable to change randomly with some frequency $f$, such that $f(x)+f(y)=1$.
\item If you choose $x$, pick a new $x$ from the conditional \hyperlink{ref:Distribution} {distribution} $\nu(x|y_i)$, i.e. the probability of $x$ given $y_i$.
\item If you choose $y$, pick a new $y$ from the conditional \hyperlink{ref:Distribution} {distribution} $\nu(y|x_i)$, i.e. the probability of $x$ given $y_i$.
\item Repeat.
\end{enumerate}
Any method to pick $x$ and $y$ from the distribution can be used, including \hyperlink{ref:MetropolisMonteCarlo} {Metropolis Monte Carlo}. The algorithm could be generalized to more than two variables, by randomly choosing each of the variables. Interestingly, one can alternate between variables deterministically (first pick $x$, then pick $y$, then pick $x$ again, or pick $x$ 100 times, then $y$, then $x$ 100 times again) and the results will not satisfy detailed \hyperlink{ref:Balance} {balance}, but they will still satisfy \hyperlink{ref:Balance} {balance}, which means the distribution $\nu(x,y)$ will still be preserved.
To be concrete, let us assume one variable is the coordinate $\vx$, and the other is the temperature $T$. Then one implementation of the Gibbs sampler, one would carry out simulations in $\vx$ using molecular dynamics in the NVT or NPT \hyperlink{ref:Ensemble} {ensemble} for a certain number of steps, then perform a move in $T$ space using an algorithm that generates a new $T$ from $\nu(T|\vx)$, while keeping the coordinates constant.

\hypertarget{ref:Balance} {\paragraph{Detailed balance and balance}}
Detailed balance is a constraint on the way moves from a given state of a system to another state of the system one are performed.  If $i$ and $j$ are two states, then detailed balance requires that
\begin{equation}
\frac{P(i)}{P(j)} = \frac{P_{i\rightarrow j}}{P_{j\rightarrow i}},
\end{equation}
where $P(i)$ and $P(j)$ are the desired probabilities of $i$ and $j$, and $P_{i\rightarrow j}$ and $P_{j\rightarrow i}$ are the probabilities of transitioning from state $i$ to $j$ and state $j$ to $i$ during some process. These states could be sets of configurations, for physical dynamics, or between different thermodynamic \hyperlink{ref:Ensemble} {ensembles}, as occurs in many nonphysical dynamical systems. Generally, physical systems obey detailed balance, and most common simulation methods, such as \hyperlink{ref:MolecularDynamics}{molecular dynamics} or \hyperlink{ref:MetropolisMonteCarlo}{Metropolis Monte Carlo} are designed to obey detailed balance, and thus preserve the overall probability \hyperlink{ref:Distribution}{distribution} of the system.

Balance is a weaker requirement~\cite{deem:jcp:1999:balance}, and is merely the requirement that the physical desired probability distribution is preserved by a set of dynamical moves. This can be done without forcing the ratio of the fluxes between states being equal to the ratio of the probabilities of two states, as is the case in detailed balance.  For example, you could have a cycle of fluxes between three states $i$,$j$, and $k$, and still preserve the probability distribution.  Proving a given set of ways to perform moves between states obeys balance is usually significantly harder than proving that a set of moves preserves detailed balance. However, studies have shown that it is often possible to obtain better sampling through states with algorithms that only obey balance rather than detailed balance~\citep{deem:jcp:1999:balance,Faizi:JCTC:2020}.

\paragraph{Temperature-based sampling} Refers to enhanced-sampling methods relying on an increased effective temperature $\tilde T$ to reduce \hyperlink{ref:metastab} {metastability}. This can be done on only some replicas within a set of replicas, on some fraction of the time within a given trajectory, or on some fraction of the system by selective scaling of potential energy terms.
In the latter case, if $U = U^\mathrm{unscaled} + U^\mathrm{scaled}$, scaling $U^\mathrm{scaled}$ to reach effective inverse temperature $\tilde  T$ means that the modified potential $\tilde U$:
\begin{equation}
\tilde U = U^\mathrm{unscaled} + \frac{T}{\tilde T} U^\mathrm{scaled}.
\end{equation}
As then the Boltzmann factor is:
\begin{equation}
e^{\frac{U^\mathrm{unscaled}}{k_BT} + \frac{U^{scaled}}{K_B \tilde T}} 
\end{equation}
And the scaled degrees of freedom have probabilities consistent with the different $\tilde T$

\hypertarget{ref:ReplEx} {\paragraph{Replica exchange}} \hyperlink{ref:GenEns} {Generalized ensemble} methods that allow for exchanging configurations between replicas, usually according to criteria that guarantee that each replica samples from a well-defined \hyperlink{ref:Distribution} {distribution}. 

\hypertarget{ref:AdiabaticDyn} {\paragraph{Adiabatic dynamics}}
A dynamics where selected degrees of freedom are assumed to not exchange energy with the rest of the system.
This can be achieved if these degrees of freedom are decoupled from the rest of the system, and evolve effectively independently.

\hypertarget{ref:Driven} {\paragraph{Driven simulations}} Refers to simulations with a time-dependent bias potential $U^{\mathrm{bias}}_t(\vx)$ or force $\vF^\mathrm{bias}_t(\vx)$ following a pre-determined schedule.

\paragraph{Out-of-equilibrium system} A system that has not reached its steady state, either because it is considered on a time-scale smaller than its relaxation time-scales, or because it is driven by time dependent forces which maintain it out of equilibrium.

\hypertarget{ref:OutOfEq} {\paragraph{Out-of-equilibrium method}}
A simulation protocol that does not generate trajectories that sample from the canonical \hyperlink{ref:Ensemble} {ensemble} associated with a given potential energy function.
Instead, initial and boundary conditions are given, as well as a possibly time- and history-dependent potential energy function or force schedule. The statistics of the generated trajectories and configurations are determined by these conditions.
Unlike the canonical \hyperlink{ref:Distribution} {distribution}, however, there is no general closed-form expression for the resulting out-of-equilibrium distributions of trajectories or configurations.

\hypertarget{ref:Seeding} {\paragraph{Seeding}}
A strategy where the configuration space covered by a set of (relatively short) simulations is governed by the choice of their starting conditions. This strategy is typically used to increase the diversity of the samples produced, despite the internal correlation of each trajectory.

\hypertarget{ref:Adaptive} {\paragraph{Adaptive method}}

All enhanced sampling methods have parameters, whose value can often be improved based on information from a simulation.
This process can be repeated, leading to an iterative form of the algorithm.
In some cases, this iteration can be built into the dynamics itself, so that parameters are updated (adapted) on the fly during the simulation. It is then called adaptive.
This can take the form of a time-dependent bias potential $U^\mathrm{bias}_t(\vx)$ or force $\vF^\mathrm{bias}_t(\vx)$, or branching decisions to stop or launch simulation instances.

Many methods can be used with fixed parameters, but also have iterative and adaptive variants. Conversely, the parameters of adaptive methods can be frozen when they are deemed sufficiently close to convergence, to continue sampling in a theoretically simpler setting.

\hypertarget{ref:FEP} {\paragraph{Free energy perturbation (FEP)}}
\label{FEP}

Free energy perturbation is the process of calculating the free energy differences due to small changes in the potential energy using reweighting or exponential averaging, hence the term ``pertubation''.  It is performed using the \hyperref[sec:fe_estimators:EXP]{exponential averaging} formula.  Alternatively, and somewhat confusingly, it can also refer to calculating a free energy difference using any sort of \hyperlink{ref:Alchemical} pathway, using any sort of \hyperlink{ref:FEestimator}{free energy estimator}.

\section{Free energy estimators}
\label{sec:fe_estimators}

Recovering the original statistical ensemble often requires estimating a \hyperlink{ref:FE} {free energy}. For some enhanced sampling methods, the \hyperlink{ref:FEestimator} {free energy estimator} is central to the enhanced sampling scheme, while for others, it is a post-processing tool. This section provides a concise presentation, but extensive reviews can be found elsewhere~\cite{cchipot07:molsim, Lelievre2010, Paliwal_comparison_2011, shirts_comparison_2005, Klimovich_Shirts_Mobley_2015}.

\hyperlink{ref:FEestimator} {Free energy estimators} are expressions that are used numerically to compute free energy differences or free energy \hyperlink{ref:FES} {surfaces} from quantities that are available in simulations (configurations, energies, and forces).
The important properties of an estimator are its accuracy or  \textit{bias}, how far off the real value it is given infinite, ideal sampling, and precision or \textit{variance}, how much the result fluctuates given a finite amount of noisy data.  Note that the \textit{bias} of the estimator is a different concept than the bias that is intentionally added to a system via a potential. 

In this section we consider estimators for:
\begin{itemize}
    \item free energy as a function of a Hamiltonian parameter (auxiliary variable) $\lambda$, which the energy $u_\lambda$ depends on, or
    \item free energy \hyperlink{ref:FES} {surface}s as a function of a \hyperlink{ref:CV} {collective variable} (or a vector of variables) $\vz = \xi(\vx)$.
\end{itemize}

For generality and cleanness of presentation, \hyperlink{ref:reduced} {reduced} units are used, which can converted back to united formulas using the definitions of reduced units.

\subsection{Directly measured ratios}

Given a set of sampled configurations that visits two states $i$ and $j$, their reduced \hyperlink{ref:FE} {free energy} difference $f_{ij}$ can be estimated by estimating free energies between states by taking the ratio of time in each states, a process called Boltzmann inversion, by approximating Equation~\eqref{eq:BoltzmannDistr}:
\begin{equation}
\Delta f_{ij} = -\ln \frac{N_j}{N_i}
\label{eq:boltzmann_inversion}
\end{equation}
where $f_{ij}$ is the \hyperlink{ref:reduced} {reduced} free energy, and $N_i$ and $N_j$ are the numbers of observed configurations in states $i$ and $j$.  This equation holds for whatever division into states one uses. However, this is only true if the simulation samples the equilibrium between states $i$ and $j$, or equivalently, if it is long enough such that many transitions between those states have been observed~\cite{No2009}. If a list of several states is defined, for example 1,2, \ldots,$K$, then then estimates from neighboring states can be chained to compute relative free energies for the entire list of states, by estimating $f_{12} + f_{23} = -\ln \frac{N_2}{N_1} + -\ln \frac{N_3}{N_2} =   -\ln \frac{N_3}{N_1} = f_{13}$, and so forth. If the states are defined as bins along \hyperlink{ref:CV} {collective variable}s, this yields a (discretized) free energy \hyperlink{ref:FES} {profile} along those variables. This idea extends to the calculation of continuous probability \hyperlink{ref:Distribution} {distributions} and \hyperlink{ref:FES} {free energy surfaces}. In that case, the probability distribution can be estimated using a kernel density estimator (KDE) or a Gaussian Mixture Model (GMM)~\cite{Westerlund2017}.

Convergence of this ratio can be accelerated by \hyperlink{ref:IS} {importance sampling}: adding \hyperlink{ref:biasingE} {biasing} potentials $u^{bias}_i$ and $u^{bias}_j$, with $\Delta u^{bias}_{ij}=u^{bias}_j - u^{bias}_i$.
If the bias $u^{bias}_i$ depends only on the state $i$, and is constant over $\vx$ within a state, the free energy can be estimated by simply subtracting the bias from the free energy estimate of the histogram in Eq.~\ref{eq:boltzmann_inversion}: \begin{equation}
\Delta f_{ij} = -\ln \frac{N_j}{N_i} - \Delta u^{bias}_{ij}
\label{eq:boltzmann_inversion_biased}
\end{equation}

\subsection{Estimating free energies from the transition count matrix\label{sec:transtion_matrix}}

The transition probabilities between states, which could be along one or more discretized \hyperlink{ref:CV} {CVs} or between different states in an \hyperlink{ref:ExpEns}{expanded ensemble}, can be used to build a \textit{transition probability matrix}. This matrix containing equivalently either the number of transitions from a state $i$ to a state $j$, or the probabilities $P_{i\rightarrow j}$ of performing the transition.

If one is considering a simulation that allows transitions between two states $i$ and $j$, then this implies the average of the transition probabilities between states is equal to the ratio of the partition functions between the states of interest \cite{deOliveira:EPJB:1998,Wang:JoSP:2002,escobedo_transition_2006}. Specifically:
\begin{equation}
e^{f_j-f_i} = \frac{Z_i}{Z_j} = \frac{\left \langle  P_{i\rightarrow j}(x)\right\rangle}{\left \langle P_{j\rightarrow i}(x) \right \rangle}
\end{equation}
where $P_{i\rightarrow j}(x)$ is the probability of accepting a move from state $i$ to state $j$ proposed from configuration $\vx$. For simplicity we assume that the probability of moves proposed from $i$ to $j$ is equal to the probability of transition from $j$ to $i$; more general transitions can also be included~\cite{escobedo_transition_2006}. We can construct a matrix of all possible transitions between any pair of states; hence we can call this a transition count matrix or \textit{transition matrix}.

Typically, this averaging is carried out over the transitions that were actually observed. However, this average can be calculations over transitions \emph{that were not actually performed}. If one counts only whether a move is made or not, one averages a number of either 0's or 1's. However, in any method that that makes transitions between states, we need to calculate the probability that the transition would occur to decide whether we move or not, as in the \hyperlink{ref:MetropolisMonteCarlo} {Metropolis Monte Carlo} algorithm (Eq.~\ref{eq:bal}).  This probability can be calculated and used in averaging even if the moves are not made. 

Interestingly, one can even compute this average of transition probability \hyperlink{ref:Distribution} {distributions} using a different transition probability formula than the one actually uses to carry out jumps between states. For example, one can show that the \hyperlink{ref:MetropolisMonteCarlo} {Metropolis Monte Carlo} criterion $(\mathrm{min} \{1,e^{-u_j+u_i})\})$ is more efficient to use to decide whether or not to move between two states, but the Barker criterion $\left(\frac{e^{-u_j}}{e^{u_j}+e^{u_i}}\right)$ is more efficient to average in order to calculate free energies~\citep{Liu:Biometrika:1996}.

Note that free energy estimated directly from population ratios will of course be inaccurate if too few transitions have been observed between states~\cite{No2009}.  More generally, the discretized master equation expresses the probability \hyperlink{ref:Distribution} {distribution} $\nu_i$ of each state $i$ over time as the sum of all of the probabilities of states moving into and out of that state, as:
\begin{equation}
   \nu_i(t + \delta t) = \nu_i(t) + \sum_{j \neq i} (\nu_j(t) P_{j\rightarrow i} - \nu_i(t) P_{i\rightarrow j}).
   \label{eq:master_equation}
\end{equation}
This is a time-dependent equation, but can be show to have an equilibrium stationary probability \hyperlink{ref:Distribution} {distribution} $\nu_i(\vx)$. Convergence of probability distributions is slow because the samples extracted from molecular dynamics simulations are correlated. However, by taking into account the conditional probability distribution (probability of an event happening given previous history), the statistical dependency of the MD trajectory is taken into account~\cite{Wu:JCP:2014,Wu:MMS:2014,Rosta2014,Wu:PNAS:2016a}. 
An even more accurate estimation of the free energies can therefore be achieved by building a Markov State Model from the transition count matrix, whereby the probability of each state is computed as the leading eigenvector of the transition matrix. In this case, it may not be necessary for single simulations to sample the entire state space, as long as each of the individual transitions is estimated accurately.  This is the basis for such methods as DHAM~\cite{Rosta2014}, dTRAM~\cite{Wu:JCP:2014,Wu:MMS:2014}, and TRAM~\cite{Wu:PNAS:2016a}, which are somewhat beyond the scope of this review, as they are advanced analysis methods, but not technically by themselves accelerated sampling methods.

\subsection{Thermodynamic integration (TI)}
\label{sec:fe_estimators:TI}

Thermodynamic integration is a family of free energy estimators that express the derivative of free energy with respect to a continuous parameter as an \hyperlink{ref:ensemble_average} {ensemble average} of the derivative of the energy with respect to the same parameter, possibly with additional terms as discussed below.

\subsubsection{TI along an alchemical parameter}

When the reduced potential energy is a function $u_\lambda$ of a smooth coupling parameter $\lambda$ that connects all states of interest (e.g. in \hyperlink{ref:Alchemical} {alchemical} perturbations), a continuous free energy $f(\lambda)$ is defined, the derivative of which is the ``mean force'' at a fixed value of $\lambda$:
\begin{equation}
\frac{\mathrm{d}f}{\mathrm{d}\lambda} =\left\langle \frac{\partial u_\lambda}{\partial \lambda} \right\rangle_\lambda
\label{eq:TI1}
\end{equation}
Where the averaging is over samples obtained with the given value of $\lambda$.

If states $i$ and $j$ correspond to values $\lambda_i$ and $\lambda_j$ of the continuous coupling parameter, then the free energy difference between states $i$ and $j$, $\Delta f_{ij}$, is
\begin{equation}
\Delta f_{ij} = \int_{\lambda_i}^{\lambda_j}  \left\langle \frac{\partial u_\lambda}{\partial \lambda} \right\rangle_\lambda \mathrm{d}\lambda
\label{eq:TI2}
\end{equation}

If $i$ and $j$ are neighboring states along $\lambda$ with a sufficiently small difference $\Delta \lambda_{ij}$, the integral can be approximated by the trapezoidal rule:
\begin{equation}
\Delta f_{ij} = \frac{\Delta \lambda_{ij}}{2}\left[\left\langle \frac{\partial u_\lambda}{\partial \lambda} \right\rangle_{\lambda_i} + \left\langle \frac{\partial u_\lambda}{\partial \lambda} \right\rangle_{\lambda_j}\right]
\label{eq:TI}
\end{equation}

Other numerical integration formulas can of course be used, although usually there has not been found to be much advantage over the straightforward trapezoidal rule~\cite{Paliwal_comparison_2011}. 


\subsubsection{TI along a collective variable}

A generalization of Equation~\ref{eq:TI1} holds for the gradient of the \hyperlink{ref:reduced} {reduced} \hyperlink{ref:FES} {FES} $a(\vz)$ over a single \hyperlink{ref:CV} {collective variable} $z = \xi(\vx)$:
\begin{equation}
\frac{\mathrm{d} a(z)}{\mathrm{d} z} = - \left\langle F_\xi(\vx)  \right\rangle_{\xi(\vx) = z}
    \label{eq:TI_CV}
\end{equation}
Where the average is over all configuration with $\xi(\vx) = z$. 
$F_\xi$ is a \textit{generalized force} that includes two terms:
\begin{itemize}
    \item the energy gradient with respect to \hyperlink{ref:CV} {collective variable}s;
    \item a geometric, \textit{Jacobian} term due to the curvature of the isosurfaces of a nonlinear CV $\xi$ (see~\cite{lelievre-rousset-stoltz-07-a, Henin2010a, Comer2015} for details).
\end{itemize}

More precisely, the ``free energy gradient with respect to the CVs'' is a partial derivative, which is not well-defined unless one specifies a complete set of generalized coordinates including the CVs~\cite{Henin2004}.
This complicated or even intractable process can be circumvented~\cite{denOtter2000, Ciccotti2005} by noting that the force may be projected onto a CV $\xi_i$ along an arbitrary vector field  $\vb_i$, verifying for all $i,j$:
\begin{equation}
    \vb_i \cdot \nabla_\vx \xi_j = \delta_{i,j}
\end{equation}
that is, each $\vb_i$ has a scalar product of 1 with the gradient of $\xi_i$ and is orthogonal to the gradients of all other CVs.
In the case of a single CV, one possible choice of $\vb$ that satisfies this condition is to make it proportional to the gradient of $\xi$ with an appropriate normalization factor:
$ \vb(\vx) = 1/(|\nabla \xi|^2) \nabla \xi$.

Then the following expression for the generalized force along coordinate $\xi_i$ holds~\cite{Ciccotti2005}:
\begin{equation}
\label{eq:TI_Ciccotti}
\vF_{\xi_i}(\vx) = - \vb_i(\vx) \cdot \nabla_\vx u(\vx) + \nabla_\vx \cdot \vb_i(\vx)
\end{equation}
where the first term is the projection of atomic forces $-\nabla_\vx u$ onto the CV, and the second is the Jacobian term described above, computed as the divergence of $\vb_i$.

Integrating the gradient to obtain the \hyperlink{ref:FES} {free energy surface} is easy when using a scalar \hyperlink{ref:CV} {collective variable}. In dimension greater than one, however, special integration methods are required~\cite{Henin2021integration}.

\subsubsection{Significance of the Jacobian term}

Take the example of a \hyperlink{ref:CV} {collective variable} measuring the distance $r$ between two particles. Here, the gradient of the \textit{free energy \hyperlink{ref:FES} {surface}} depends on the geometry of the CV. Given a complete coordinate transform from Cartesian to generalized coordinates including $r$, which is necessary to define partial derivatives with respect to $r$, one may write~\cite{Henin2010a}:
\begin{equation}
    \frac{\mathrm{d}a}{\mathrm{d}r} = \left\langle \frac{\partial u}{\partial r}
    - \frac{2}{r} \right\rangle_{\xi(\vx) = r}
    \label{eq:fes_grad}
\end{equation}
As mentioned in Section~\ref{sec:glossary}, the historic definition of the \hyperlink{ref:PMF} {``potential of mean force'' (PMF)} for $r$ is simply the potential arising from the mean force and corresponds to the first term in Equation~\ref{eq:fes_grad}:
\begin{equation}
    \frac{\mathrm{d} w}{\mathrm{d} r} = \left\langle \frac{\partial u}{\partial r} \right\rangle_{\xi(\vx) = r}
    \label{eq:pmf_grad}
\end{equation}

\subsubsection{Comparison between alchemical and configurational TI}

Let us consider extended configurations $(\vx, \lambda)$ that include the \hyperlink{ref:Alchemical} {alchemical} parameter, and define an \hyperlink{ref:Alchemical} {alchemical} CV as a simple projection of those onto $\lambda$:
\begin{equation}
    \xi(\vx, \lambda) = \lambda .
\end{equation}
Then Equation~\ref{eq:TI1} can be seen as a special case of Equation~\ref{eq:TI_Ciccotti}, in which the second term is zero because no nonlinear coordinate transform is involved.

\subsection{Exponential averaging}
\label{sec:fe_estimators:EXP}

The following sections apply to the free energy difference between two discrete states $i$ and $j$, when they can be characterized by different \hyperlink{ref:reduced} {reduced} energies $u_i$ and $u_j$, with $\Delta u_{ij} = u_j - u_i$.
These estimators (exponential averaging, BAR, MBAR and WHAM) rely on the overlap between the thermodynamic states $i$ and $j$, i.e. that configurations have a significant probability under both states. They cannot be used when the states do not have overlap, i.e. the states are defined by each having a different value of some collective variable.  In contrast, free energy calculated by visitation ratios, transition matrices, and thermodynamic integration can.

The total free energy difference between two such states is, by definition:
\begin{align}
    e^{-\Delta f_{ij}} &=  \frac{\int e^{-u_j} \, d\vx}{\int e^{-u_i} \, d\vx}\\
    &= \frac{\int e^{-u_i}  e^{-(u_j-u_i)} \, d\vx}{\int e^{-u_i} \, d\vx}
\end{align}
On the right-hand side, one can recognize the expression for an \hyperlink{ref:ensemble_average} {ensemble average} in state $i$.
Thus the free energy difference can be written:
\begin{equation}
    \Delta f_{ij} = -\ln \langle e^{-\Delta u_{ij}}\rangle_i.
\end{equation}

This average can be estimated numerically as:
\begin{equation}
\Delta f_{ij} = -\ln \frac{1}{N_i}\sum_{n=1}^{N_i} e^{-\Delta u_{ij}(x_n)},
\label{eq:expav}
\end{equation}
where the $N_i$ samples are from the $i$th state.
While this expression is formally exact, convergence of the exponential average is critically dependent on the tail of the distribution of $\Delta u_{ij}$, and as a result, on the overlap between states $i$ and $j$.
This can be alleviated by collecting energy differences going both ways ($\Delta u_{ij}$ sampled in state $i$, and $\Delta u_{ji}$ sampled in state $j$), and combine them using the BAR estimator.
This general strategy is sometimes also referred to as ``Overlap Sampling"~\cite{Lu2003}.

\subsection{Bennett's acceptance ratio (BAR)}

The Bennett acceptance ratio method is the lowest variance method to estimate the free energy difference between two states using the energies sampled at those states. Specifically we
obtain an optimal value of the free energy difference $\Delta f_{ij}$ given $N_i$ samples performed with \hyperlink{ref:reduced} {reduced} energy $u_i$ and $N_j$ samples performed with reduced energies $u_j$. There are many ways to write
the method~\cite{bennett:jcp:1976:fe-estimate,shirts_comparison_2005,fenwick-escobedo:jcp:2003:replica-exchange-expanded-ensembles}; see these references for the detailed derivations.

One standard approach is to derive the thermodynamic identity:
\begin{eqnarray}
\Delta f_{ij} = \ln \frac{\left\langle \frac{1}{1 + \exp(-\Delta u_{ij}+c))}\right\rangle_j}{\left \langle \frac{1}{1 + \exp(\Delta u_{ij}-c)}\right\rangle_i} + c-\ln\frac{N_j}{N_i}
\end{eqnarray}
Where $c$ is an arbitrary constant. This expression is an asymptotically unbiased estimator (meaning, if enough data is collected, it will converge to the correct answer) for any choice
of $c$. However, one can prove that the lowest variance estimate of free energy is obtained when $c=\Delta f_{ij}$~\cite{bennett:jcp:1976:fe-estimate}. To find $\Delta f_{ij}$, this equation needs to be solved self-consistently, resulting in the equation one can solve for $c$, and therefore $f_{ij}$ as well:
\begin{equation}
\sum_{i=1}^{N_j} \frac{1}{1 + \exp(-\Delta u_{ij}+c))} - \sum_{j=1}^{N_i} \frac{1}{1 + \exp(\Delta u_{ij})-c} = 0
\end{equation}
This can be solved by a number of numerical techniques as implemented in many codes, such as $\texttt{pymbar}$~\cite{shirts-chodera:jcp:2008:mbar} and $\texttt{gmx bar}$~\cite{lindahl_2021}.

Several variants, which can be useful in specific cases, use fixed $c$~\cite{fenwick-escobedo:jcp:2003:replica-exchange-expanded-ensembles,Paliwal_comparison_2011}, but for almost all standard uses, the optimized, lowest variance version is best.

\subsection{Multistate Bennett acceptance ratio (MBAR)}

If we carry out simulations at $K$ different thermodynamic states,
then the free energies of all of the states can be estimated as:
\begin{equation}
f_i = -\ln \left[\sum_{n=1}^N \frac{e^{-u_i(\vx_n)}}{\sum_k N_k e^{f_k-u_k(\vx_n)}}\right]
\label{eq:MBAR}
\end{equation}
Where $N$ is a sum over all of the samples collected at any of the $K$ states.
Since there is one equation for each of the free energies $f_i$ from each of the $K$ states, this leads to a series of $K$ that must be solved for the set of $f_i$~\footnote{Effectively, there are only $K-1$ independent free energies since the set of free energies has an arbitrary reference zero.}. This is a set of implicit equations since the $f_k$ appear on both sides of the equations. Similarly to BAR, there are a standard ways to solve this set of equations implemented in a number of different codes~\cite{shirts-chodera:jcp:2008:mbar,tan_binless_2012,Zhang:JPCL:2015}.

MBAR also allows the computation of high precision uncertainties in the $\Delta
f_{ij}$'s, as it can take into account the correlations in $f_i$ and
$f_j$ thanks to their simultaneous estimation.  A key fact is that the MBAR system of equations reduces precisely to the equation for BAR for estimating the free energy difference between two states~\cite{shirts-chodera:jcp:2008:mbar}.

MBAR can also be seen as exponential averaging to \hyperlink{ref:targetdist} {target distributions} described by the \hyperlink{ref:reduced} {reduced} potential $u_i(\vx)$ from the \emph{mixture distribution} $\nu_{mix}(\vx)$~\cite{reweighting_mixture_distribution}. Specifically, the mixture distribution is formed by combining all the samples from all simulations that are put into the mixture, proportional to the number of samples $N_k$ from each simulation.  Using eq.~\ref{eq:expav}, we can then rewrite Equation~\ref{eq:MBAR} as:
\begin{eqnarray}
f_i &=& -\ln \left[\frac{1}{N}\sum_{n=1}^N \frac{e^{-u_i(\vx_n)}}{\nu_{mix}(\vx_n)}\right] \nonumber\\
\nu_{mix}(\vx) &=& \sum_k \frac{N_k}{N} e^{f_k-u_k(\vx)} \nonumber \\
               &=& \sum_k \frac{N_k}{N} \nu_k(\vx)
\end{eqnarray}

\subsection{Weighted histogram analysis method (WHAM)}

In its original formulation, WHAM consists in applying Boltzmann inversion (i.e. calculating free energies from probabilities) to histograms collected under a localizing potential (Section~\ref{sec:localization}), and joining the bias-corrected histograms in an iterative and statistically optimal way to reconstruct the global histogram, from which the global free energy \hyperlink{ref:FES} {profile} is calculated~\cite{kumars:WHAM}.  However, the same idea can apply to histograms collected at different temperatures or other sets of histograms, as long as the histograms overlap. 

One can start from the MBAR equations to derive WHAM as well. If one can histogram the data by energy into $i$ bins, then the sum over $n$ \textit{samples} in MBAR becomes a sum over $M$ energy \textit{bins} instead. The iterative equations of WHAM can be written as:
\begin{eqnarray}
P_{i,k} = \frac{\sum_{k=1}^K n_{k,i} e^{-u_{k,i}}}{\sum_{j=1}^{K}n_k {e^{f_j-u_{j,i}}}}\qquad
e^{-f_k} = \sum_{m=1}^M P_{m,k}
\end{eqnarray}
Where $n_{k,i}$ are the counts in each bin from the $k$th state, $P_{i,k}$ is the probability distribution of bin $i$ in the $k$th state,  $u_{k,i}$ is the energy of the system in the $i$th bin and $k$th state~\cite{kumars:WHAM}. These are equivalent; in the WHAM equations, instead of directly summing over all $n$ samples, we first sum over all of the samples in each bin $n_{k,i}$ from each state to get $P_{i,k}$, and then sum over all bins to get the free energy. When the bins in WHAM are shrunk to have zero width (i.e. down to $\delta$ functions), then the WHAM equations become equivalent to MBAR, as the sum again becomes a sum over samples. If there are many samples, WHAM can be significantly faster than MBAR. However, WHAM only gives similar results to MBAR if the bins are narrow enough for the energy $u$ and probability $P_i$ to both be approximately constant within a bin.

\section{Out-of-equilibrium / driven methods}
\label{sec:Out-of-equilibrium_driven}

The idea of \hyperlink{ref:OutOfEq} {out-of-equilibrium} \hyperlink{ref:Driven} {driven} methods is to force the system to follow a given schedule of a \hyperlink{ref:CV} {collective variable}, or of an \hyperlink{ref:AuxVar} {\hyperlink{ref:Alchemical} {alchemical} parameter} $\lambda$, in order to explore configuration space. This class of methods yields simulations in which the original \hyperlink{ref:Distribution} {distribution}  is modified, and which do not converge to an equilibrium ensemble (Figure~\ref{fig:scheme}). The equilibrium \hyperlink{ref:Distribution} {distribution}  may be retrieved under specific circumstance, as detailed below. The schedule followed may be fast, so that even if the system starts at equilibrium (a usual assumption), it does not remain at equilibrium.
An early version was targeted MD~\cite{Schlitter1994}, which consists in a moving constraint on the RMSD between current Cartesian coordinates and a target.
Steered molecular dynamics, introduced shortly thereafter to mimic Atomic Force Microscopy experiments, introduces a fictitious 3D particle moving at constant velocity, and connected to a molecule by a harmonic spring~\cite{Grubmueller1996}.
These two methods can be considered to have converged, as moving harmonic restraints can be applied to arbitrary \hyperlink{ref:CV} {collective variable}s $\vz$ using software tools such as Colvars~\cite{Fiorin2013} or PLUMED~\cite{Tribello2014}.

\hyperlink{ref:OutOfEq} {Out-of-equilibrium} pulling behaves differently depending on the rate of the transformation.
In the limit of infinitely slow (quasistatic) switching, all orthogonal degrees of freedom are fully relaxed at all times, so that equilibrium properties are recovered.
This is the case of the ``slow growth" approach \cite{Postma1982}, which uses very slow switching from energy $U_A$ to $U_B$. The work $W$ performed along the way is an approximation of the reversible work, that is, the free energy difference from $A$ to $B$.

At the other end of the spectrum, infinitely fast switching amounts to comparing the energy of a given configuration (in absence of any relaxation) for two different Hamiltonians, using a \hyperlink{ref:FEP} {FEP} approach~\cite{Kirkwood1935,Zwanzig1954}.

In intermediate cases, the free energy difference can be estimated~\cite{jarzynski-97} by weighting the non-equilibrium trajectories.
Calling $\mathcal W_\lambda$ the total non-equilibrium work exerted by the bias over a trajectory up to a value $\lambda$, the so-called Jarzynski identity states:
\begin{equation}
\label{eq:jarz}
e^{-\beta (F_\lambda - F_0)} = \left\langle e^{-\beta \mathcal W_\lambda}\right\rangle \text{ ,}
\end{equation}
where the average is taken over the equilibrium ensemble of initial conditions at $\lambda = 0$.

\hyperlink{ref:OutOfEq} {Out-of-equilibrium} methods are not frequently used to \hyperlink{ref:FEestimator} {estimate free energy differences} because the variance of the exponential averaging free energy estimator is in general plagued by a large variance.
Nevertheless, variants of the Jarzynski identity have been successfully applied to rare examples of sufficiently fast-relaxing systems~\cite{park-khalili-araghi-tajkhorshid-schulten-03}.
The high variance of the Jarzynski estimator is due to the fact that the work values that contribute the most to the average have small probability (as discussed in Section~\ref{sec:fe_estimators:EXP}).
Improved estimators~\cite{minh-adlib-08} and modified algorithms~\cite{VJ08,hartmann-schuette-zhang-19,rousset2006equilibrium} can thus take advantage of running simulations in the forward and backward direction~\cite{hummer-07}. Although less common, there have been important applications of these principles for applications such as ligand binding free energies~\citep{alchemy_Gapsys_2020}.

As a result, most simulations that aim to recover free energies resort to equilibrium or near-equilibrium sampling (as is often the case with \hyperlink{ref:Adaptive} {adaptive} \hyperlink{ref:biasingE} {biasing} algorithms).

\section{Localization methods}
\label{sec:localization}

In the broad sense, localization refers to sampling only in a small, well-defined volume of configuration space. In this class of methods, the original configurational \hyperlink{ref:Distribution}{distribution} is modified, the simulation converges to an equilibrium ensemble, and sampling is enhanced by specifying the starting and ending coordinates of the simulation, and localizing them to a well-defined region of space (Figure~\ref{fig:scheme}). Localized sampling within region $i$ can be achieved by either imposing constraints around $\vz_i$ (constraining strategy) or a confining potential $U^{\mathrm{bias}}_i(\vx)$ (restraining strategy), which can result in sampling from overlapping regions.

Note that this strategy is also sometimes referred to as ``stratification''. In statistics, stratification strictly means partitioning configuration space along one or more \hyperlink{ref:CV} {collective variable}, and collecting samples separately in each discrete section (\textit{stratum}). This approach can be used in conjunction with ABF (\ref{sec:abf_hybrids}).
In contrast, the most common restraining strategies mentioned in this section yield overlapping samples, which is why we prefer the term ``localization''.

Restraining or constraining a simulation is equivalent
to convolving the original probability density with a localizing function: a Dirac distribution $\delta(\vz)$ (constrained case),
a Gaussian kernel, or a (possibly smoothed) indicator function of an interval (exponential of a flat-bottom potential) in the restrained case.
The most used restraining strategy involves \hyperlink{ref:biasingE} {biasing} the distribution by imposing harmonic potential restraints $U^{\mathrm{bias}}_i(\vx) = \frac{k}{2} |\xi(\vx)-\vz_i|^2$, and is equivalent to convolving the original probability distribution with a Gaussian kernel.
Nowadays this is generally referred to as ``umbrella sampling", although it is quite different from the historic umbrella sampling method~\cite{TORRIE1977187} which was not localized (see Section~\ref{sec:biasing_potential} for a further description of this approach and how it relates to other methods described in this review).
Based on umbrella sampling trajectories, the free energy \hyperlink{ref:FES} {landscape} can be estimated using the WHAM, BAR, or MBAR estimators (see Section~\ref{sec:fe_estimators}). The constraining strategy was introduced as ``Blue Moon" sampling~\cite{doi:10.1080/0892702042000270214}. In that case, the free energy can be reconstructed by thermodynamic integration (Section~\ref{sec:fe_estimators:TI}). Using flat-bottom potentials, equivalent to convolving the probability distribution with an indicator function, is referred to as, for example, ``boxed MD"~\cite{doi:10.1021/jp9074898}, and is frequently used in combination with ABF (Section~\ref{sec:ABF}).

These methods have been improved upon by slightly more complicated variants that include transitions between the restrained or constrained windows can improve sampling by avoiding kinetic traps. Successful (and thus popular) approaches involve introducing transitions between the restraining potentials, in which case the methods fall under \hyperlink{ref:ExpEns} {expanded ensemble} or \hyperlink{ref:ReplEx} {replica exchange} (see Sections~\ref{sec:generalized-ensemble}).

\section{Non-adaptive biasing potential methods}
\label{sec:biasing_potential}

A range of methods are designed to flatten the energy \hyperlink{ref:FES} {landscape} in a static way.

This class of methods modifies the original configurational \hyperlink{ref:Distribution} {distribution} , lets the simulations converge to an equilibrium ensemble, but does not do so by localizing the simulations to specific regions of configuration space. Instead, sampling is biased by employing an external bias potential. Contrary to the adaptive methods presented in the next section, in this case the bias potential is pre-determined and static (Figure ~\ref{fig:scheme}).

This is the case of the original Accelerated MD~\cite{Hamelberg2004} and Gaussian-accelerated MD~\cite{Miao2017, Wang2021} methods. In these methods, the form of the modified potential energy is determined by user-controlled parameters, lowering the energy barriers either for specific transitions (e.g. along dihedral angles), or within a given energy range. Note that these methods can also be run in an \hyperlink{ref:Adaptive} {adaptive} manner (Section~\ref{sec:abp_energy}). 

Following the principle of \hyperlink{ref:IS} {importance sampling}, unbiased statistics can be recovered by \hyperlink{ref:Reweighting} {reweighting}.
Success of this, however, depends crucially on good overlap between the \hyperlink{ref:BiasedDist} {biased} and unbiased distributions of configurations.

Custom-designed static \hyperlink{ref:biasingE} {biasing} potentials combined with the idea of localization (Section~\ref{sec:localization}) are the principle behind modern Umbrella Sampling. In the seminal ``Umbrella Sampling'' paper from 1977 by Torrie and Valleau~\cite{TORRIE1977187}, the authors incorporate an external bias (called a weighting function in their language) that is designed to lead to flat sampling. Note that in this original ``Umbrella Sampling'' paper, a single simulation was used. The idea of multiple windows (localized, ``stratified'') umbrella sampling with fixed harmonic bias potential, as described in Section~\ref{sec:localization}, was introduced later on. The authors constructed the bias potential by hand using trial and error but suggest that the computer could be programmed to perform this task, as is now done routinely in \hyperlink{ref:Adaptive} {adaptive} methods (see Section~\ref{sec:AdaptiveBiasSimulations}).

\section{Adaptive bias simulations}
\label{sec:AdaptiveBiasSimulations}

This class of the method is related to the previous one in that sampling is biased, but contrary to the methods presented above, the bias is learned during the simulation and adapted on-the-fly (Figure ~\ref{fig:scheme}).

\subsection{Adaptation and the adaptation rate}

In \hyperlink{ref:Adaptive} {adaptive} bias simulations, the external forces needed to enhance sampling are learned based on information from the trajectory itself, and updated as the simulation progresses.
The bias always depends on statistical properties of the trajectory, so that it cannot be adapted instantaneously, but more or less progressively as information becomes available.
This is controlled by some external parameter, generically called \emph{adaptation rate}, or \emph{learning rate} in the machine learning community.
For example, in well-tempered metadynamics (\ref{sec:wtmetad}), the adaptation rate results from the deposition rate $N_G$, the height $H_0$ of the Gaussian kernels, and the bias factor $\gamma$, whereas in ABF (\ref{sec:ABF}), it depends on the number of samples collected locally before applying the estimated \hyperlink{ref:biasingE} {biasing} force.

The choice of adaptation rate is essential to the success of an \hyperlink{ref:Adaptive} {adaptive} bias simulation: if it is too slow, the process will not learn efficiently, but setting it too fast will also negatively affect convergence.
Typically, adaptation should be slow enough that some orthogonal degrees of freedom have time to relax to changes in biased degrees of freedom, or at least, remain within a short relaxation time of their equilibrium \hyperlink{ref:Distribution} {distribution}  under the current state of the bias. Details depend on the specific method, and the optimal value of the adaptation rate is system-dependent.
Choosing a high adaptation rate may favor rapid initial exploration, while slower adaptation optimizes long-term convergence~\cite{Invernizzi2022}.

\subsection{Adaptive biasing potential (ABP) methods}
\label{sec:ABP}
In \hyperlink{ref:Adaptive} {adaptive} biasing potential (ABP) methods, an external bias potential in a space of some chosen \hyperlink{ref:CV} {collective variable}s (\hyperlink{ref:CV} {CV}s) is added to bias the dynamics of the system. The purpose of the bias potential is to counteract the free energy \hyperlink{ref:FES} {surface} and lead to more uniform sampling in CV space. In other words, the bias potential leads to sampling of a biased CV-distribution that is easier to sample. As the \hyperlink{ref:FES} {FES} is naturally \textit{a priori} unknown, the bias potential is generally constructed in an \hyperlink{ref:Adaptive} {adaptive} manner through some kind of iterative scheme. ABP methods  differ in how the bias potential is constructed and which kind of sampling is obtained at convergence.

Note that most ABP methods can also be used in conjunction with \hyperlink{ref:MetropolisMonteCarlo} {Monte Carlo} simulations. In this case, the bias potential needs to be taken into account in the Monte Carlo acceptance probability (Equation~\ref{eq:bal}).

The performance of ABP methods depends critically on the choice of the CVs, the CVs need to be chosen carefully and they should properly separate the relevant metastable states and correspond to essential slow degrees of freedom. Furthermore, most ABP methods are limited in the number of CVs that they can handle and generally we can use no more than three to four CVs, though there are ABP methods that can handle a larger number of CVs~\cite{Piana2007_bemeta,Pfaendtner2015_pbmetad}. 

A wide range of ABP methods have been introduced throughout the years. In the following two Sections, we discuss two of these methods in detail, metadynamics~\cite{Laio-PNAS-2002,Barducci-PRL-2008,Valsson2016_ARPC_MetaD} and variationally enhanced sampling~\cite{Valsson_VES_PRL_2014,Valsson2020Handbook_VES}. Metadynamics is the most widely used ABP method and it has spurred the development of a great number of variants. Variationally enhanced sampling is a more recent ABP method that is based on the variational principle. For other ABP methods, it would be beyond the scope of this review to discuss all of them in detail, so we limit ourselves to give an non-exhaustive list of ABP methods. We refer the reader to the original references and review papers~\cite{Dickson_ABP-Review_2017,Shalini_Review_2019,Allison_Review_2020} on the subject for further details regarding these methods.

Among methods that we can  categorise as ABP methods are
local elevation~\cite{Huber1994},
energy landscape paving~\cite{Hansmann-PRL-2002},
self-healing umbrella sampling~\cite{Marsili2006,Gersende_SelfHealing_2017},
adaptive biasing MD (ABMD)~\cite{Babin2008},
Gaussian-mixture umbrella sampling~\cite{Maragakis-JPCB-2009}, the adaptive biasing potential method~\cite{Dickson2010},
basis function sampling~\cite{Whitmer_BFS_2014}, Green's function sampling~\cite{Whitmer_GFS_2015}, flying Gaussian method~\cite{Sucur2016},
artificial neural network sampling~\cite{Sidky_ANNSampling_2018},
on-the-fly probability-enhanced sampling (OPES)~\cite{Invernizzi2020opus,invernizzi2020unified},
reweighted autoencoded variational Bayes for enhanced sampling (RAVE)~\cite{Tiwary_RAVE_2018},
targeted adversarial learning optimized sampling (TALOS)~\cite{Zhang_TALOS_2019},
Gaussian mixture-based enhanced sampling (GAMBES)~\cite{Debnath_GAMBES_2020},
adaptive topography of landscapes for accelerated sampling
(ATLAS)~\cite{giberti2021atlas},
and reweighted Jarzynski sampling~\cite{Bal_ReweightedJarzynski_2021}.

For mathematical analysis of the convergence and efficiency of ABP methods, we refer to the series of works available in the literature~\cite{fort-jourdain-lelievre-stoltz-18,fort-jourdain-kuhn-lelievre-stoltz-15,fort-jourdain-kuhn-lelievre-stoltz-14,fort-jourdain-lelievre-stoltz-17}.

\subsubsection{Metadynamics}
\begin{figure*}[!ht]
    \centering
    \includegraphics[width=0.95\textwidth]{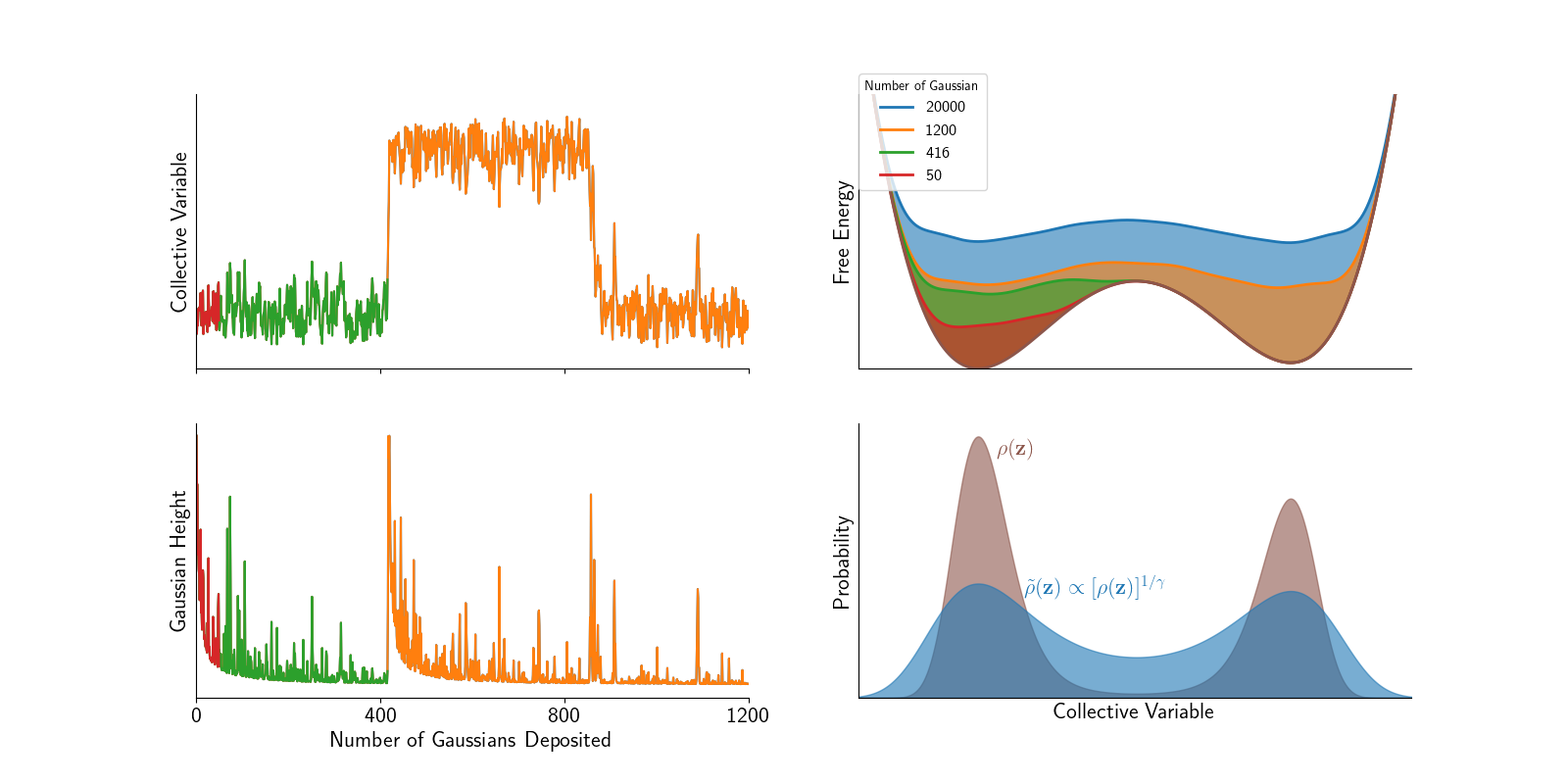}
    \caption{Prototypical behaviour of a well-tempered metadynamics simulations. Shown are results for a model energy landscape given by $A(z)=2z^4-8z^2+0.1918z$ where the metadynamics simulation is performed using $\beta=0.5$ and $\gamma=4$.
    \textbf{(Left top panel)} Time series of the added Gaussian kernels \textbf{(Left bottom panel)}. The height $H_{k}$ of the added Gaussians goes down as the simulations proceeds (see equations~\ref{wtmetad_update} and~\ref{wtmetad_sum}). \textbf{(Right top panel)} The FES $A(z)$ and FES with added bias potential $A(z)+U_{\mathrm{bias}}(z)$ shown for different number of added Gaussians (colors to correspond to the time series on the left side). The bias potential only partially cancel out the FES at convergence, leading to sampling on an effective FES $A_{\gamma}(\vz) = A(\vz) + U_{\mathrm{bias}}(\vz) =  \frac{1}{\gamma} A(\vz)$ where the barriers have been reduced by a factor of $\gamma$. \textbf{(Right bottom panel)} At convergence, the CV are sampled according to the well-tempered distribution where fluctuations are enhanced as compared to the unbiased distribution (see Equation~\ref{eq:wtmetad_wtdist}). The bias factor $\gamma$ determines how much we enhance the fluctuations. The figure is inspired by Figure~1 in Ref~\cite{Bussi2020}}
    \label{fig:MetaD}
\end{figure*}

One of the most widely used ABP method is metadynamics~\cite{Laio-PNAS-2002,Barducci-PRL-2008,Valsson2016_ARPC_MetaD}, and a large number of metadynamics variants have been developed and introduced through the years. In metadynamics methods, we enhance the sampling along a few selected \hyperlink{ref:CV} {collective variable}s (\hyperlink{ref:CV} {CV}s) that correspond to slow degrees of freedom. Metadynamics methods are based on adding a time-dependent external bias potential that counteracts the free energy \hyperlink{ref:FES} {surface}. This external \hyperlink{ref:Adaptive} {adaptive} \hyperlink{ref:biasingE} {biasing} potential is composed as a sum of repulsive Gaussian kernels that are periodically deposited at the current location in the \hyperlink{ref:CV} {CV} space, see Figure~\ref{fig:MetaD}. From the bias potential, one can directly estimate the free energy \hyperlink{ref:FES} {surface} as a function of the selected \hyperlink{ref:CV} {CV}s. Furthermore, it is possible to obtain the \hyperlink{ref:FES} {FES}, both for the biased \hyperlink{ref:CV} {CV}s and for any other set of \hyperlink{ref:CV} {CV}s, via \hyperlink{ref:Reweighting} {reweighting}. Under certain conditions, one can also rescale simulation time to obtain rare-event kinetics from biased metadynamics simulations.

There are two main variants of metadynamics that are in common usage nowadays, well-tempered metadynamics~\cite{Barducci-PRL-2008} and conventional (non-well-tempered) metadynamics~\cite{Laio-PNAS-2002}, that we will discuss below. Metadynamics methods and their applications have been discussed in various reviews~\cite{Barducci-WIREsCMS-2011,10.1080/08927022.2014.923574,10.1107/s2052252514027626,Valsson2016_ARPC_MetaD,10.1007/978-1-4939-9608-7_8,Bussi2020,BussiLaio_ReviewMetaD_2020}.

\paragraph{Well-tempered metadynamics}
\label{sec:wtmetad}
In well-tempered metadynamics~\cite{Barducci-PRL-2008}, the bias potential $U^{\mathrm{bias}}(\vz)$ is updated through the following stochastic iteration scheme where every $N_G$ simulation steps (i.e., MD steps) we add a Gaussian biasing kernel $G(\vz,\vz_n)$ at the current \hyperlink{ref:CV} {CV} value $\vz_n$
\begin{align}
\label{wtmetad_update}
U^{\mathrm{bias}}_{n}(\vz) = U^{\mathrm{bias}}_{n-1}(\vz) +
\exp \left[-\frac{1}{\gamma-1} \beta  U^{\mathrm{bias}}_{n-1}(\vz_n)   \right]
\, G(\vz,\vz_{n}). 
\end{align}
Here $n$ is the current step number in the recursive updating of the bias potential (i.e., number of added Gaussian kernels; note that this is not the same as MD steps), $U^{\mathrm{bias}}_{0}(\vz)=0$, and the Gaussian kernels are scaled by the factor $\exp \left[-\frac{1}{\gamma-1}
\beta U^{\mathrm{bias}}_{k-1}(\vz_n)   \right]$, with $\gamma$ a parameter called the bias factor. The Gaussian kernels are given by
\begin{equation}
G(\vz,\vz_n)=H_{0}\exp \left[-\frac{1}{2} \left(\vz-\vz_n\right)^{\mathrm{T}}\boldsymbol{\Sigma}^{-1}\left(\vz-\vz_n\right)\right],
\end{equation}
where $H_{0}$ is the height and $\boldsymbol{\Sigma}$ is the variance matrix of the kernel. Generally the variance matrix is taken as diagonal, $\Sigma_{ij}= \delta_{ij} \sigma_{i}^{2}$ where $\delta_{ij}$ is the Kronecker delta function ($\delta_{ij}=1$ if $i=j$ and 0 otherwise) and $\boldsymbol{\sigma}=[\sigma_1,\ldots,\sigma_d]$ is a vector of the standard deviations (i.e., widths) corresponding to the CVs. In this case, the Gaussian kernels can be written as
\begin{equation}
G(\vz,\vz_n)=H_{0} \exp \left[-\frac{1}{2}\sum^{d}_{i=1} \frac{(z_i-z_{n,i})^2}{\sigma^2_i} \right],
\end{equation}
where $d$ is the number of \hyperlink{ref:CV} {CV}s

In between bias potentials updates, after depositing $n$ Gaussian kernels, the bias potential is given by
\begin{align}
\label{wtmetad_sum}
U_{t}^{\mathrm{bias}}(\vz) &=
\sum_{k=1}^{n}
\exp \left[-\frac{1}{\gamma-1} \beta  U^{\mathrm{bias}}_{k-1}(\vz_k)\right]
G(\vz,\vz_{k})
\nonumber
\\
& =
\sum_{k=1}^{n}
H_{k} \,
\exp \left[-\frac{1}{2}\sum^{d}_{i=1} \frac{(z_i-z_{k,i})^2}{\sigma^2_i} \right]
\end{align}
where we write $H_{k} = H_{0} \, \exp \left[-\frac{1}{\gamma-1} \beta U^{\mathrm{bias}}_{k-1}(\vz_k)\right]$ as the height of the $k$-th added Gaussian.
In the long time limit, the height $H_k$ goes to zero as the scaling factor $\exp \left[-\frac{1}{\gamma-1} \beta U^{\mathrm{bias}}_{k-1}(\vz_k)   \right]$ decreases as $\frac{1}{k}$~\cite{Barducci-PRL-2008,Dama-PRL-2014} (see Equation 3 in Ref~\cite{Barducci-PRL-2008}). Therefore, as the metadynamics simulations progresses, the change of the bias potential is smaller and it becomes quasi-stationary.
It has been proven~\cite{Dama-PRL-2014} that updating the bias potential according to Equation~\ref{wtmetad_update} leads to an asymptotic solution given by
\begin{equation}
\label{eq:wtmetad_asymptotic}
U_{t}^{\mathrm{bias}}(\vz) = - \left(1-\frac{1}{\gamma} \right)
A(\vz) + K(t)
\end{equation}
where $K(t)$ is a time-dependent constant.
Note that in the metadynamics literature, the \hyperlink{ref:FES} {FES} is generally denoted as $F(\vz)$ (or $F(\mathbf{s})$) instead of $A(\vz)$.

The bias potential in Equation~\ref{eq:wtmetad_asymptotic} only partially cancels out the \hyperlink{ref:FES} {FES} and the \hyperlink{ref:CV} {CV}s  are sampled according to a so-called well-tempered distribution
\begin{equation}
\label{eq:wtmetad_wtdist}
\tilde{\rho}(\vz) =
\frac
{[\rho(\vz)]^{1/\gamma}}
{\int [\rho(\vz)]^{1/\gamma} \, d\vz},
\end{equation}
which can be viewed as a distribution where \hyperlink{ref:CV} {CV} fluctuations are enhanced as compared to the equilibrium \hyperlink{ref:Distribution} {distribution} , see Figure~\ref{fig:MetaD}. The bias factor determines the magnitude of the fluctuation enhancements and also how fast the height $H_{k}$ of the deposited Gaussian kernels goes to zero.

By taking the logarithm on both sides of Equation~\ref{eq:wtmetad_wtdist}, we can view the well-tempered distribution as sampling an effective \hyperlink{ref:FES} {FES},  $A_{\gamma}(\vz) = \frac{1}{\gamma} A(\vz)$ (up to a constant), where the barriers have been reduced by a value corresponding to the selected bias factor, as we can see in Figure~\ref{fig:MetaD}. This gives a rule of thumb for selecting the bias factor: it should be chosen such that the barriers become on the order of the thermal energy so that the system can easily migrate between \hyperlink{ref:metastab} {metastable} states on the simulation timescale.

In the metadynamics literature, a temperature parameter $\Delta T$ is often used instead of the bias factor $\gamma$. Their relation is given by $\gamma = (T + \Delta T)/T$. We can re-write the well-tempered distribution in Equation~\ref{eq:wtmetad_wtdist} as 
\begin{equation}
\label{eq:wtmetad_wtdist_deltaT}
\tilde{\rho}(\vz) =
\frac
{\exp\left[-\frac{1}{k_\mathrm{B}(T+\Delta T)}A(\vz)\right]}
{\int \exp\left[-\frac{1}{k_\mathrm{B}(T+\Delta T)}A(\vz)\right] d\vz}=
\frac
{e^{-\bar{\beta} \, A(\vz)}}
{\int e^{-\bar{\beta} \, A(\vz)} \, d\vz},
\end{equation}
where $\bar{\beta}=\left[k_\mathrm{B}(T+\Delta T)\right]^{-1}$. This introduces another interpretation of well-tempered metadynamics. We can view the well-tempered distribution as sampling the \hyperlink{ref:CV} {CV}s at a higher temperature $T+\Delta T$ (but with the \hyperlink{ref:FES} {FES} $A(\vz)$ fixed). 

Like most \hyperlink{ref:CV} {CV}-based enhanced sampling methods, metadynamics cannot work with too many \hyperlink{ref:CV} {CV}s. In practical applications, we are generally limited to biasing three to four \hyperlink{ref:CV} {CV}s. However, there are variants of metadynamics that allow us to employ a large number of \hyperlink{ref:CV} {CV}s, see Sections~\ref{sec:pb-metad} and~\ref{sec:be-metad}.

\paragraph{Obtaining the free energy surface}
\label{sec:metad_obtaining_fes}
The \hyperlink{ref:FES} {FES} can be estimated directly from the bias potential at time $t$ through Equation~\ref{eq:wtmetad_asymptotic} by summing up the deposited Gaussians
\begin{align}
\label{eq:metad_sumhills}
A(\vz) &=
- \left(\frac{\gamma}{\gamma-1}\right) U_{t}^{\mathrm{bias}}(\vz) + 
\frac{1}{\beta}
\ln 
\int \exp
\left[
\frac{\gamma}{\gamma-1} \beta U_{t}^{\mathrm{bias}}(\vz)
\right] \, d\vz 
\nonumber
\\
& =
- \left(\frac{\gamma}{\gamma-1}\right)
\sum_{k=1}^{n}
H_{k} \,
\exp \left[-\frac{1}{2}\sum^{d}_{i=1} \frac{(z_i-z_{k,i})^2}{\sigma^2_i} \right] 
\nonumber \\
& \phantom{=}
+ 
\frac{1}{\beta}
\ln 
\int \exp
\left[
\frac{\gamma}{\gamma-1} \beta U_{t}^{\mathrm{bias}}(\vz)
\right] \, d\vz
,
\end{align}
where the second term is a time-dependent constant and allows us to assess the \hyperlink{ref:FES} {FES}'s convergence and measure error bars~\cite{tiwary_rewt,Valsson2016_ARPC_MetaD}. Here, the two terms will increase in magnitude with time, the first term with the sum over the Gaussians will become more negative, while the second term with the integral will become more positive. However, their sum will converge. Alternately, a common procedure to assess convergence is to compare \hyperlink{ref:FES} {FES}s obtained at different times by aligning their minimum to zero.

We can also obtain the \hyperlink{ref:FES} {FES} via various post-processing \hyperlink{ref:Reweighting} {reweighting} procedures~\cite{bonomi_rewt,tiwary_rewt,Branduardi-JCTC-2012,Schafer_RewMetaD_2020,Giberti_IterRew_JCTC2019,10.1063/1.5123498,10.1016/j.cplett.2020.137384}. The most common way to achieve this task is the so-called $c(t)$ reweighting procedure~\cite{bonomi_rewt,tiwary_rewt} that takes the time-dependence of the bias potential into account. 

Starting from Equation~\ref{eq:biased_x_distribtion_rewritten}, we can re-write the biased configurational \hyperlink{ref:Distribution} {distribution} at time $t$ as 
\begin{equation}
\label{eq:biased_prob_dist_with_c_of_t}
\tilde{\nu}_{t}(\vx) = \nu(\vx) \, \exp
\left(-\beta\left[
U_{t}^{\mathrm{bias}}(\xi(\vx))-c(t)
\right] \right),
\end{equation}
where the time-dependent constant $c(t)$ is the logarithm of the ratio of the unbiased and biased partition functions (see Equation~\ref{eq:fraction_of_partition_functions}),
\begin{equation}
c(t) = 
\frac{1}{\beta} \ln \frac{Z}{\tilde{Z}(t)}.
\end{equation}
Here the biased partition function $\tilde{Z}(t)$ is explicitly time-dependent due to dependence on the \hyperlink{ref:Adaptive} {adaptive} metadynamics bias potential. 
The unbiased and biased partition functions can be written as integrals over either the coordinates $\vx$ or the CVs $\vz$, 
\begin{align}
& Z = \int e^{-\beta U(\vx) } \, d\vx = \int e^{-\beta A(\vz) } \, d\vz \\
& \tilde{Z}(t) = \int e^{-\beta\left [ U(\vx) + U_{t}^\mathrm{bias}(\xi(\vx)) \right]} \, d\vx = \int e^{-\beta\left [ A(\vz) + U_{t}^\mathrm{bias}(\vz) \right]} \, d\vz.
\end{align}
We can thus write the time-dependent constant $c(t)$ as \begin{equation}
\label{eq:wtmetad_coft2}
c(t) = 
\frac{1}{\beta} \ln \frac{Z}{\tilde{Z}(t)} = 
\frac{1}{\beta} \ln \frac
{\int e^{-\beta A(\vz)} \,  d\vz }
{\int e^{-\beta \left[ A(\vz) + U_{t}^{\mathrm{bias}}(\vz) \right]} \, d\vz }.
\end{equation}
In practice, $c(t)$ can be estimated using
\begin{equation}
\label{eq:wtmetad_coft_estimate}
c(t) = \frac{1}{\beta} \ln
\frac
{\int \exp \left[ \frac{\gamma}{\gamma-1} \beta U_{t}^{\mathrm{bias}}(\vz)  \right] d\vz }
{\int \exp \left[ \frac{1}{\gamma-1} \beta U_{t}^{\mathrm{bias}}(\vz)  \right] d\vz },
\end{equation}
which is obtained by inserting Equation~\ref{eq:metad_sumhills} into Equation~\ref{eq:wtmetad_coft2}~\cite{tiwary_rewt,Valsson2016_ARPC_MetaD}. 

The biased configurational \hyperlink{ref:Distribution} {distribution}  (Equation~\ref{eq:biased_prob_dist_with_c_of_t}) and the $c(t)$ (Equation~\ref{eq:wtmetad_coft_estimate}) allows us to reweight and calculate any average of an observable as
\begin{equation}
\label{wtmetad_reweighting}
\langle O(\vx) \rangle = \frac{\langle O(\vx)
\,
e^{
\beta\left[
U_{t}^{\mathrm{bias}}(\xi(\vx))-c(t)
\right]}\rangle_{\tilde U}}
{\left\langle
e^{\beta \left[
U_{t}^{\mathrm{bias}}(\xi(\vx))-c(t)
\right]} \right\rangle_{\tilde U}}, 
\end{equation}
where $\langle \cdots \rangle_{\tilde U}$ indicates an \hyperlink{ref:ensemble_average} {ensemble average} in the biased metadynamics simulation (see Equation~\ref{eq:reweighting_bias_potential}). To perform the $c(t)$ reweighting in practice, we weight each configuration $\vx$, taken at time $t$, with weighting factor $e^{\beta\left[U_{t}^{\mathrm{bias}}(\xi(\vx))-c(t)\right]}$ acting at time $t$. We can think of the sum $U_{t}^{\mathrm{bias}}(\xi(\vx))-c(t)$ as a relative (or re-normalized) bias potential (the relative bias potential will converge while the two terms will still increase). This \hyperlink{ref:Reweighting} {reweighting} procedure assumes that the relative bias potential is quasi-stationary. Therefore, in practice, we ignore a short initial transient part of the simulation where the relative bias potential is not converged and still changing considerably.

Another \hyperlink{ref:Reweighting} {reweighting} procedure that avoids the calculation of the time-dependent constant $c(t)$ is the so-called last bias \hyperlink{ref:Reweighting} {reweighting}~\cite{Branduardi-JCTC-2012}. In this case, we use weights obtained using the bias potential at the end of the simulation, in other words, the last bias potential. An average of an observable is then calculated as
\begin{equation}
\label{wtmetad_reweighting_lastbias}
\langle O(\vx) \rangle = \frac{\langle O(\vx)
\,
e^{
\beta
U_{t=t_F}^{\mathrm{bias}}(\xi(\vx))
}\rangle_{\tilde U}}
{\langle
e^{\beta
U_{t=t_F}^{\mathrm{bias}}(\xi(\vx))
} \rangle_{\tilde U}},
\end{equation}
where $t_F$ is the time at the end of the simulation. 

Using $O(\vx) = \delta(\vz-\xi(\vx))$ in Equation~\ref{wtmetad_reweighting} or~\ref{wtmetad_reweighting_lastbias} (see Equation~\ref{eq:fes_definition}), allows us to use \hyperlink{ref:Reweighting} {reweighting} to estimate the \hyperlink{ref:FES} {FES} as a function of the biased \hyperlink{ref:CV} {CV}s, or for any other set of \hyperlink{ref:CV} {CV}s. It is a good practice to compare the \hyperlink{ref:FES} {FES} estimated from the bias potential via Equation~\ref{eq:metad_sumhills} to the \hyperlink{ref:FES} {FES} estimated using \hyperlink{ref:Reweighting} {reweighting}. They should give the same results and a major disagreement would indicate that the simulation is not converged or that the \hyperlink{ref:CV} {CV}s might not be good enough. Furthermore, we can estimate error bars for reweighted \hyperlink{ref:FES} {FES}s using block averaging (see PLUMED tutorials~\cite{plumed_masterclass}).

Other post-processing methods for obtaining the \hyperlink{ref:FES} {FES} include mean force integration~\cite{10.1063/1.5123498} and a weighted histogram analysis method adapted to metadynamics~\cite{10.1016/j.cplett.2020.137384}, which both allow for estimating the \hyperlink{ref:FES} {FES} by combining different independent metadynamics simulations.  Metadynamics has also been combined with Gaussian Process Regression for obtaining the \hyperlink{ref:FES} {FES}~\cite{Mones2016}. See the references for further details on these post-processing methods.

\paragraph{Conventional metadynamics}
\label{sec:meta-classic}
In conventional (non-well-tempered) metadynamics~\cite{Laio-PNAS-2002}, which is the original formulation of metadynamics, the height of Gaussian kernels is kept fixed throughout the simulations. We can view this as the limit $
\gamma \to \infty$ for well-tempered metadynamics. At convergence, this leads to a bias potential that oscillates around the negative of the \hyperlink{ref:FES} {FES}. Thus, on average the bias potentials cancels out the \hyperlink{ref:FES} {FES} and a uniform \hyperlink{ref:CV} {CV} sampling is obtained. In this case, the \hyperlink{ref:FES} {FES} should be estimated as the negative of a time-average of the bias potentials~\cite{BussiLaio_ReviewMetaD_2020}
\begin{equation}
A(\vz) = - \frac{1}{n-n_0} \sum^{n}_{k=n_0} U^{\mathrm{bias}}_k(\vz).
\end{equation}
The accuracy of this estimate will depend on how well the biased \hyperlink{ref:CV} {CV}s are \hyperlink{ref:AdiabaticDyn} {adiabatically} separated from the dynamics of the other variables~\cite{laio-gervasio-08,jourdain-lelievre-zitt-21}, see Ref~\cite{BussiLaio_ReviewMetaD_2020} for further discussion regarding this point.

In conventional metadynamics simulations, it is a common practice to introduce restraining walls (i.e., bias potential) to avoid exploring unimportant free energy regions. This is generally not needed in well-tempered metadynamics simulations as the bias factor and the well-tempered distribution naturally limits the extent of CV space exploration.

\paragraph{Multiple walkers metadynamics}
\label{sec:mwmeta}
A way to reduce the wall-clock time for convergence and make a better usage of modern parallel HPC resources is to employ multiple walkers or replicas~\cite{Raiteri-JPCB-2006}. The walkers collaboratively sample the free energy \hyperlink{ref:FES} {landscape} and share a bias potential that is constructed by considering the Gaussians deposited by all the walkers.

\paragraph{Well-tempered ensemble}
\label{sec:wtensemble}
The well-tempered ensemble~\cite{Bonomi-PRL-2010} is obtained by using the potential energy $U$ as \hyperlink{ref:CV} {CV} within well-tempered metadynamics. It is an example of a category of ABP methods that use the potential energy as a CV that we will discuss in more detail below in Section~\ref{sec:abp_energy}. In the well-tempered ensemble, we obtain at convergence a statistical \hyperlink{ref:Ensemble} {ensemble} defined by $\tilde{\rho}(U) \propto [\rho(U)]^{1/\gamma}$ where the potential energy $U$ has the same average as the canonical one but mean square fluctuations amplified by a factor of $\gamma$ (assuming that $\rho(U)$ is roughly Gaussian). By tuning $\gamma$, we can interpolate between the canonical ensemble ($\gamma=1$) and the multicanonical~\cite{Berg1992_Multicanonical} ensemble ($\gamma \to \infty$). In Section~\ref{sec:pt-wte}, we show how this property of the well-tempered ensemble can be used to our benefit when combined with parallel-tempering (see Section~\ref{sec:generalized-ensemble}). The well-tempered ensemble can also be used to obtain thermodynamic properties like the \hyperlink{ref:density_of_states}{density of states}~\cite{Valsson-JCTC-2013}. It has been shown~\cite{Junghans2014wte-wl} that the well-tempered ensemble can be mathematically related to the Wang-Landau method~\cite{wang-landau:prl:2001:wang-landau} and statistical temperature molecular dynamics~\cite{Kim2006_PRL_STMD}.

\paragraph{Parallel-bias metadynamics}
\label{sec:pb-metad}
Parallel-bias metadynamics~\cite{Pfaendtner2015_pbmetad} allows for \hyperlink{ref:biasingE} {biasing} a large number of \hyperlink{ref:CV} {CV}s simultaneously. This is done by considering many low-dimensional bias potentials (generally one-dimensional), each biasing a separate \hyperlink{ref:CV} {CV}s, all acting within a single simulation.  The metadynamics bias potential update step is modified to account for the effect of the bias potentials on each other. Thus, the method yields the correct low-dimensional free energy \hyperlink{ref:FES} {profile} for each \hyperlink{ref:CV} {CV}. The method has been extended to bias \hyperlink{ref:CV} {CV} sets where the \hyperlink{ref:CV} {CV}s can be considered identical or indistinguishable~\cite{Prakash2018_pbmetad-families}.

\paragraph{Infrequent metadynamics}
\label{sec:infreq_meta}
Well-tempered metadynamics has been extended for obtaining kinetics of rare events in a so-called infrequent metadynamics method~\cite{Tiwary-PRL-2013}. The idea is to reduce the deposition rate, in other words increase $N_{G}$, the number of simulation steps between depositing Gaussians. In this way, we can avoid adding bias to the transition state region. If there is no bias acting on the transition state, we can rescale the time using the ideas of hyperdynamics~\cite{Voter-PRL-1997} and conformational flooding~\cite{Grubmuller-PRE-1995}. The physical time for a rare event to happen $t_b$ is obtained by summing up the MD steps and scaling the MD time step $dt$ by the bias acting at each step
\begin{equation}
\label{wtmetad_hyperdynamics}
t_b = \sum_{i} dt \, e^{\beta U_{t=t_{i}}^{\mathrm{bias}}(\vz(t_{i}))},
\end{equation}
where $t_{i} = i\, dt$. By performing multiple simulations, one can obtain a distribution of escape times from a long-lived \hyperlink{ref:metastab} {metastable} state. It is possible to assess the reliability of the kinetics by performing a statistical analysis to test how well the obtained escape time distribution follows the expected time-homogeneous Poisson distribution~\cite{KS_Test_JCTC_2014}. An extension on the idea of infrequent metadynamics is frequency adaptive metadynamics~\cite{Wang2018_FA-MetaD} where the deposition rate is adjusted on the fly during the simulation. In addition, various other approaches and strategies for obtaining rare-event kinetics from metadynamics simulations have been introduced in the literature~\cite{
10.1021/acs.biochem.8b00977,
10.1021/ar500356n,
10.1371/journal.pcbi.1000452,
10.1021/ja903045y,
10.1103/physrevlett.110.108106,
10.1063/1.5027728,
10.1103/physreve.98.052408, 
10.1063/5.0019100,
10.1021/acs.jpclett.2c01807}

\paragraph{Adaptive Gaussian metadynamics}
\label{sec:AGmeta}
In the \hyperlink{ref:Adaptive} {adaptive} Gaussian variant~\cite{Branduardi-JCTC-2012}, the shape of the deposited Gaussian biasing kernel is not kept fixed but changes with time and adapts to the features of the underlying \hyperlink{ref:FES} {FES}, by employing an off-diagonal variance matrix that is dynamically adjusted according to the estimated shape of the underlying free energy landscape. Employing adaptive Gaussian can improve the performance in some cases, for example, if the \hyperlink{ref:FES} {FES}'s \hyperlink{ref:metastab} {metastable} states differ considerably in their shape, or if the \hyperlink{ref:CV} {CV}s are highly coupled.

\paragraph{Other variants of metadynamics}
\label{sec:metad_variants}
An non-exhaustive list of other extensions and variants of metadynamics that have been introduced throughout the years includes
reconnaissance metadynamics~\cite{10.1073/pnas.1011511107},
$\lambda$-metadynamics~\cite{10.1021/jz200808x},
flux-tempered metadynamics~\cite{Singh-JStatPhys-2011,Singh-JCTC-2012},
path-metadynamics~\cite{10.1103/physrevlett.109.020601,10.1063/1.5027392},
funnel metadynamics~\cite{10.1073/pnas.1303186110,10.1038/s41596-020-0342-4},
algorithms for boundary corrections~\cite{McGovern-JCP-2013},
transition-tempered metadynamics~\cite{Dama-JCTC-2014}, metabasin metadynamics~\cite{10.1021/acs.jctc.5b00907},
experiment directed metadynamics~\cite{White_EDM_2015}
ensemble-biased metadynamics~\cite{Marinelli_EnsembleBiased_2015},
target metadynamics~\cite{GilLey_TargetMetaD_2016},
$\mu$-tempered metadynamics~\cite{10.1063/1.4937939},
adaptive-numerical-bias metadynamics~\cite{10.1002/jcc.25066},
altruistic metadynamics~\cite{10.1021/acs.jpcb.6b00087,10.1063/1.4978939},
metadynamics for automatic sampling of quantum property manifolds~\cite{Lindner_ASQPM-MetaD_2019},
and
metadynamics with scaled hypersphere search for high-dimensional FES~\cite{Mitsuta_SHS-MetaD_JCTC2020}.
We refer the reader to the references for further details. Metadynamics has also been used in various types of hybrid methods as discussed in Section~\ref{sec:hybrids}.

\paragraph{Public implementations of metadynamics}
\label{sec:meta-impl}
Various variants of metadynamics are implemented in the PLUMED 2 enhanced sampling plug-in~\cite{Bonomi-CPC-2009,Tribello2014,plumed-nest}. PLUMED also includes numerous practical tutorials showing how to perform and analyze metadynamics simulations~\cite{plumed_masterclass}. Furthermore, a large number of example input files are
available in the PLUMED-NEST~\cite{plumed-nest,plumed_nest_url}

Metadynamics are also available in the external libraries SSAGES and the Colvars module (see Table~\ref{Table:Libraries})., and also is natively implemented in some MD codes codes such as CP2K and DESMOND, see Table~\ref{Table:Codes} and and also Table 2 in Ref~\cite{BussiLaio_ReviewMetaD_2020}. 

\subsubsection{Variationally enhanced sampling}
\label{sec:ves}
A more recently introduced ABP method is variationally enhanced
sampling (VES) that is based on a variational principle~\cite{Valsson_VES_PRL_2014,Valsson2020Handbook_VES}.
In variationally enhanced sampling, a bias potential $U^{\mathrm{bias}}(\vz)$ is constructed by minimizing a convex functional given by
\begin{equation}
\label{ves_omega}
\Omega [U^{\mathrm{bias}}] =
\frac{1}{\beta} \ln
\frac
{\int e^{-\beta \left[ A(\vz) + U^{\mathrm{bias}}(\vz)\right]} \, d\vz }
{\int e^{-\beta A(\vz)} \, d\vz}
+
\int p_{\mathrm{tg}}(\vz) \, U^{\mathrm{bias}}(\vz) \, d\vz,
\end{equation}
where $p_{\mathrm{tg}}(\vz)$ is a so-called \hyperlink{ref:targetdist}{target distribution} that is chosen by the user. The $\Omega [U^{\mathrm{bias}}]$ functional has a global minimum given by
\begin{equation}
\label{ves_biaspotential}
U^{\mathrm{bias}}(\vz) = -A(\vz)- \frac{1}{\beta} \ln {p_{\mathrm{tg}}(\vz)} + C,
\end{equation}
where $C$ is an unimportant constant. This bias potential results in a biased \hyperlink{ref:CV} {CV} distribution that is equal to the \hyperlink{ref:targetdist}{target distribution}, $\tilde{\rho}(\vz) = p_{\mathrm{tg}}(\vz)$. Therefore, the \hyperlink{ref:targetdist}{target distribution} determines the \hyperlink{ref:CV} {CV} sampling that is obtained when minimizing $\Omega [U^{\mathrm{bias}}]$. By choosing a \hyperlink{ref:targetdist}{target distribution} that is easier to sample than the equilibrium \hyperlink{ref:Distribution} {distribution} , we enhance the sampling of the CVs. Furthermore, we can directly obtain the \hyperlink{ref:FES} {FES} from the bias potential through Equation~\ref{ves_biaspotential}.

To obtain a better understanding of the functional in Equation~\ref{ves_omega}, we can rewrite it as~\cite{Valsson2020Handbook_VES,Invernizzi-PNAS-2017}
\begin{equation}
\beta \Omega [U^{\mathrm{bias}}] = D_{\mathrm{KL}}(p_{\mathrm{tg}}||\tilde{\rho})-D_{\mathrm{KL}}(p_{\mathrm{tg}}||\rho),
\end{equation}
where $D_{\mathrm{KL}}(p||q)=\int p(\vx) \ln\frac{p(\vx)}{q(\vx)} \, d\vx$ is the Kullback-Leibler divergence between a probability distribution $p(\vx)$ and a probability distribution $q(\vx)$ (or more correctly, from  $q(\vx)$ to  $p(\vx)$ as the Kullback-Leibler divergence is not symmetric with respect to its arguments). Therefore, minimizing $\Omega [U^{\mathrm{bias}}]$ is equivalent to minimizing the Kullback-Leibler divergence (or the relative entropy or cross entropy) between the \hyperlink{ref:targetdist}{target distribution} $p_{\mathrm{tg}}(\vz)$ and the \hyperlink{ref:BiasedCVDist} {biased distribution} $\tilde{\rho}(\vz) \propto e^{-\beta \left[A(\vz) + U^\mathrm{bias}(\vz)\right]}$~\cite{Rubinstein-MetCompAppProb-1999,Shell-JCP-2008,Bilionis-JCompPhys-2012,Zhang_SIAM-JSC-2014}. Note that the second term, $D_{\mathrm{KL}}(p_{\mathrm{tg}}||\rho)$, is independent of $U^{\mathrm{bias}}(\vz)$ and is a constant for a given \hyperlink{ref:targetdist}{target distribution} $p_{\mathrm{tg}}(\vz)$.

In practice, $\Omega [U^{\mathrm{bias}}]$ is minimized by introducing a functional form for the bias potential $U_{\boldsymbol{\alpha}}^\mathrm{bias}(\vz)$ that depends on a set of variational parameters $\boldsymbol{\alpha}$. We then go from an abstract functional minimization to a minimization of the multidimensional function $\Omega(\boldsymbol{\alpha}) = \Omega [U_{\boldsymbol{\alpha}}^{\mathrm{bias}}]$. The elements of the gradient $\nabla \Omega(\boldsymbol{\alpha})$ are defined as
\begin{equation}
\frac
{\partial \Omega({\boldsymbol{\alpha}})}
{\partial \alpha_{i}}
 = -
\left<
\frac
{\partial U_{\boldsymbol{\alpha}}^{\mathrm{bias}}(\vz)}
{\partial \alpha_{i}}
\right>_{\tilde{U}_{\boldsymbol{\alpha}}}
+
\left<
\frac
{\partial U_{\boldsymbol{\alpha}}^{\mathrm{bias}}(\vz)}
{\partial \alpha_{i}}
\right>_{p_{\mathrm{tg}}}.
\end{equation}
The first term is an \hyperlink{ref:ensemble_average} {ensemble average} obtained in the biased ensemble given by the
\hyperlink{ref:biasingE}{biased potential energy} $\tilde{U}_{\boldsymbol{\alpha}}(\vx) = U(\vx) + U_{\boldsymbol{\alpha}}^{\mathrm{bias}}(\xi(\vx))$, and the second term is an average over the \hyperlink{ref:targetdist}{target distribution} $p_{\mathrm{tg}}(\vz)$.  The gradient is noisy due to the need of estimating the first term from a biased simulation. Therefore, it is better to employ stochastic optimization methods to minimize $\Omega({\boldsymbol{\alpha}})$ by iteratively updating the bias potential.

The most general bias representation is to use a linear expansion in some set of basis functions
\begin{equation}
U_{\boldsymbol{\alpha}}^{\mathrm{bias}}(\vz) = \sum_{\mathbf{k}} \alpha_{\mathbf{k}} \cdot f_{\mathbf{k}}(\vz).
\end{equation}
This could for example be a tensor product of one-dimensional basis functions like Chebyshev or Legendre polynomials, or wavelets. In particular, localized wavelet basis functions have been shown to perform the best~\cite{ValssonPampel_Wavelets_2022}. A neural network bias potential has also been used~\cite{Bonati2019_NN-VES}. Furthermore, one can use bespoke bias potentials like a model of the free energy \hyperlink{ref:FES} {profile}~\cite{McCarty2016_JCTC,Piaggi2016_Faraday}.

The \hyperlink{ref:targetdist}{target distribution} $p_{\mathrm{tg}}(\vz)$ can be chosen freely by the user. However, one needs to keep in mind that it should be chosen such that the sampling is easier than in the equilibrium \hyperlink{ref:Distribution} {distribution} . The most straightforward choice would be a uniform \hyperlink{ref:targetdist}{target distribution}, such that the aim to completely cancel out the \hyperlink{ref:FES} {FES} and flatten the sampling in the \hyperlink{ref:CV} {CV} space. However, that is generally not optimal. Instead it is better to only enhance \hyperlink{ref:CV} {CV} fluctuations to a certain degree and thus only partly cancel out the \hyperlink{ref:FES} {FES}~\cite{Valsson-JCTC-2015}. We can achieve this by employing a well-tempered distribution as in Equation~\ref{eq:wtmetad_wtdist} (see Figure~\ref{fig:MetaD}). The bias factor $\gamma$ determines how much we enhance \hyperlink{ref:CV} {CV} fluctuations. The well-tempered distribution is unknown \textit{a priori} so it needs be determined iteratively~\cite{Valsson-JCTC-2015}.

Note that in variationally enhanced sampling, we generally do not need to account for time-dependent constants as in metadynamics when performing \hyperlink{ref:Reweighting} {reweighting}. It is sufficient to include the bias potential in the weights (see Equation~\ref{eq:reweighting_bias_potential}). In other words, we can obtain the unbiased equilibrium average of an observable as 
\begin{equation}
\label{ves_reweighting}
\langle O(\vx) \rangle = \frac{\langle O(\vx)
\,
e^{
\beta
U_{\boldsymbol{\alpha}}^{\mathrm{bias}}(\xi(\vx))
}\rangle_{\tilde U_{\boldsymbol{\alpha}}}}
{\left\langle
e^{\beta 
U_{\boldsymbol{\alpha}}^{\mathrm{bias}}(\xi(\vx))
} \right\rangle_{\tilde U_{\boldsymbol{\alpha}}}}, 
\end{equation}
where $\langle \cdots \rangle_{\tilde U_{\boldsymbol{\alpha}}}$ indicates an \hyperlink{ref:ensemble_average} {ensemble average} in the biased simulation  This \hyperlink{ref:Reweighting} {reweighting} procedure assumes a quasi-stationary bias potential and thus is valid after a short initial transient (i.e., once the variational parameters $\boldsymbol{\alpha}$ are no longer changing substantially). 

The \hyperlink{ref:FES} {FES} can be estimated directly from the bias potential though Equation~\ref{ves_biaspotential}. Furthermore, the \hyperlink{ref:FES} {FES}, both for the biased \hyperlink{ref:CV} {CV}s and for any other \hyperlink{ref:CV} {CV}s, can be obtained through \hyperlink{ref:Reweighting} {reweighting} by using $O(\vx) = \delta(\vz-\xi(\vx))$ in Equation~\ref{ves_reweighting} (see Equation~\ref{eq:fes_definition}). For the reweighted FES, we can also estimate error bars. As for metadynamics, it is a good practice to estimate the \hyperlink{ref:FES} {FES} both from the bias potential and through \hyperlink{ref:Reweighting} {reweighting} and compare the results. Furthermore, in cases where the bias potential might not have sufficient variational flexibility to represent the underlying \hyperlink{ref:FES} {FES}, the reweighted results can be more accurate.

Variationally enhanced sampling has been extended to obtaining kinetics of rare events~\cite{McCarty-PRL-2015}, using the same principles of hyperdynamics~\cite{Voter-PRL-1997} as used in infrequent metadynamics. The idea is to use variationally enhanced sampling to construct a bias potential that only floods the \hyperlink{ref:FES} {FES} up to some given cutoff value. If the flooding bias potential does not touch the transition state, we can rescale the biased simulation times using Equation~\ref{wtmetad_hyperdynamics} as in infrequent metadynamics. It is convenient and useful to combine the flooding bias potential with infrequent metadynamics as done in Ref~\cite{Palazzesi2017_JPCL}.

Furthermore, variationally enhanced sampling has been extended in various ways, for example:
to obtain parameters for phenomenological coarse-grained model~\cite{Invernizzi-PNAS-2017};
to perform Monte Carlo renormalization group simulations~\cite{Wu_VES-RGMC_PRL2017,Wu_VES-RGMC_PRE2019,Wu_VES-RGMC_PRL2020};
to perform multithermal-multibaric simulations~\cite{Piaggi_MultiVES_2019};
for enhanced sampling targeting transition states~\cite{Debnath_VES-TS_2019}; and
multiscale simulations for sub-optimal CVs~\cite{Invernizzi_VES_DelteF_2019}.

Variationally enhanced sampling is implemented in the VES Code module of PLUMED 2. The VES Code also includes practical tutorials showing how to employ the method~\cite{plumed_masterclass}.

\subsubsection{Wang-Landau and other adaptive biasing potential methods using the potential energy as a collective variable}
\label{sec:abp_energy}

A certain category of enhanced sampling methods can be loosely defined as \hyperlink{ref:Adaptive} {adaptive} biasing potential methods using the potential energy $U$ as a CV, even though they might not be explicitly formulated in terms of a bias potential. The most well-known example in this category is the Wang-Landau method~\footnote{The Wang-Landau procedure for converging weights can also be used with other algorithms with discrete states, such as the \hyperlink{ref:ExpEns} {expanded ensemble} algorithms; see Section~\ref{sec:singlestate} for further details}~\cite{wang-landau:prl:2001:wang-landau}. 

What unites many of these methods is that they aim to sample a potential energy distribution $p_{\mathrm{tg}}(U)$ (i.e., a target potential energy distribution) that is broadened or expanded as compared to the unbiased distribution corresponding to the simulation temperature. Thus, they can be considered as sampling from a ensemble where the sampling should be easier~\footnote{sometimes called a generalized ensemble in the literature, see Section~\ref{sec:generalized-ensemble}.}. A well known example of this is the multicanonical (or multithermal) ensemble where the aim to sample the potential energy near uniformly in a given range~\cite{Berg1992_Multicanonical}. The idea can also be extended to multithermal-multibaric simulations by incorporating the volume as a variable whose dynamics is biased or modified~\cite{Okumura_MultiTP_2004,Shell_MultiTP_2002}. Furthermore, sometime only a subset of the potential energy is used to better focus the sampling enhancement~\cite{Yang_SITS_2009}. Other methods in this category include for example
the multicanonical ensemble~\cite{Berg1992_Multicanonical},
statistical temperature MD~\cite{Kim2006_PRL_STMD},
metadynamics using the energy as CV~\cite{Micheletti_MetaE_Energy_2004,Bonomi-PRL-2010}
(which is the the well-tempered ensemble discussed in Section~\ref{sec:wtensemble} when using well-tempered metadynamics, see also Ref~\cite{Valsson-JCTC-2013}),
integrated tempering sampling~\cite{Gao_ITS_2008,Gao_ITS_Review_2015}.
Furthermore, variationally enhanced sampling and on-the-fly probability-enhanced sampling have been extended to perform simulations in the multithermal and the multithermal-multibaric ensembles as described in Refs~\citep{Piaggi_MultiVES_2019,Piaggi_MultiVES+CV_2019} and~\citep{invernizzi2020unified}, respectively.

To achieve the given target potential energy distribution $p_{\mathrm{tg}}(U)$, these methods employ some kind of \hyperlink{ref:Adaptive} {adaptive} or iterative scheme to estimate \textit{a priori} unknown quantities. For example, these unknown quantities can be the configurational \hyperlink{ref:density_of_states}{density of states}, which is the case in the Wang landau method~\cite{wang-landau:prl:2001:wang-landau,Kim2006_PRL_STMD,DePablo_DOS_2012}, factors or weights in sum over Boltzmann factors at different temperatures~\cite{Gao_ITS_2008,invernizzi2020unified}, or a free energy as in the case of well-tempered metadynamics. As noted above, these methods are not necessarily explicitly formulated in terms of a bias potential. Nevertheless, we can still associate an effective bias potential as function of the potential energy to them
\begin{equation}
U_\mathrm{bias}(U) = - A(U) -\frac{1}{\beta} \ln p_{\mathrm{tg}}(U) = - U + \frac{1}{\beta} \ln \Omega(U) -\frac{1}{\beta} \ln p_{\mathrm{tg}}(U),
\end{equation}
where $\Omega(U) = \int \delta(U-U(\vx)) \, d\vx$ is the configurational \hyperlink{ref:density_of_states}{density of states} that is related to the free energy as function of $U$ through $A(U)=U - \frac{1}{\beta} \ln \Omega(U)$, and we ignore unimportant constants. Therefore, we can loosely define these methods as adaptive \hyperlink{ref:biasingE} {biasing} potential methods using the potential energy $U$ as a \hyperlink{ref:CV} {CV}. From this equation, we can also see a certain theoretical equivalence between methods that work with the density of states, such as Wang-Landau sampling and statistical temperature MD, and methods that work with the free energy, such as metadynamics, see Ref~\cite{Junghans2014wte-wl} for further discussion on this point.

Furthermore, closely related to this category are accelerated MD~\cite{Hamelberg2004} and Gaussian-accelerated MD~\cite{Miao2017, Wang2021}. These methods employ a bias potential acting on the potential energy (or generally a subset of it, e.g., only the dihedral angle terms of the potential energy) to lower the energy barriers of specific transitions. These methods do not directly aim to sample a given \hyperlink{ref:targetdist} {target distribution}, but rather employ a bias potential (e.g., a harmonic potential in the case of Gaussian-accelerated MD) that is designed to lower energy barriers of specific transitions. The bias potential depends on a few so-called boost parameters that are generally \hyperlink{ref:Adaptive} {adaptively} determined through a series of equilibration runs.

\subsection{Adaptive biasing force (ABF)}
\label{sec:ABF}

\begin{figure}
    \centering
    \includegraphics{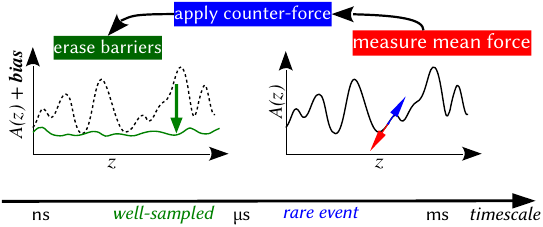}
    \caption{Principle of the ABF method. The native free energy surface (solid black line, right) shows high barriers, resulting in long transition time scales.
    In ABF the mean force (red arrow) is measured and countered by the biasing force (blue arrow). As a result, the biased free energy surface (dashed black line, left) approaches a uniform one, and the transition time scales become short enough to be well-sampled in simulations.}
    \label{fig:ABF}
\end{figure}

The \hyperlink{ref:Adaptive} {adaptive} biasing force method (ABF)\cite{Darve2001, Comer2015} belongs to the same category of methods as the adaptive \hyperlink{ref:biasingE} {biasing} potential methods~\ref{fig:scheme}. Contrary to those, however, it is rooted in free energy estimation by thermodynamic integration (Section~\ref{sec:fe_estimators:TI}).
In ABF, the gradient of the free energy with respect to selected \hyperlink{ref:CV} {CV}s $\vz = \xi(\vx)$ is estimated, and that estimate is used to apply a time-dependent external force $\vF^\mathrm{ABF}_t(\vz)$ that counteracts the estimated free energy gradient, enhancing sampling along those coordinates.

The principle is the following (Figure ~\ref{fig:ABF}):
\begin{enumerate}
 \item A small number of \hyperlink{ref:CV} {collective variables} slow degrees of freedom $\vz=\xi(\vx)$  are chosen.
 \item In the simulation, the gradient of the free energy \hyperlink{ref:FES} {surface} $A(\vz)$  is estimated as:
\begin{equation}
    \nabla_\vz A(\vz) = - \left\langle \vF_\xi(\vx)  \right\rangle_{\xi(\vx) = \vz}
\end{equation}
That is, as the conditional \hyperlink{ref:ensemble_average} {ensemble average} of a collective force $\vF_\xi(\vx)$ at a given value of $\xi(\vx)$.
 \item When sufficient sampling (determined by the adaptation rate) is collected to have a reliable estimate of the average force at the current value of $\vz$, a \hyperlink{ref:biasingE} {biasing} force $\vF^\mathrm{ABF}_t(\vz)$ equal to the opposite of this average is applied.
 \item This force cancels out, on average, the forces acting along $\vz$, leveling the free energy barriers and accelerating diffusion in \hyperlink{ref:CV} {collective variable} space.
 \item At convergence, the \hyperlink{ref:biasingE} {biasing} force is the negative of the gradient of the free energy, and $\vz$ experiences a flat effective free energy \hyperlink{ref:FES} {landscape}. The gradient can be integrated numerically to estimate the free energy \hyperlink{ref:FES} {landscape} itself.
\end{enumerate}

Thus the core of ABF is an adaptation method to calculate a time-dependent \hyperlink{ref:biasingE} {biasing} force $\vF^\mathrm{ABF}_t$ that, at long times, converges towards the free energy gradient $\nabla_\vz A$, or an approximation of it.

The following optional components of an ABF method can be used to obtain reliable free energies:
\begin{enumerate}
\item a different \textit{a posteriori} estimator of the free energy gradient (especially if the one used for \hyperlink{ref:biasingE} {biasing} is approximate, see Sections~\ref{sec:eABF} and \ref{sec:ABF_variants});
\item a method to integrate the free energy gradients and obtain the free energy \hyperlink{ref:FES} {surface}~\cite{Henin2021integration}.
\end{enumerate}

In what we will call standard ABF (Section~\ref{sec:ABF_standard}), a single simulation is run at a time and the exact free energy derivative with respect to the coordinate of interest is estimated directly as an \hyperlink{ref:ensemble_average} {ensemble average}.
Variants may involve multiple interacting replicas of the simulation, fictitious proxy coordinates, approximate gradient estimators, and additional \hyperlink{ref:biasingE} {biasing} forces or potentials. ABF methods can be shown to converge quickly to equilibrium for well chosen reaction coordinates, see for example the following mathematical analysis for rigorous formulations~\cite{lelievre-rousset-stoltz-08,benaim-brehier-monmarche-20}.

\subsubsection{Standard ABF}
\label{sec:ABF_standard}
In the original formulation of ABF~\cite{Darve2000, Darve2001, Darve2002}, the collective force is calculated based on a constraint force calculation. That is, a constraint solver algorithm such as SHAKE or RATTLE~\cite{Ryckaert1977,Andersen1983} is executed as if to keep the coordinate of choice constant.
However, the calculated constraint force is not applied and the coordinate remains unconstrained; instead, this force is used to determine the collective force $\vF_\xi$, which is averaged over time to estimate the free energy derivative~\cite{Darve2001}.
Whereas a usual constraint algorithm would cancel the \textbf{instantaneous} collective force acting on the coordinate, the ABF algorithm cancels the \hyperlink{ref:ensemble_average} {ensemble average} of this force, so that the coordinate ``sees'' no force \textbf{on average}, but simply zero-mean fluctuations. This is equivalent to evolving on a locally flat free energy \hyperlink{ref:FES} {surface}.

An alternate formulation that does not need an iterative constraint solver was proposed~\cite{Henin2004}, using a projection of Cartesian forces~\cite{denOtter2000}. This was the basis of the first public implementation of ABF in NAMD.
These were completed with vector formulations that allow for estimation of free energy gradients along several \hyperlink{ref:CV} {collective variable}s, either using time derivatives~\cite{Darve2008}, or projected forces~\cite{Henin2010a}.
The latter is implemented in the Colvars Module~\cite{Fiorin2013}.
In this multidimensional projected force version~\cite{Henin2010a}, the free energy gradient is estimated using Equation~\ref{eq:TI_Ciccotti}.

In dimension greater than 1, obtaining the free energy \hyperlink{ref:FES} {surface} knowing an estimate of its gradients is not trivial.
In the Colvars implementation, this is done by solving a Poisson equation~\cite{Lelievre2010, Alrachid2015, Henin2021integration}.

For a detailed review of ABF, see Ref~\cite{Comer2015}.

ABF dynamics has also been successfully combined with other enhanced sampling methods to accelerate the relaxation of orthogonal degrees of freedom. See Section~\ref{sec:abf_hybrids} for details.

\subsubsection{Multiple-walker ABF}
\label{sec:mwABF}

In shared or multiple-walker ABF (mwABF)~\cite{Lelievre2007, Minoukadeh2010, Comer2014mwABF}, several ABF simulations of the same system using the same \hyperlink{ref:CV} {collective variable} run concurrently, sharing their ABF data at periodic intervals, thus benefitting from the exploration of all other walkers.
In the selection variant~\cite{Minoukadeh2010, Comer2014mwABF}, walkers are replicated or killed using a selection criterion that promotes undersampled regions of \hyperlink{ref:CV} {collective variable} space.

\subsubsection{Extended-system ABF}
\label{sec:eABF}

Extended-system ABF (eABF) is an \hyperlink{ref:ExtL} {extended-Lagrangian} formulation of ABF.
In eABF (\cite{Lelievre2010} p. 368, \cite{Zheng2012}) the coordinate is not a \hyperlink{ref:CV} {collective variable} $z = \xi(\vx)$, but an additional variable $\lambda$, separate from Cartesian coordinates, so that the dynamics is now propagated in the extended space $(\vx, \lambda)$.
$\lambda$ is coupled to the \hyperlink{ref:CV} {collective variable} by a harmonic restraint:
\begin{equation}
    U^\mathrm{ext}(\vx, \lambda) = \frac{1}{2} k (\lambda - \xi(\vx))^2 .
\end{equation}
The main benefit of eABF is that the biasing coordinate $\lambda$ is not a function of Cartesian coordinates, so all geometric considerations raised by the projected force formalism (Section~\ref{sec:fe_estimators:TI}) become moot. This makes eABF easy to implement for any combination of \hyperlink{ref:CV} {collective variable}s, as long as their gradients can be calculated.
One limitation is that the free energy associated to $\lambda$ is close to, but not identical to that associated to $z$, which is the quantity of interest.
The implementation of eABF in the Colvars Module~\cite{Lesage2017} is complemented by two \hyperlink{ref:FEestimator} {free energy estimators}, the corrected z-averaged restraint (CZAR)~\cite{Lesage2017}, and an Umbrella Integration estimator~\cite{Kastner2005, Zheng2012, Fu2016}.
For pointers on choosing parameters for eABF, refer to \cite{Fu2016, Lesage2017}.

\subsubsection{Other ways to calculate the biasing force}
\label{sec:ABF_variants}

Besides the classic ways mentioned above (constraint force, force projection, time derivatives, eABF), other estimators have been proposed:
\begin{enumerate}
\item by projection of the (noisy) gradient estimate on the space of ``true gradients" to reduce their variance: projected ABF (pABF)~\cite{lelievre-rousset-stoltz-07-a,Alrachid2015}, which proved to lead to slower exploration in practice~\cite{Henin2021integration};
\item by ABF with Gaussian Process Regression~\cite{Mones2016} to reconstruct the free energy in a smoother, less local way (currently no accessible, high-performance implementation);
\item by a neural network (ABF-FUNN)~\cite{Guo2018}, which is implemented only in SSAGES (\ref{sec:software});
\item using an implicit, \hyperlink{ref:AdiabaticDyn} {adiabatic} extended coordinate: ABF with adiabatic \hyperlink{ref:Reweighting} {reweighting} (ABF-AR)~\cite{Cao2014} to bypass constraints of the extended dynamics and minimize variance (no public high-performance implementation yet);
\item approximated as the sum of independent \hyperlink{ref:biasingE} {biasing} forces on individual \hyperlink{ref:CV} {collective variable}s: generalized ABF (gABF)~\cite{Chipot2011, Zhao2017} to handle higher-dimensional CV spaces (this is most effective for loosely-coupled CVs).
\end{enumerate}

\subsubsection{Estimating convergence and diagnosing issues in ABF simulations}

In ABF simulations, convergence of the calculated free energy gradients is reflected in convergence of the sampling histogram towards uniformity.
A more demanding criterion is the occurrence of numerous transition events between \hyperlink{ref:metastab} {metastable} basins.
This indicates not only that the free energy gradients have converged, but that the chosen CVs capture the major slow degrees of freedom at play.
Conversely, the presence of slow-relaxing orthogonal degrees of freedom will cause a slow drift of the estimated gradients, and will manifest itself by the trajectory getting stuck in a region of CV space for a long time.
Remedies include improving the set of CVs, or complementing ABF dynamics with other methods that help overcome orthogonal barriers, such as multiple walkers, metadynamics, or Gaussian-accelerated MD (Section~\ref{sec:abf_hybrids}).

\subsubsection{Public implementations of ABF}

The first public implementation of ABF~\cite{Henin2004} was a scripted extension to NAMD~\cite{Phillips2020}.
This was superseded by a more flexible, multi-dimensional implementation~\cite{Henin2010a} which is part of the Collective Variables (Colvars) Module ~\cite{Fiorin2013}.
The Colvars Module is interfaced with NAMD~\cite{Phillips2020}, LAMMPS~\cite{Plimpton1995}, GROMACS~\cite{Abraham2015}, and VMD~\cite{Humphrey1996} for colvar analysis~\cite{Henin2022dashboard}.
An interface of the Colvars Module with Tinker-HP~\cite{Lagardere2018} is in preparation.
ABF is implemented in PMFlib~\cite{kulhanek2011pmflib} for use in the sander version of AMBER.
Dynamic Reference Restraining~\cite{Zheng2012} (equivalent to eABF) is implemented in PLUMED~\cite{Tribello2014}.
ABF-FUNN is part of SSAGES~\cite{Sidky2018}.

\section{Generalized ensemble and replica exchange methods}
\label{sec:generalized-ensemble}

A broad category of simulation methodologies known as \hyperlink{ref:GenEns} {generalized ensemble}~\cite{okamoto:biopolymers:2001:generalized-ensemble} (also sometimes referred to \emph{extended ensemble}~\cite{iba:intl-j-mod-phys-c:2001:extended-ensemble}) algorithms have become popular over the last two decades. These methods follow a strategy that is orthogonal to the methods presented so far: in this class of methods, the original configurational \hyperlink{ref:Distribution} {distribution}  is preserved, and the sampling is enhanced by exploiting transitions to other ensembles (Figure~\ref{fig:scheme}). 
The main  algorithmic classes in this category are \hyperlink{ref:ReplEx} {replica exchange},~\cite{geyer:conference-proceedings:1991:replica-exchange} which includes parallel tempering~\cite{hukushima-nemoto:j-phys-soc-jpn:1996:parallel-tempering,hansmann:chem-phys-lett:1997:parallel-tempering-monte-carlo,sugita-okamoto:chem-phys-lett:1999:parallel-tempering-md} and Hamiltonian exchange~\cite{sugita-kitao-okamoto:jcp:2000:hamiltonian-exchange,fukunishi-watanabe-takada:jcp:2002:hamiltonian-exchange,jang-shin-pak:prl:2003:hamiltonian-exchange,kwak-hansmann:prl:2005:hamiltonian-exchange}, among others, and the serial equivalent, the method of \hyperlink{ref:ExpEns} {expanded ensemble}s~\cite{lyubartsev:jcp:1992:expanded-ensembles}, which includes simulated tempering~\cite{marinari-parisi:europhys-lett:1992:simulated-tempering,geyer-thompson:j-am-stat-assoc:1995:expanded-ensembles} and simulated scaling~\cite{li-fajer-yang:jcp:2007:simulated-scaling}.

In both \hyperlink{ref:ReplEx} {replica exchange} and \hyperlink{ref:ExpEns} {expanded ensemble} algorithms, a mixture of thermodynamic states are sampled within the same simulation framework. Simulations are able to access all of the  thermodynamic states through a stochastic hopping process between these thermodynamic states.  In the rest of the discussion of both \hyperlink{ref:ReplEx} {replica exchange} and expanded ensembles, we will often use ``states" as shorthand for thermodynamically defined \hyperlink{ref:Macrostate} {macrostate}s, which all share the same configuration space $\Sigma$, but each of which have different probabilities due to the differences in $T$, $P$, or Hamiltonian parameters.

In \hyperlink{ref:ExpEns} {expanded ensemble} simulations, the states are explored in a single simulation via a biased random walk in state space; in \hyperlink{ref:ReplEx} {replica exchange} simulations, multiple coupled simulations are carried out in parallel, and periodically the simulations exchange thermodynamic states with each other, keeping the same number of simulations at each state. Both methods, if implemented correctly, and potentially after an initial equilibration stage, allow estimation of equilibrium expectations at each state as well as free energy differences between states.

The primary reason for introducing switching between thermodynamic states is that the transitions between these different thermodynamic states can reduce correlation times in configurational sampling at any given thermodynamic state and increase sampling efficiency relative to straightforward sampling of a single state. This acceleration is because the simulations can ``go around"  kinetic barriers within any of these single states. This is done by escaping to a neighboring state where, by chance or better yet, by design, the free energy barriers are lower.  For this switching between states to work, each thermodynamic state must have neighbors which have a moderate overlap (perhaps 5-20\%, depending on the method) in the configuration space each samples. Overall state space must be connected, which means that there must exist pathways between all the states. 

This class of methods therefore only works when the states are designed to have overlap, such as \hyperlink{ref:Alchemical} {alchemical} intermediates, temperatures, or harmonic \hyperlink{ref:biasingE} {biasing} potentials, and not when the states are defined by values of a collective variable. One can however approximate the computation of a collective variable while still using an overlapping states method.  One can use series of harmonic \hyperlink{ref:biasingE} {biasing} functions with that are closely spaced enough to still overlap in sampled configurations. For example, if one was interested in the free energy as a function of the center of mass of two molecules, one could put a series of harmonic biases on these distance, with spring constants that allowed the distances to fluctuate by enough that neighboring simulations would share some visited distances with each other. Unlike the methods use partitioning of the collective variable, one is not guaranteed to get good sampling at each value of the collective variable if the harmonic potentials are too spread apart, or the spring constants are too strong. But because the states have overlap, all of the methods of analysis and simulation used for such simulations are the ones used for overlapping state methods.  This parallel between methods demonstrates again how many ways the different ``ingredients`` in free energy calculations can be combined.   

Because of their popularity, these algorithms for simulating multiple simulations with different states and their properties have been the subject of intense study over recent years.
For example, given optimal weights, \hyperlink{ref:ExpEns} {expanded ensemble} simulations have been shown to have provably higher exchange acceptance rates than \hyperlink{ref:ReplEx} {replica exchange} simulations using the same set of thermodynamic states~\cite{park:pre:2008:simulated-tempering}.
Higher exchange attempt frequencies have been demonstrated to improve mixing for replica exchange simulations~\cite{sindhikara-meng-roitberg:jcp:2008:exchange-frequency,sindhikara-emerson-roitberg:jctc:2010:exchange-often-and-properly}.
Alternative velocity rescaling schemes have been suggested to improve exchange probabilities~\cite{nadler-hansmann:pre:2007:optimized-replica-exchange-moves}.
Other work has examined the degree to which \hyperlink{ref:ReplEx} {replica exchange} simulations enhance sampling relative to straightforward molecular dynamics simulations~\cite{rhee-pande:biophys-j:2003:multiplexed-replica-exchange,zuckerman-lyman:jctc:2006:replica-exchange-efficiency,gallicchio-levy:pnas:2007:replica-exchange,nymeyer:jctc:2008:replica-exchange-efficiency,tavan:cpl:2008:pseudoconvergence,rosta-hummer:jcp:2009:replica-exchange-efficiency,rosta-hummer:jcp:2010:simulated-tempering-efficiency}.
Numerous studies have examined the issue of how to optimally choose thermodynamic states to enhance sampling in systems with second-order phase transitions~\cite{kofke:2002:jcp:acceptance-probability,katzberger-trebst-huse-troyer:j-stat-mech:2006:feedback-optimized-parallel-tempering,trebst-troyer-hansmann:jcp:2006:optimized-replica-selection,nadler-hansmann:pre:2007:generalized-ensemble,gront-kolinski:j-phys-condens-matter:2007:optimized-replica-selection,park-pande:pre:2007:choosing-weights-simulated-tempering,shenfeld-xu:pre:2009:thermodynamic-length}, though systems with strong first-order-like phase transitions (such as two-state protein systems) remain challenging~\cite{neuhaus-magiera-hansmann:pre:2007:parallel-tempering-first-order,straub:jcp:2010:generalized-replica-exchange}.
A number of combinations~\cite{fenwick-escobedo:jcp:2003:replica-exchange-expanded-ensembles,mitsutake-okamoto:jcp:2004:rest} and elaborations~\cite{mitsutake-sugita-okamoto:2003:remuca,rhee-pande:biophys-j:2003:multiplexed-replica-exchange,simmerling:jctc:2007:reservoir-replica-exchange,gallicchio-levy-parashar:j-comput-chem:2008:asynchronous-replica-exchange,hansmann:physica-a:2010:replica-exchange} of these algorithms have also been explored.
A few publications have examined the mixing and convergence properties of \hyperlink{ref:ReplEx} {replica exchange} and \hyperlink{ref:ExpEns} {expanded ensemble} algorithms with mathematical rigor~\cite{madras-randall:annals-appl-prob:2002:decomposition-theorem,bhatnagar-randall:acm:2004:torpid-mixing,woodard_conditions_2009,woodard_sufficient_2009}, but there remain many unanswered questions about these sampling algorithms, both in terms of theoretical bounds and practical guidelines for how much these methods accelerate sampling for complex molecular systems.

In these methods, we label the different $K$ thermodynamic states either using an \hyperlink{ref:AuxVar} {auxiliary variable} $\mathbf{\lambda}$  or by an index $k$. We will assume that the simulation can visit the same configurations for each choice of $\mathbf{\lambda}$~\footnote{Note however that the Boltzmann weights of any given sample with coordinates $\vx$ can vary significantly between choices of $\mathbf{\lambda}$. For example, if the $K$ thermodynamic states are defined by different temperatures, then the simulation will visit all of the same sets of configurations, but low-energy configurations will have much higher probability at lower temperatures.}. Each choice of $\lambda$ results in a sub-ensemble within this \hyperlink{ref:ExpEns} {expanded ensemble}, in which we can carry out a perfectly reasonable simulation in absence of any switching. Finally, note that in this section, we use \hyperlink{ref:reduced} {reduced} units, as it makes it possible to use a common framework and a coherent notation for the different \hyperlink{ref:ReplEx} {replica exchange} and \hyperlink{ref:ExpEns} {expanded ensemble} methods, easing the comparison between the schemes.

\subsection{Replica exchange}
\label{sec:ReplicaExchange}
In a replica exchange simulation, we consider $K$ simulations, with one simulation in each of the $K$ thermodynamic states. In many cases, the data gathered in all of the states is important to calculate observables. However, in other cases, only one simulation actually samples a state of interest, and the other simulations are added exclusively to aid the sampling in one way or another.

The current state of the \hyperlink{ref:ReplEx} {replica exchange} simulation at any time is given by $(X,S)$, where $X$ is a vector of the configurations of \emph{all} of the replicas, $X \equiv \{\vx_1, \vx_2, \ldots, \vx_K\}$, and $S \equiv\{s_1,\ldots,s_K\} \in \mathcal{S}_K$ is a permutation of the state labels $\{1, \ldots, K\}$ associated with each of the replica configurations $\{\vx_1, \ldots, \vx_K\}$. Then the joint probability density of the entire set of all simulations $\mathcal{Q}$ is given by
\begin{eqnarray}
\mathcal{Q}(X, S) &\propto& \prod_{i=1}^{K} \nu_{s_i}(\vx_i) \propto \exp\left[-\sum_{i=1}^K u_{s_i}(\vx_i)\right]
\label{eq:parallereplica}
\end{eqnarray}
Where $\mu_{s_i}$ and $u_{s_i}$ are the normalized \hyperlink{ref:Distribution}{probability distributions} and the \hyperlink{ref:reduced}{reduced} energies of the $s_i$ state, respectively.
The conditional densities, upon specifying a particular order of the replicas $S$, is given by:
\begin{eqnarray}
\mathcal{Q}(X | S) &=& \prod_{i=1}^K \left[ \frac{e^{-u_{s_i}(\vx_i)}}{\int_\Omega dx \, e^{-u_{s_i}(\vx_i)}}\right]
\end{eqnarray}
and
\begin{eqnarray}
\mathcal{Q}(S | X) &=& \frac{\exp\left[- \sum\limits_{i=1}^K u_{s_i}(\vx_i) \right]}{\sum\limits_{S' \in \mathcal{S}_K} \exp\left[- \sum\limits_{i=1}^K u_{s'_i}(\vx_i) \right]}
\end{eqnarray}
Note that this is more complicated than the equations for a single simulation: we need an equation that describes the state of all of the replicas, since they all are coupled together. With the conditional densities, we can then describe how to make jumps in coordinate space (which just ends up being independent, standard dynamics in each replica) and jumps in permutations of the replica (described in more detail below).

In the most standard \hyperlink{ref:ReplEx} {replica exchange} simulation algorithms, a proposed new permutation $S$ of the state of the set of all systems $(X,S)$  only considers exchanges between states that are currently neighboring each other~\cite{hukushima-nemoto:j-phys-soc-jpn:1996:parallel-tempering,hansmann:chem-phys-lett:1997:parallel-tempering-monte-carlo,sugita-okamoto:chem-phys-lett:1999:parallel-tempering-md,sugita-kitao-okamoto:jcp:2000:hamiltonian-exchange,fukunishi-watanabe-takada:jcp:2002:hamiltonian-exchange,jang-shin-pak:prl:2003:hamiltonian-exchange,kwak-hansmann:prl:2005:hamiltonian-exchange}.
For example, one such scheme involves attempting to exchange either the set of state index pairs $\{(1,2), (3,4), \ldots\}$ or $\{(2,3), (4,5), \ldots\}$, chosen with equal probability. Each state index pair $(i,j)$ exchange attempt is carried out independently, with the exchange of states $i$ and $j$ associated with configurations $\vx_i$ and $\vx_j$, respectively, accepted with the standard \hyperlink{ref:MetropolisMonteCarlo} {Metropolis} probability
\begin{eqnarray}
\label{eq:re_exchange_prob}
P_\mathrm{accept}(\vx_i, i, \vx_j, j) &=& \min\left\{ 1, \frac{e^{-[u_i(\vx_j)+u_j(\vx_i)]}}{e^{-[u_i(\vx_i) + u_j(\vx_j)]}}\right\}
\label{eq:metropolis-replica}
\end{eqnarray}
However, it is also possible to sample from the space of all possible permutations, which can in some cases improve sampling~\cite{shirts_gibbssamp}.

Because \hyperlink{ref:ReplEx} {replica exchange} is so general, only requiring that the $K$ states have the same coordinates but any reasonable reduced potentials $u_i$ that have overlap with each other, then there are many different variants that attempt to solve different sampling problems by defining different $u_i$.

The most straightforward versions are parallel implementations of algorithms that already use $K$ simulations. For example, if one is computing some property as a function of temperature, then one can run a number of simulations at different temperatures, with the simulations at higher temperature providing faster kinetics to all of the simulations. If one is performing an \hyperlink{ref:Alchemical} {alchemical} protein-ligand binding simulation, then one typically has simulations at $K$ values of $\lambda$, which describe the degree of interaction of the ligand with the rest of the system, which can be run in \hyperlink{ref:ReplEx} {replica exchange}. In this case, the fully interacting ligand can escape by moving to the fully uncoupled state. If one is performing umbrella sampling with $K$ umbrellas, they can be run in \hyperlink{ref:ReplEx} {replica exchange}, allowing simulations to move between umbrellas as long as they are placed sufficiently closely with sufficiently weak restraints to allow overlap between states.

However, even if one is only interested in a single state, one can add higher $T$ replicas just to escape energy barriers, or add \hyperlink{ref:Alchemical} {alchemical} states to allow parts of the system to move. The possibilities are virtually endless to create different states where configuration sampling happens faster.

For example, in the Replica Exchange Solute Tempering~\cite{REST1_Liu_2007} (REST) and REST2~\cite{REST2_Wang_2011} variants, the ``temperature" is adjusted for only a part of the system. This is a misnomer, as temperature is only rigorously defined for an entire system. What this means in practice is that the the potential energy of a select part of the system is scaled by  $T_m/T_0$ in replica $m$, where $T_0$ is the temperature of the system. In the case of REST2, the terms of the energy function corresponding interactions between this designated part of the system and the rest of the system is additionally scaled by the $T_m/T_0$, which can be shown to sample better than the original REST~\citep{REST2_Wang_2011} by keeping the subsystem $m$ better coupled to the rest of the simulation. A combination of different approaches can be used in the same set of simulations; one of the popular protocols that the computational chemistry software company Schr\"{o}dinger has implemented for relative free energy binding is to simultaneously perform \hyperlink{ref:Alchemical} {alchemical} ligand transformations, apply REST2 to the area around the ligand, and reduce the torsional potentials of side chains in the binding site~\cite{Wang:JCTC:2013}.

Various \hyperlink{ref:ReplEx} {replica exchange} techniques have a long history of investigation over the last 20 years. For other analyses of the issues, subtleties, and variants of \hyperlink{ref:ReplEx} {replica exchange}, see a number of reviews such as~\cite{Abrams:E:2014,Gallicchio:CPC:2015,Itoh:JCTC:2013,kofke:2002:jcp:acceptance-probability,Liu:CPL:2018,nadler-hansmann:pre:2007:optimized-replica-exchange-moves,Qi:PSMaP:2018,shenfeld-xu:pre:2009:thermodynamic-length,sindhikara-emerson-roitberg:jctc:2010:exchange-often-and-properly}.

\subsubsection{Estimating convergence and diagnosing issues in replica exchange simulations}

\begin{figure*}[!ht]
    \centering
\begin{tabular}{c|c|c}
\includegraphics[width=0.33\textwidth]{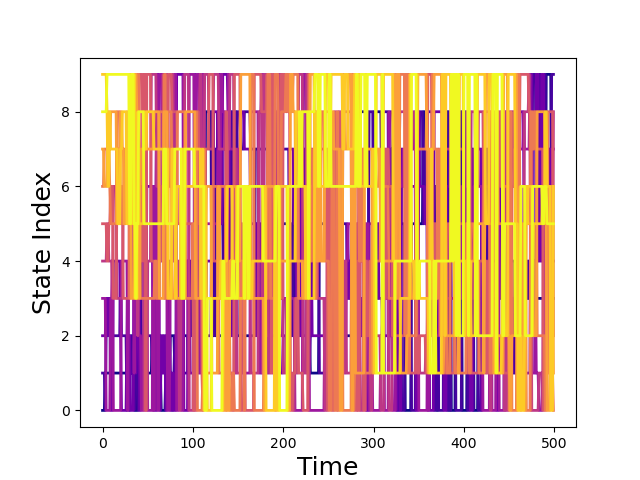} & \includegraphics[width=0.33\textwidth]{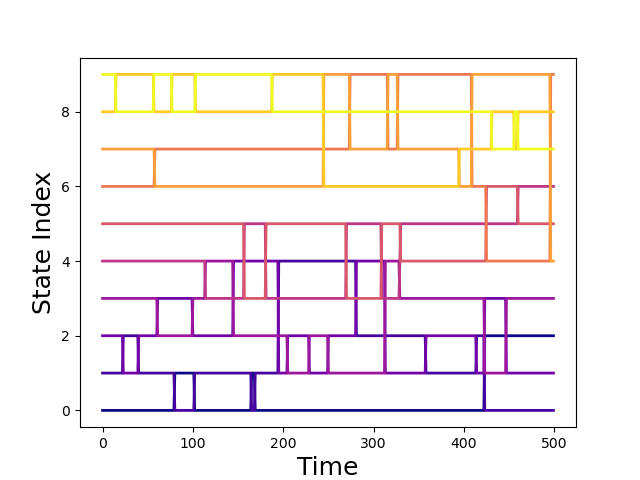}&
\includegraphics[width=0.33\textwidth]{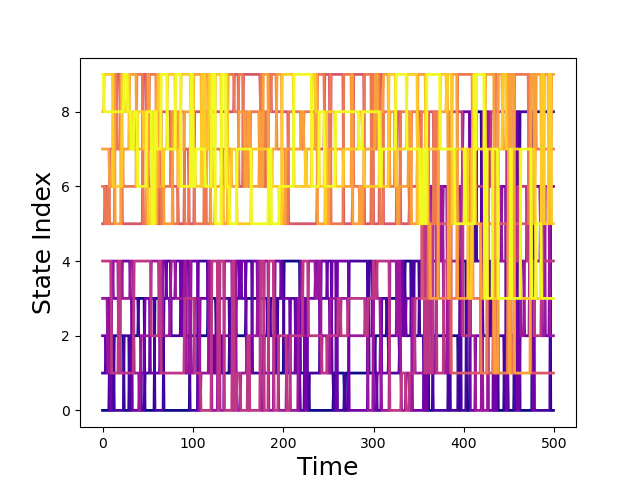} 
\end{tabular}
    \caption{Three examples of state indices plotted as a function of time for replica exchange simulations, plotted with a color gradient on the state index.  On the left, all replicas travel between all states multiple times, suggesting good mixing of states on this timescale.  In the center, exchanges are occurring slowly and not making the trip from top to bottom, suggesting that simulation might be more efficient if additional replicas were added, reducing the spacing between replicas were run. On the right, we have a bottleneck in the middle of the state space, preventing mixing of all the states together, even though states 0--4 and 5--9 can mix readily. For the similar expanded ensemble cases, see Figure~\ref{fig:diagnostic_expanded}.}
    \label{fig:diagnostic_replica}
\end{figure*}

One of the main limitations of \hyperlink{ref:ReplEx} {replica exchange} is the need to have simulations with some overlap with each other that are arranged to exchange. Otherwise, the swaps will occur with too low of a probability, resulting in an inefficient exchange scheme (i.e. $P_\mathrm{accept}$ between some pairs of replicas in eq.~\ref{eq:metropolis-replica} will tend to $0$). If the replica spacing in the auxiliary variable is chosen poorly, it is very common for many of the replicas to remain in the same few states for the entire simulation, a clear indication of poor global overlap. A necessary (though not sufficient) check on the global overlap of the simulation is to make sure that each individual simulation can travel between all of the different states, preferably multiple times in the same simulation (see Figure~\ref{fig:diagnostic_replica}).

If one is using temperature replica simulations, then the width of the energy distribution of each state will scale as roughly $N^{-1/2}$, where $N$ is the number of particles. Because the overlap in energy decreases as systems get larger, a tighter spacing is required for large systems and \hyperlink{ref:ReplEx} {replica exchange} becomes increasingly less efficient. Indeed, not only does each individual simulation become more expensive, but more simulations are needed to span the same range of $T$s. For the other types of \hyperlink{ref:ReplEx} {replica exchange} described here, such as those in Hamiltonian variables, the changes in $u_i$ affect a smaller portion of the simulation or smaller number of atoms, and thus this scaling problem is not as severe, but this limitation must be kept in mind when deciding how different the differences in are $u_i$ are between systems.

A common problem, even when the average overlap between simulations is reasonable, is to have a "bottleneck" where the set of simulations separates into two essentially independent set of simulations that only exchange between themselves. This usually defeats the purpose of the \hyperlink{ref:ReplEx} {replica exchange} simulation, since the simulations cannot move between all of the states, especially since the system of interest is often at one end of the chain of replicas, and the ``fastest" system at the other. Plotting the state index of each of the replicas versus time can help reveal these sorts of issues.

A related diagnostic quantity to look at is the matrix of transitions between states, where each entry correspond to the transitions from state $i$ to state $j$ (which ones are the rows and which ones are columns is a matter of convention).  One generally wants each replica to transition to another state at least 30\% of the time that exchanges are proposed; if replicas are transitioning at a slower rate, then sampling can be improved by increasing the spacing; or if some transitions are occurring more frequently that that, reallocating the spacing. However, if the spacing is \textit{too} small, then it will take too long for each replica visit all of the states, so there is a delicate balance.

One important note is that in the development of many \hyperlink{ref:ReplEx} {replica exchange} methods, there is frequently an assumption that the states contain some natural ordering, so that one can definitively say the configurations that result from simulations at $i$ are more similar to the configurations generated in state $i-1$ and $i+1$ than they are similar to any other states.  A number of methods of choosing how to exchange therefore assume this ordering, but a natural ordering may not exist in the general case.  Such an ordering is straightforward when states are selected points along a single \hyperlink{ref:Alchemical} {alchemical} variable $\lambda$ or temperature $T$, but when instead thermodynamic states are defined in a multidimensional space, say \emph{both} $T$ and $\lambda$, no such ordering may exist, and some of the schemes for exchange in replica exchange may not apply.

\subsubsection{Public implementations of replica exchange methods}

\hyperlink{ref:ReplEx} {replica exchange} is perhaps one of the most common \hyperlink{ref:ExpEns} {expanded ensemble} methods, and is implemented natively in GROMACS, AMBER, OpenMM, CHARMM, LAMMPS, and other MD packages. PLUMED provides some additional tools to run and analyze replica exchange simulations as well, but it builds on top of the native replica exchange packages.

\subsection{Expanded Ensemble}
\hyperlink{ref:ExpEns} {Expanded ensemble} simulations~\cite{lyubartsev:jcp:1992:expanded-ensembles} use many of the same concepts as \hyperlink{ref:ReplEx} {replica exchange} simulations, but a single simulation moves between $K$ states. Specifically, a single replica or ``walker'' samples pairs $(\vx,k)$ from a joint \hyperlink{ref:Distribution} {distribution}  of configurations $\vx \in \Gamma$ and state indices $k \in \{1,\ldots,K\}$ given by,
\begin{eqnarray}
\nu(\vx,k) &\propto& e^{g_k-u_k(\vx)},
\end{eqnarray}
where $g_k$ is an optional (but usually necessary for effective sampling) state-dependent weighting factor that adjusts the relative probability of the simulation visiting each of the $k$ states.

The \emph{conditional} distribution of the state index $k$ given $\vx$ is specifically:
\begin{eqnarray}
\nu(k | \vx) &=& \frac{e^{g_k - u_k(\vx)}}{\sum\limits_{k'=1}^K e^{g_{k'} - u_{k'}(\vx)}}.
\end{eqnarray}

If we were to sample independently in the joint $(\vx,k)$ space with all of the $g_k=0$
we would find that we would be spending more time in the states with the lowest free energies $f_k$.  If we wish to visit
  lower probability (i.e. higher free energy) states, we need to adjust the compensating biases $g_k$ to increase the time spent at these states. It turns out that the states will all be visited equally in the long time limit if the $g_k$ are equal to the \hyperlink{ref:reduced} {reduced} free
    energy $f_k$ (so-called ``perfect weights''~\cite{park-pande:pre:2007:choosing-weights-simulated-tempering}).
    Of course, the goal of our simulations
    themselves is often to calculate $f_k$, so it all
    becomes somewhat circular; we need to know $f_k$ to calculate $f_k$.
    Thus we need some sort of \hyperlink{ref:Adaptive} {adaptive} method to gradually learn $f_k$
    as we go. The choice of the appropriate iterative procedure is usually the main topic of research in \hyperlink{ref:ExpEns} {expanded ensemble} simulations~\cite{lyubartsev:jcp:1992:expanded-ensembles,marinari-parisi:europhys-lett:1992:simulated-tempering,wang-landau:prl:2001:wang-landau,park-ensign-pande:pre:2006:bayesian-weight-update,park-pande:pre:2007:choosing-weights-simulated-tempering,li-fajer-yang:jcp:2007:simulated-scaling,chelli:jctc:2010:optimal-weights-expanded-ensembles}, and is discussed below.

\begin{figure*}[ht]
\captionsetup{justification=centering}
\begin{subfigure}[b]{0.33\textwidth}
\includegraphics[width=\textwidth]{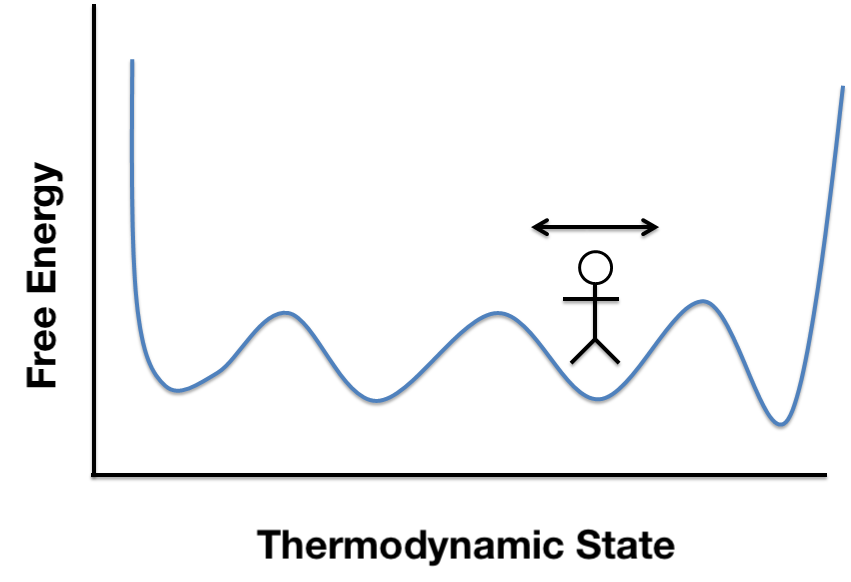}
\caption{Beginning of Simulation}
 \label{fig:emptywell}
\end{subfigure}%
~
\begin{subfigure}[b]{0.33\textwidth}
\includegraphics[width=\textwidth]{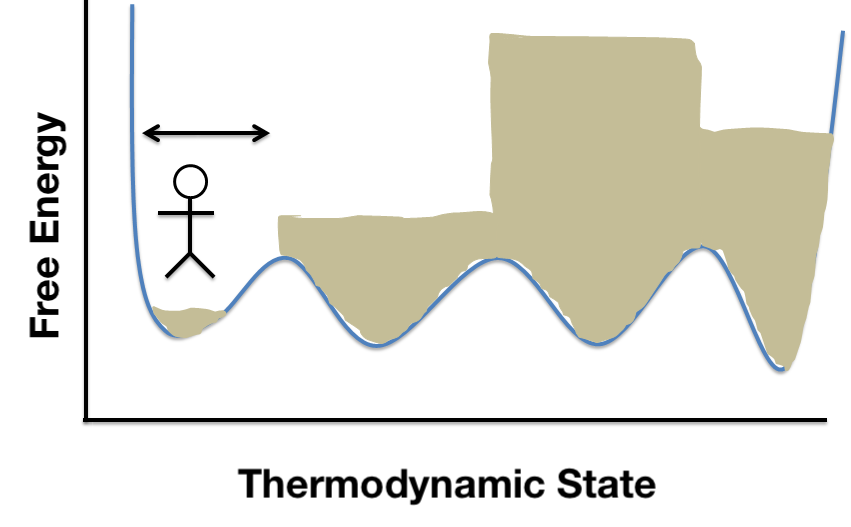}
\caption{During Simulation}
\label{fig:halfwell}
\end{subfigure}
~
\begin{subfigure}[b]{0.33\textwidth}
\includegraphics[width=\textwidth]{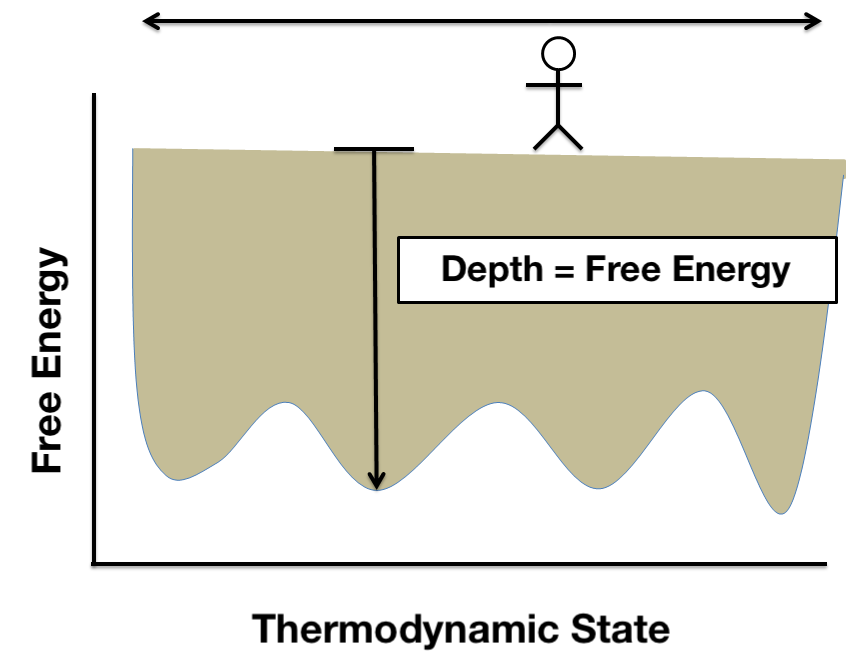}
\caption{End of Simulation}
\label{fig:fullwell}
\end{subfigure}
\caption{The expanded ensemble hole analogy. (\ref{fig:emptywell}) At the beginning the biasing weights as a function of thermodynamic state are unknown. (\ref{fig:halfwell}) As the simulation proceeds, a random walker samples the states and deposits ``dirt'' (probability of visitation) in each location.  The short arrow over the random walker signifies uneven sampling of the states as the the biases are built adaptively.  (\ref{fig:fullwell}) At the end of the simulation, the weights, given by the height of ``dirt'', in each well are equal to the free energy.  The random walker now samples all states equally, as illustrated by the long bar over the walker which now extends over all of state space.} \label{fig:EXEanalogy}
\end{figure*}

Generally, such algorithms are classified as ``visited states''
algorithms, because we collect statistics about the states as we visit
them, and then update our information (Figure~\ref{fig:EXEanalogy}). Consider the free energies of
the physical states as holes/wells of some initially unknown depth.  A
fictitious random walker visits the different states labeled by
$k$ as the simulation proceeds, dropping ``dirt'' into the
wells, thus gradually building up the importance weights.  At the end
of the simulation, when all states are visited equally, one counts how
much much ``dirt'' the walker has added to each state's weight to
achieve equal sampling. The negative logarithm of this weight is simply
the free energy of the state. The basic principle is similar to metadynamics, so expanded ensemble, unlike \hyperlink{ref:ReplEx} {replica exchange}, can be considered in some ways an ``adaptive bias'' method, though rather than bias being calculated as a function of some collective coordinate, the bias is calculated as a function of the parameter that $k$ labels (such as $\lambda$ or temperature) that controls the thermodynamic state.

Transitions between the subensembles labeled by indices $k$ or equivalently discrete values of a vector $\mathbf{\lambda}$ can most simply be performed by \hyperlink{ref:MetropolisMonteCarlo} {Monte Carlo}~\footnote{Molecular dynamics is also possible, but we will restrict our discussion of transitions
between states to Monte Carlo for now, as dynamics in a continuous $\mathbf{\lambda}$ has a number of additional subtleties and is much less common.}.  We can use any proposal/acceptance scheme that ensures this conditional distribution is sampled in the long run for any fixed $\vx$. At each step, we can choose to sample in either $k$ or $\vx$ according to some fixed probability $p$.  We can also alternate $N_k$ and $N_x$ steps of $k$ and $\vx$ sampling, respectively.  Although this Gibbs sampling algorithm does not satisfy detailed \hyperlink{ref:Balance} {balance}, it does satisfy the weaker condition of \hyperlink{ref:Balance} {balance}~\cite{deem:jcp:1999:balance} which is sufficient to preserve sampling from the joint stationary probability \hyperlink{ref:Distribution} {distribution}  $\nu(\vx,k)$.  When proposal probabilities are based on past history, however, the algorithm will not preserve the equilibrium distribution~\cite{reinhardt:cpl:2000:step-size-adjustment}, though in some cases the deviations caused by the history dependence can be mitigated with proper choices of parameters (as shown in the parallel case of metadynamics)~\cite{bussi.equilibriummetadynamics}.

One possibility is to make steps to neighboring states, much like in \hyperlink{ref:ReplEx} {replica exchange}; if the states have a natural ordering, this is perfectly reasonable. However, they may not be a natural ordering, for example when there are multiple neighboring states in dimension higher than one, or if temperature and external biases are combined.

If we think of sampling this joint space in configuration and state in the context of Gibbs sampling, an \hyperlink{ref:ExpEns} {expanded ensemble} simulation can proceed by alternating between sampling from the two conditional \hyperlink{ref:Distribution} {distribution} s,
\begin{eqnarray}
\nu(\vx | k) &=& \frac{q_k(\vx)}{\int_\Gamma  \, q_k(\vx) \, d\vx}  = \frac{e^{-u_k(\vx)}}{\int_\Gamma  \, e^{-u_k(\vx)} \, d\vx }  \\
\nu(k | \vx) &=& \frac{e^{g_k}q_k(\vx)}{\sum\limits_{k'=1}^K e^{g_{k'}}q_{k'}(\vx)} = \frac{e^{g_k - u_k(\vx)}}{\sum\limits_{k'=1}^K e^{g_{k'} - u_{k'}(\vx)}} .\label{equation:expanded-ensemble-gibbs-update}
\end{eqnarray}

Sampling from the conditional probability \hyperlink{ref:Distribution} {distribution} 
$\nu(\vx | k)$ is just standard molecular dynamics sampling that generates time-correlated samples. Fortunately, the sampling in $\nu(k | \vx)$ is often much simpler.   If we have
discrete states which can be enumerated, we can simply calculate
$u_k(\vx)$ for each state and select randomly the state $k$ to move to according to $\nu(k|\vx)$.  The
probabilities of transition to state $k$ only depend on the \hyperlink{ref:reduced} {reduced} energy differences $\Delta
u_{ik}(\vx) = u_i(\vx) - u_k(\vx)$, which is often much cheaper to calculate
than the entire $u_k(\vx)$ is. Sampling in this way is often significantly easier with \hyperlink{ref:ExpEns} {expanded ensemble} than with \hyperlink{ref:ReplEx} {replica exchange}. In \hyperlink{ref:ReplEx} {replica exchange}, often each state is only stored on a single replica, and communicating the energy between states is complicated. With \hyperlink{ref:ExpEns} {expanded ensemble} sampling, all of the states are being kept track of by the same simulation, so calculating the conditional probabilities of the other states is requires no additional communication.

\subsubsection{\label{sec:singlestate} Methods updating biases one state at a time}
Given ways to sample between the states, we now need methods that \hyperlink{ref:Adaptive} {adaptively} change the biases $g_k$ until sampling of the different states reaches the desired ratio. The simplest way to update biases is to do it one state at a time, where the state that is currently visited is the one that is changed. Other schemes will introduce updating the bias of several schemes at a time.

\paragraph{Wang-Landau updating}
As noted above, if subensemble weighting factors $g_k$ are equal to \hyperlink{ref:reduced} {reduced} free energies then one will eventually obtain even sampling of all the states
\cite{lyubartsev:jcp:1992:expanded-ensembles}.  One procedure for the
iterative calculation of these biasing weights is the Wang-Landau
algorithm~\footnote{Not to be confused with the Wang-Landau method of enhanced sampling, see Section~\ref{sec:abp_energy}}~\cite{wang-landau:prl:2001:wang-landau}. While this algorithm was originally proposed to calculate biasing weights where the different values of the total energy are the different thermodynamic states, in which case these biasing weights $g_k$ are equal to the \hyperlink{ref:density_of_states} {density of states} $\Omega(U)$ (see Section~\ref{sec:abp_energy}), it has also been used as a general approach to calculate biasing weights associated with other ensembles.

The approach is as follows. We keep track of a histogram $h(i)$ of visits to each thermodynamic state during the \hyperlink{ref:ExpEns} {expanded ensemble} simulation.  When a state is visited, the corresponding
histogram is updated by 1.  The weight $g_i$ of that state is updated by some user-chosen increment $-\delta$.  $\delta$ itself is reduced by a monotonically decreasing function as the simulation proceeds until the changes in the $g_i$ weights go to zero. The choice of how $\delta$ decreases is discussed later. The
weight of the reference state is subtracted from all states after each step, as only the weight differences matter physically. As an algorithm, this is written as:
\begin{eqnarray}
h_{\mathrm{new}}(i) = h_{\mathrm{old}}(i) + 1 \\
g_{i,\mathrm{new}} = g_{i,\mathrm{old}} - \delta
\label{eq:wang-landau}
\end{eqnarray}
The Wang-Landau algorithm is self-correcting; if a state is visited more frequently than it should, its weight decreases, resulting in fewer visits to that state.  Eventually, the weight falls back to the correct range.  In the original Wang-Landau scheme, the $\delta$ increment is decreased during the simulation when the histogram $h(i)$ reaches a specified flatness criteria, meaning none of the histograms is lower than the average occupancy, often set at 80\%~\cite{wang-landau:prl:2001:wang-landau}.  When a sufficiently flat histogram is reached, $\delta$ is reduced by multiplying by a scaling factor $0<s<1$, typically $s=1/2$ as proposed in the original algorithm~\cite{wang-landau:prl:2001:wang-landau}, and the histograms are set to zero again.

\paragraph{$1/t$ modifications to Wang-Landau}
The Wang-Landau updating scheme can lead to saturation in the error,
meaning that frequently, simulations weights will converge too quickly, and the system will get stuck in only a subset of the possible states.  Even if it does visit all states, the updating scheme could be too slow, and thus the simulation will still  never reach the correct answer in the alloted amount of time~\cite{Belardinelli2007, Belardinelli2008}.  Taking $s$ closer to one will delay this saturation at the cost of slower convergence.

To avoid this saturation of error, Belardinelli and Pereyra proposed a
power law update to $\delta$, independent of histogram flatness at
long time scales. This update to Wang-Landau, called the $1/t$ method
scales $\delta$ as $1/t$, where $t$ is the \hyperlink{ref:MetropolisMonteCarlo} {Monte Carlo} time, \textit{e.g.} the
number of attempted state transitions.  Belardinelli and Pereyra
suggest starting with a standard Wang-Landau algorithm, and then
switching to using weights of $1/t$ when $\delta \leq 1/t$~\cite{Belardinelli2008, Belardinelli2007}. This need to have the increments decrease by $1/t$ has also been noted by statisticians~\cite{wl_convergence}, and most improved versions of updating schemes also have this feature, even if it is not clear in the formulation.

\subsubsection{\label{sec:multiplestatemethods}Multiple state reweighting methods}

The Wang-Landau approach and its $1/t$ variant only update the free energy of one state at one time, i.e. when it is visited during the random walk in the \hyperlink{ref:ExpEns} {expanded ensemble}.  However, in theory, it should be possible to update the weights of multiple states at the same time, based on information obtained at any given state.

A number of closely related methods have been developed that update multiple states simultaneously. The key to these approaches is recognizing that the same information needed to correctly calculate the probability of  transitions between states (i.e. the transition matrix) is the same information that is needed to calculate the free energy differences between the same states~\cite{escobedo_transition_2006,Wang:JoSP:2002} (see Section~\ref{sec:transtion_matrix}).

\paragraph{Transition matrix approaches}
The simplest way to use this concept is to directly compute the transition matrix and calculate free energies, and thus the weights $g_k$ to use from this matrix~\cite{siderius_2013}. Assuming one uses \hyperlink{ref:MetropolisMonteCarlo} {Metropolis Monte Carlo}, rather than collecting histograms, we collect the transitions from each state $i$ to $j$ in a matrix $C$

It can be updated with any of the following:
\begin{enumerate}
\item Transitions actually performed (adding 1 if the transition occurs, 0 if it does not)\label{item:actual}
\item The probability of acceptance of a proposed move whether or not it was accepted.\label{item:proposal}
\item The probability of acceptance of any move transition that could have been proposed, independent of the algorithm actually used to perform the move. \label{item:transition}
\end{enumerate}

The free energy difference between any two states $i$ and $j$ can then be estimated, either at each step, or at some interval, as
\begin{equation}
f_{ij} = - \ln \langle C_{ij}/C_{ji} \rangle
\label{eq:transitionmc}
\end{equation}
And the weights to apply can simply be set from $f$.

\paragraph{Asymptotically optimal weights}
We can also update all weights simultaneously by \hyperlink{ref:Reweighting} {reweighting} the information gathered at the current state. Several variants of this approach have been proposed. These include the accelerated weight histogram (AWH) method~\cite{Lidmar2012} and the independently developed self-adjusted mixture sampling (SAMS)~\cite{tan_optimally_2017}.

The basic idea of these variants is to use the \hyperlink{ref:GibbsSampler} {Gibbs sampler} to transition between states. After $N_x$ configurational updates with fixed state $i$, a new state $j$ is proposed using the \hyperlink{ref:GibbsSampler} {Gibbs sampler} probabilities $\alpha(j|\vx, i)$:
\begin{equation}\label{eq:gibbsproposal}
  \alpha(j|\vx, i) = \nu(i|\vx)
\end{equation}
where $\nu(i|\vx)$ is given in Equation~\ref{equation:expanded-ensemble-gibbs-update}.

In the self-adjusted mixture sampling variant, the weights are updated at each step. Using the transition rule defined in Equation~\ref{equation:expanded-ensemble-gibbs-update}, the rule for updating is:
\begin{eqnarray}
g_{i,new} = g_{i,old} - t^{-1}\nu(\vx|k)
\end{eqnarray}
where $t$ is a timescale that is ideally the number of uncorrelated steps in $(k,\vx)$ space that have been taken in the algorithm so far~\footnote{$t$ is not necessarily the number of timesteps taken so far in the simulation: if the correlation time in configuration space is slow, then updating the weights at every step, or updating them every step without a scaling factor to make their contribution smaller might lead to premature convergence of the biases on a subset of the conformational states.}.

In the accelerated weight histogram approach, a histogram of the weights for each state is maintained during the simulation. The weight histogram is then updated using the computed $\nu(i|\vx)$ weights:
\begin{equation}
h_{\mathrm{new}}(i) = h_{\mathrm{old}}(i) + \nu(i|\vx)
\label{eq:awh-weight-histogram}
\end{equation}

Every $N_I$ iterations, the simulation weights are updated according to:
\begin{equation}
g_{i,\mathrm{new}} = g_{i,\mathrm{old}} - \ln{\frac{h_{\mathrm{new}}(i)K}{N}}
\label{eq:awh_free_energy_up}
\end{equation}
where $N$ is the total number of samples collected up to that point and $K$ is the number of states.

Although we do not prove it here, AWH is actually an close approximation to SAMS. The difference is that in AWH, the free energies are only updated every $N_I$ steps instead of every step. The $1/t$ dependence in SAMS shows up as a $1/N$ dependence in AWH.

\subsubsection{Estimating the biasing weights near the beginning of the simulation}

One common problem with all of these near-optimal methods is that they are only optimal in the asymptotic limit, when a large number of uncorrelated samples have been collected.  If only a small number of samples have been collected, they may not converge quickly, and indeed may tend to keep samples in the same thermodynamic state for quite a while.

It is still not clear what the best approach is to initialize the biasing weights until the number of samples reach the asymptotic limit.  The current general approach is to run the original Wang-Landau approach with a large initial increment size without reducing the weights until some predetermined point. This predetermined point varies between approaches and implementations.  The simulation is kept in this initial phase until the histograms are roughly equal using weights somewhat near $k_BT$, and then switching to a variant of the asymptotically optimal methods. However, it is not really known if this is optimal, and there are still choices to make, such as the initial bias added to each step.

\begin{figure*}[!ht]
    \centering
\begin{tabular}{c|c|c}
\includegraphics[width=0.33\textwidth]{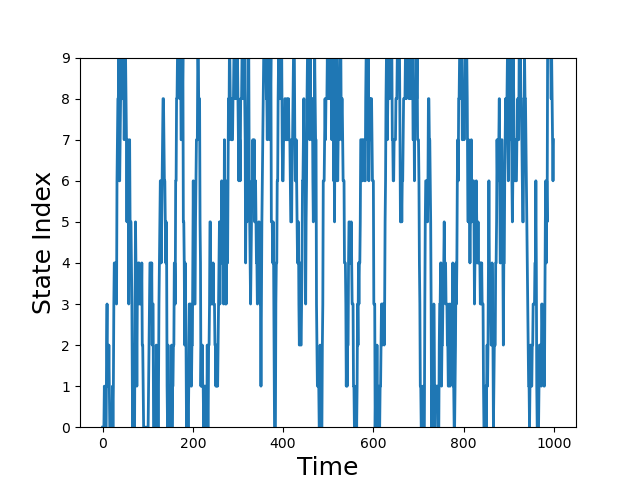} & \includegraphics[width=0.33\textwidth]{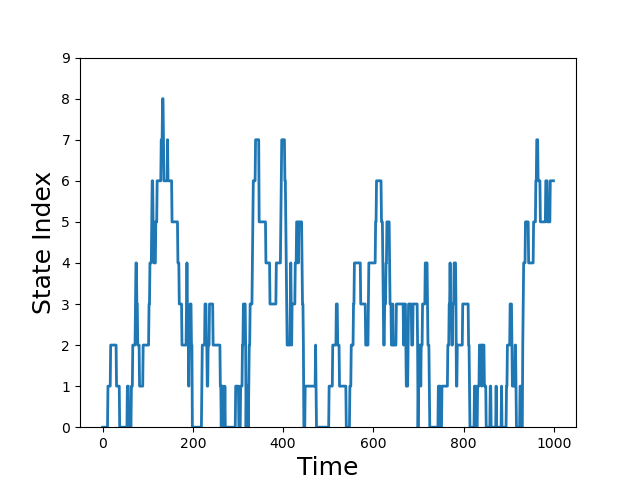}&
\includegraphics[width=0.33\textwidth]{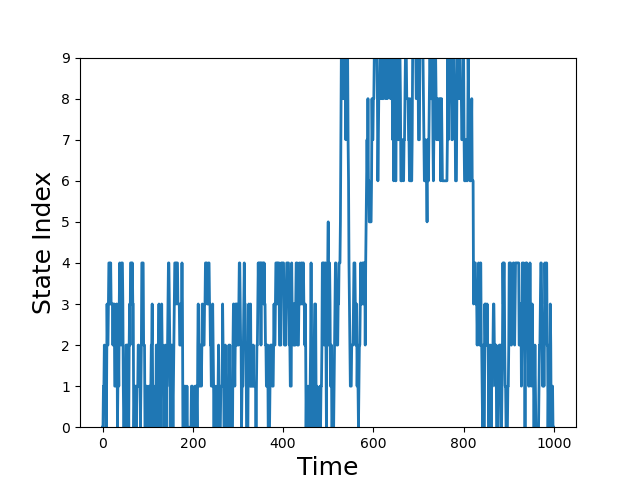} 
\end{tabular}
    \caption{Three examples of the state index trajectory in an expanded ensemble simulation in one dimension. On the left, the simulation travels from top to bottom state multiple times, suggesting good mixing of states on this timescale.  In the center, jumps between states are occurring slowly, and the simulation is not spanning the entire replica range, suggesting the spacing between states may need to be decreased (i.e. overlap between states increased), or weights may need to be improved. On the right, we have a bottleneck in the middle of the states, preventing the overall simulation from mixing easily through the entire range of states. Although states and 0--4 and 5--9 can mix readily, the overall simulation timescales are slow, and the overlap and weights of that transition should be examined to improve sampling. For similar  replica exchange cases, see Figure~\ref{fig:diagnostic_replica}.}
    \label{fig:diagnostic_expanded}
\end{figure*}

\subsubsection{Convergence and diagnosing problems in expanded ensemble simulations}

Unsurprisingly, measuring convergence and diagnostics for expanded ensemble are very similar to those of replica exchange.  A necessary (but not sufficient) requirement for expanded ensemble simulations to have converged is if all thermodynamic states are being visited evenly (or, if the target distribution is some other uneven distribution, according to that distribution). This equal visitation does not have to be exact; if the most visited states are visited twice as much as the least visited states, the free energies will still be accurately estimated. However, if not all states are visited, or some are visited significantly less (say, an order of magnitude) than others, the free energies of the unvisited states will most likely not be accurately estimated.  Note that there is an additional problem that can occur with expanded ensemble compared to replica exchange that can lead to poor transitions between states; not only can overlap between states be low, but the biases on each state can be poorly estimated, so both of these criteria must be checked. 

As with replica exchange, another important quantity to measure is the movement of the auxiliary state variable over the total possible values; in one dimension, this means exploring the range from the lowest to the highest values (see Figure~\ref{fig:diagnostic_expanded}) If there is not a substantial number of transitions across the entire range, perhaps 20--40, then it is likely that the weights along the variable are not well-estimated. It is not uncommon to see good mixing of states in both the upper and lower range, but very few transitions in between these, which usually means that the free energy differences within the upper and lower ranges might be much better estimated than the overall free energy difference between these two groups of states.  

Similarly to replica exchange, one can also check the transition matrix generated from the counts or probabilities of going from state $i$ to state $j$, and one should decrease the spacing (or reallocate the spacing from better transitioning intervals to worse transitioning intervals) so that transitions occur approximately 30\% of the time.  As with replica exchange, tight spacing does increase the amount of time it takes for each replica to move back and forth between all states, so the proper balance must be struck. 

Note that equal visitation of all states does not of course guarantee sampling in the coordinate space. The hope is that some expanded ensemble states can travel through the slow degrees of freedom faster than others, but this depends highly on how useful definition of the auxiliary states is. One must thus explicitly check the slow degrees of freedom of the system to see if enhanced sampling actually occurred. 

\subsubsection{Public implementations of expanded ensemble methods}

Of these different variants of expanded ensemble, the SAMS version is implemented in OpenMM~\cite{10.1371/journal.pcbi.1005659}, and GROMACS~\cite{lindahl_2021} implements both the AWH and a version of SAMS as ``weighted Wang-Landau`` option in expanded ensemble sampling.

\section{Adaptive seeding methods}
\label{sec:seeding}

These methods are related to localization techniques in that the sampling is enhanced by specifying starting coordinates, in order to focus sampling on productive or undersampled regions of configuration space. Contrary to localization methods, however, the simulations are not restricted to a region with restraints or constraints, but instead are merely instantiated -- and sometimes terminated -- strategically (Figure~\ref{fig:scheme}).

\subsection{Adaptive sampling}
\hyperlink{ref:Adaptive} {Adaptive} sampling seeks to sample the configuration space by focusing simulation efforts in regions that will lead to a more accurate description of the ensemble. It generally takes advantage of analysis methods such as Markov State Modeling (MSM) to then stitch together trajectories and build a coherent model of, not only the equilibrium population of \hyperlink{ref:metastab} {metastable} states, but also the rates of conversion between these states~\cite{10.1007/978-94-007-7606-7,10.1021/jacs.7b12191}.

\hyperlink{ref:Adaptive} {Adaptive} \hyperlink{ref:Seeding} {seeding} methodologies were mostly developed with the problem of protein folding in mind, a process that is characterized by rare transitions along a pathway to a single folded state. In such an application, observing a rare event for which a free energy barrier needs to be crossed only depends on the aggregated simulation time, not the length of the simulation so far, because the probability of overcoming a free energy barrier depends on the total number of attempts made at crossing it \cite{PhysRevLett.86.4983,doi:10.1021/acs.jctc.8b00500}. Adaptive \hyperlink{ref:Seeding} {seeding} thus seeks to seed simulations from regions of space that are likely to lead to free energy barrier crossing in order to sample the entire configuration space, using the information acquired through the sampling so far. In other words, sampling over and over regions of space that have already been explored does not add valuable information, whereas crossing into regions of space not previously sampled adds valuable information to reconstruct probability \hyperlink{ref:Distribution} {distributions}.

MSM building involves discretizing the configuration space into states (usually called microstates in the MSM literature, at odds with \hyperlink{ref:Microstate} {the definition} used in this review) and estimating their probability distribution as well as the probability of transitions between these states for a given time lag $\tau$. Discretization of the space can in principle be done based on geometric criteria. However, since the focus of MSM building has been to extract kinetic properties, discretization is often performed based on kinetic proximity, using time-lagged independent component analysis (TICA). The transition probabilities are gathered in a transition matrix $T_{ij}$ which records the probabilities of transitions from state $i$ to $j$ given a time lag $\tau$. Estimating the probability density directly from counting the number of configurations falling in a state would be incorrect for a strategically seeded ensemble. However, the fact that the transition probabilities are taken into account enables to reconstruct the probability density, see Section~\ref{sec:fe_estimators}. These microstates are generally then clustered into \hyperlink{ref:Macrostate} {macrostate}s using a distance metric.

Different \hyperlink{ref:Adaptive} {adaptive} \hyperlink{ref:Seeding} {seeding} strategies have been put forward. They all follow the same basic algorithm: a single or a set of relatively short MD simulations are launched. Configurations from these simulations are grouped into ``states". New simulations are then started from selected states. The different strategies then differ according to the principle they follow to select which states to seed new simulations from.
Arguably the first proposed strategy relied on selecting randomly a fixed number of structures from each \hyperlink{ref:Macrostate} {macrostate}~\cite{doi:10.1063/1.2740261,Huang19765,doi:10.1021/ct900620b}. This method was coined \emph{\hyperlink{ref:Adaptive} {adaptive} seeding}. In contrast, all the more refined methods derived therefrom and listed below are referred to as \emph{adaptive sampling}. Another natural proposal has been to seed simulation from states that contribute the most to the statistical uncertainty of MSMs built after each iteration \cite{doi:10.1021/ct500827g}. Several variants have proposed to seed from low-population microstates, and have been called ``counts" methods. This strategy is well-suited to enhance the exploration of space, but not necessarily to accurately estimate the probability distribution of low free energy states \cite{doi:10.1021/ct2004484,doi:10.1021/ct400919u,lecina_adaptive_2017,shamsi_enhanced_2017,6114444}. \hyperlink{ref:Seeding} {Seeding} from low-population \hyperlink{ref:Macrostate} {macrostate}s has also been suggested. This is better suited to converge the free energy calculation but will not lead to as an extensive space exploration \cite{doi:10.1021/acs.jctc.6b00762,doi:10.1063/1.5053582}. There are also methods that do not rely on MSM analysis. iMapD relies on clustering in a low-dimensional manifold inferred by \hyperlink{ref:DimRed} {dimensionality reduction} and selecting states to seed from the boundaries of a diffusion map in diffusion coordinates \cite{ChiavazzoE5494}~\footnote{iMapD does not enable the direct estimation of probability distributions.}. Following a similar approach, it has been suggested to pick configurations to reseed from using \hyperlink{ref:DimRed} {dimensionality reduction} algorithms such as sketch-map \cite{doi:10.1021/acs.jctc.6b00503}. PIGS, for Progress Index Guided Sampling, uses an unsupervised heuristic to avoid re-sampling the same region of space by organizing simulation frames along a progress index that connects configuration to existing configuration by finding the one to which a chosen distance is minimal. Weakly connected snapshots that have a large distance to other configurations are used to start \hyperlink{ref:Seeding} {seeding} new trajectories \cite{BACCI2015889}. Methods introducing directionality into the sampling have also been suggested and are particulary interesting when the target state (or set of states) is known. Broadly speaking, those suggest \hyperlink{ref:Seeding} {seeding} from states that are close to the end state in terms of a target property:
    \begin{enumerate}
    \item AdaptiveBandit expands on the methods described above by proposing to formulate the \hyperlink{ref:Adaptive} {adaptive} sampling problem in terms of reinforcement learning. This offers the promise and the computational platform needed to increased performance and flexibility of the algorithm across different systems \cite{doi:10.1021/acs.jctc.0c00205}.
    \item Reinforcement Learning Based Adaptive Sampling (REAP) proposes to efficiently explore configuration space by using reinforcement learning to choose new states. It does so by learning the relative importance of candidate \hyperlink{ref:CV} {collective variable}s as it makes progress along the \hyperlink{ref:FES} {landscape}, in a framework that rewards actions that lead to further exploration of the \hyperlink{ref:FES} {landscape}. Here too, states to learn from are selected as the least sample microstates. In that sense, it is a derivative of the counts method \cite{doi:10.1021/acs.jpcb.8b06521}.
    \item The Fluctuation Amplification of Specific Traits (FAST) method proposes to choose new states based on an objective function that balances tradeoffs between exploring novel regions of space (exploration) and focusing on regions that are important to lead to a converged estimate of the target properties (exploitation) \cite{doi:10.1021/acs.jctc.5b00737,doi:10.1021/acs.jctc.8b00500} The parameter that balances these two aspects needs to be tuned, a non-trivial aspect of using this method. This method has been recognized to be a specific case of a multi-armed bandit problem.
    \item Specifically for protein folding, or conformational changes in biological molecules, a common target property is inter-residue contacts, or its opposite, minimizing non-desirable contacts \cite{doi:10.1063/1.5053582}.
    \item In the same vein, it has been suggested to use evolutionary information by deriving inter-residue contacts from a multiple-sequence alignment, in a method called evolutionary couplings-guided \hyperlink{ref:Adaptive} {adaptive} sampling \cite{shamsi_enhanced_2017}.
    \end{enumerate}

Attempts at a quantitative comparison of several of these \hyperlink{ref:Adaptive} {adaptive} sampling methods have been published \cite{doi:10.1063/1.5053582,doi:10.1021/acs.jctc.8b00500}, but a systematic comparison is still missing. Preliminary work indicates that accelerating rare event is better achieved with \hyperlink{ref:Macrostate} {macrostate} count or directed methods, while exploring the space is most efficient using microstate count~\cite{doi:10.1063/1.5053582}. Methods explicitly taking into account the tradeoff between exploitation and exploration can be more versatile in their usage but their success (measured as the convergence of the configurational \hyperlink{ref:Distribution} {distribution}, or of the MSM) will depend on hyperparameter choice for a specific application.

\subsection{Weighted ensemble simulations - splitting/replication strategies}

The weighted ensemble (WE) method, also referred to as a splitting/replication approach, is particularly well-suited to exhaustively find pathways between \hyperlink{ref:Macrostate} {macrostate}s and evaluate transition rates between states~\cite{HUBER199697}. The approach relies on running relatively short MD simulation and terminating simulations that are not making progress towards the target state while replicating simulations that are instead progressing towards the end state~\cite{doi:10.1146/annurev-biophys-070816-033834}. By keeping track of the total number of trajectories, the approach is statistically rigorous. Clustering into pathway ensembles can provide a rigorous estimate of the relative importance of the different pathways.

Because of their \hyperlink{ref:OutOfEq} {non-equilibrium} nature, this variety of enhanced sampling methods is not particularly well-suited to calculate equilibrium population distributions. However, even if weighted ensemble simulations do not achieve steady state, frameworks have been proposed to recover equilibrium properties~\cite{doi:10.1063/1.3456985,doi:10.1021/ct401065r,doi:10.1021/jacs.8b10735}.

We note that weighted ensemble methodologies, while falling under the umbrella of \hyperlink{ref:Adaptive} {adaptive} \hyperlink{ref:Seeding} {seeding} strategies, can be categorized under transition path-finding methods, along with the string method with swarms of trajectories~\cite{doi:10.1021/jp0777059}, milestoning~\cite{doi:10.1063/1.1738640}, transition interface sampling~\cite{doi:10.1063/1.1562614}, forward flux sampling~\cite{PhysRevLett.94.018104}, adaptive multilevel splitting~\cite{cerou-guyader-lelievre-pommier-11,teo-mayne-schulten-lelievre-16} and supervised unbiased MD~\cite{doi:10.1021/acs.jcim.9b01094, doi:10.1021/acs.jcim.5b00702}. We choose here to not review these methods in detail given the focus of this review on estimating configurational averages (see instead reviews~\cite{doi:10.1146/annurev.physchem.040808.090412, doi:10.1146/annurev.physchem.53.082301.113146, doi:10.1063/1.5127780}).
The Weighted Ensemble method is implemented in the WESTPA software. Given the versatility of the framework, strategies and schedules can be easily explored~\cite{Bogetti2019Suite}.

\section{Selective acceleration methods}
\label{sec:selective_accel}

In selective acceleration methods, the dynamics of slow degrees of freedom is directly modified to accelerate transitions.
Contrary to \hyperlink{ref:Adaptive} {adaptive} \hyperlink{ref:biasingE} {biasing} methods, this is not achieved through a modification of the statistical \hyperlink{ref:Distribution} {distribution} sampled by the dynamics -- indeed, these methods are designed so that the sampled distribution is as close as possible (identical, in recent variants) to the unbiased \hyperlink{ref:targetdist} {target distribution}.
Such methods differ from \hyperlink{ref:ExpEns} {expanded ensemble} methods in that they modify the local dynamics instead of going through transitions to other ensembles to enhance the sampling (Figure~\ref{fig:scheme}).

Timescale separation between the dynamics along the \hyperlink{ref:CV} {collective variable} (or the external parameter in an \hyperlink{ref:Alchemical} {alchemical} setting) is assumed, or artificially enforced, in order to obtain (almost) Markovian dynamics along the \hyperlink{ref:CV} {collective variable} (respectively the auxiliary variable).
This was first introduced as the adiabatic free energy dynamics (AFED) method~\cite{doi:10.1063/1.1448491}.

For example, the temperature accelerated molecular dynamics (TAMD)~\cite{MV06} consists in adding an extended degree of freedom $\lambda$, with mass $m_\lambda$, and a harmonic coupling potential $k^\mathrm{ext} \left( \xi(\vx) - \lambda \right)$.
The extended \hyperlink{ref:Langevin} {Langevin dynamics} reads:
\begin{equation}
\left\{
\begin{array}{ll}
    d\vx &= M^{-1} \vp \,  dt \\
    d\vp &= \left(-\nabla_\vx U(\vx) + k^\mathrm{ext} ( \lambda -\xi(\vx)) - \gamma \vp \right) dt
    + \sqrt{ \frac{2 \gamma M}{\beta}} \; d\mathbf{W}_t \\
    d\lambda &= m_\lambda^{-1} p_\lambda \, dt\\
    d p_\lambda &= \left( k^\mathrm{ext} (\xi(\vx) - \lambda)  - \overline\gamma p_\lambda \right) dt
    + \sqrt{ \frac{2 \bar\gamma m_\lambda}{ \overline\beta}} dW_t
\end{array}
\right.
\end{equation}

Note that the Langevin equation on $\lambda$ is based on a separate temperature factor $\overline \beta$ and friction coefficient $\overline{\gamma}$.

TAMD relies on the regime where $\overline{\gamma} \gg \gamma$ (the extended dynamics is slower than the original one) and $\beta \gg \frac{1}{k^\mathrm{ext}}$ (tight coupling). Then the dynamics on $\lambda$ becomes uncoupled from the dynamics of $\vx$ and converges~\cite{MV06} to an effective dynamics of the form (in the overdamped case $\gamma \gg 1$):
\begin{equation}\label{eq:eff_dyn_z_bar}
d\lambda = - \frac{1}{ m_\lambda \overline{\gamma}} \nabla A(\lambda) \, dt + \sqrt{ \frac{2  m_\lambda}{\overline{\beta} \, \overline{\gamma}}} \, dW_t,
\end{equation}
so that the probability distribution sampled by $\lambda$ is proportional to $\exp(- \overline{\beta} A(\lambda))$. The artificial inverse temperature $\overline{\beta}$ is then chosen so that the effective dynamics~\eqref{eq:eff_dyn_z_bar} is less \hyperlink{ref:metastab} {metastable}.
The practical difficulty of such a method lies in  choosing the numerical parameters $(\gamma,\overline{\gamma},k^\mathrm{ext},\overline{\beta})$ for the algorithm to be efficient while keeping the \hyperlink{ref:AdiabaticDyn} {adiabatic} separation between the extended coordinate and the physical system -- failure to do so introduces a bias in the simulation.

In the ``single sweep" method, TAMD is used to estimate local free energy gradients, followed by estimation of the \hyperlink{ref:FES} {free energy surface}~\cite{Maragliano2008}.
Driven-AFED (d-AFED)~\cite{doi:10.1021/jp805039u}, unified free energy dynamics (UFED)~\cite{doi:10.1063/1.4733389}, canonical adiabatic free energy sampling (CAFES)~\cite{doi:10.1021/jp013346k} and on-the-fly free energy parameterization (OTFP)~\cite{ABRAMS2012114} are all related to this scheme.

The family of methods known as self-guided Molecular/Langevin Dynamics (SGMD~\cite{Wu1999SGMD} and SGLD~\cite{Wu2003_SGLD}) shares with these ``adiabatic'' methods the idea of selectively enhancing the dynamics while preserving (at least approximately) the statistical distribution sampled by the trajectory.
In SGLD, Langevin Dynamics is augmented with a \hyperlink{ref:biasingE} {biasing} force that accelerates slow degrees of freedom, computed as a running time average of the momenta, which acts as a low-pass filter. This process pinpoints slow degrees of freedom without having to define them \textit{a priori}.

This accelerates sampling, but creates a bias in configurational statistics that increases with the strength of the acceleration applied, and has to be corrected a posteriori to recover accurate averages.
However, SGLD with a generalized Langevin equation (where stochastic forces are not white noise but obey a given memory kernel)~\cite{Wu2015_SGLD-GLE} recovers the detailed \hyperlink{ref:Balance} {balance} property of standard Langevin dynamics, and as a result, is able to sample the NVT and NPT \hyperlink{ref:Ensemble} {ensembles} in an unbiased way.
Recently, variants of SGMD and SGLD have been proposed to combine optimally biases based on momenta and on forces~\cite{Wu2020_SGLDg}.
SGMD and SGLD are implemented in the CHARMM and AMBER simulation packages.

Note also that conventional (non well-tempered) metadynamics (see Section~\ref{sec:meta-classic}) implicitly makes an \hyperlink{ref:AdiabaticDyn} {adiabatic} hypothesis to get an unbiased \hyperlink{ref:FEestimator} {free energy estimator}~\cite{laio-gervasio-08, jourdain-lelievre-zitt-21}.

\section{Hybrid methods}
\label{sec:hybrids}

Enhanced sampling methods leveraging different principles can be combined together leading to hybrid schemes.
For example, a common theme is to combine an enhanced sampling method that focuses on \hyperlink{ref:biasingE} {biasing} specific degrees of freedom or CVs (e.g., ABP methods such as metadynamics) with a enhanced sampling method that more generally enhances the sampling of a large number of, or even all, degrees of freedom (e.g., \hyperlink{ref:ReplEx} {replica exchange} methods). In this way, one can better sample slow orthogonal degrees of freedom that are missing in the biased CV set.
Another common combination is to complement a sampling method with an external \hyperlink{ref:FEestimator} {free energy estimator}.
In addition, path finding methods like the string-of-swarms method can be combined with, e.g. umbrella sampling along the discretized minimum path obtained by the path finding method~\cite{doi:10.1021/jp0777059}. This is followed by free energy estimation along path coordinates.

There are so many possible combinations of enhanced sampling methods that it is difficult to offer a comprehensive discussion of all hybrid methods. Thus, we limit ourselves to discuss a few notable combinations below.

\subsection{Combination of replica exchange and external biasing potential methods}

Many hybrid enhanced sampling methods are a combination of Hamiltonian replica exchange and methods incorporating an external bias potential, which can be static or \hyperlink{ref:Adaptive} {adaptively} updated (e.g., umbrella sampling and metadynamics). Each replica $i$ includes its own bias potential $U^{\mathrm{bias}}_{i}(\vz)$, which depends on a \hyperlink{ref:CV} {collective variable} set $\vz_{i} = \xi_{i}(\vx)$ that can differ between replicas. Within each replica, the bias potential is kept static or evolved according to the \hyperlink{ref:Adaptive} {adaptive} \hyperlink{ref:biasingE} {biasing} potential method. When calculating the acceptance probability $P_\mathrm{accept}(\vx_i, i, \vx_j, j)$ for an exchange of configurations $\vx_{i}$ and $\vx_{j}$ between two replicas $i$ and $j$ given in in Equation~\ref{eq:re_exchange_prob}, we need to take the bias potentials into account. We limit ourselves to the, still rather general, case that all replicas have same potential energy function $U(\vx)$ but can have different temperatures.

We start by rewriting the exchange acceptance probability given in Equation~\ref{eq:re_exchange_prob}
as~\footnote{Note that we use here full quantities rather than \hyperlink{ref:reduced} {reduced} quantities as in Section~\ref{sec:generalized-ensemble}. However, the discussion is also valid for canonical and isothermal-isobaric \hyperlink{ref:Ensemble} {ensembles}.}
\begin{align}
\label{eq:replexc_acceptance probability}
P_\mathrm{accept}(\vx_i, i, \vx_j, j) &=
\min\left\{1,\frac{\exp(-[\beta_{i}U(\vx_j)+\beta_{j}U(\vx_i)])}{\exp(-[\beta_{i}U(\vx_i)+\beta_{j}U(\vx_j)])}\right\}
\nonumber \\
&=
\min\left\{1,\exp\left(\left(\beta_{i} - \beta_{j}\right)
\left[U(\vx_{i}) - U(\vx_{j})\right]\right)\right\}
\nonumber \\
&=
\min\left\{1,\exp \Delta_{i,j}\right\},
\end{align}
where we have defined $\Delta_{i,j}=\left(\beta_{i} - \beta_{j}\right)
\left[U(\vx_{i}) - U(\vx_{j})\right]
$ as the exponential term for conventional replica-exchange.

Incorporating the effect of the bias potentials, the exchange acceptance probability is calculated using the exponential term
\begin{align}
\Delta_{i,j} = &
\left(\beta_{i} - \beta_{j}\right)
\left[U(\vx_{i}) - U(\vx_{j})\right]
\nonumber \\ & +
\beta_{i} \left[
U^{\mathrm{bias}}_{i}(\xi_{i}(\vx_{i})) - U^{\mathrm{bias}}_{i}(\xi_{i}(\vx_{j}))
\right]
\nonumber \\ & +
\beta_{j} \left[
U^{\mathrm{bias}}_{j}(\xi_{j}(\vx_{j})) - U^{\mathrm{bias}}_{j}(\xi_{j}(\vx_{i}))
\right],
\end{align}
where the two last terms originate from the effect of the bias potentials acting on the two replicas~\cite{Bussi-JACS-2006}. In the following we discuss a few specific cases.

\subsubsection{Replica exchange umbrella sampling}
\label{sec:repex_umbrsampl}
In replica exchange umbrella sampling~\cite{Sugita2000_REUS}, all the replicas are simulated at the same temperature but differ in the fact that the different replicas have their umbrella potential centered at different locations. Thus, by allowing exchanges between neighboring umbrella windows, the convergence is improved. The exchange also helps with sampling orthogonal slow degrees of freedom not included in the \hyperlink{ref:CV} {CV} set. The exchange probability is obtained using
\begin{align}
\Delta_{i,j} = &
\beta \left[
U^{\mathrm{bias}}_{i}(\xi(\vx_{i})) - U^{\mathrm{bias}}_{i}(\xi(\vx_{j})) +
U^{\mathrm{bias}}_{j}(\xi(\vx_{j})) - U^{\mathrm{bias}}_{j}(\xi(\vx_{i}))
\right],
\end{align}
where the bias potential correspond to different umbrella sampling windows (generally neighbouring windows) that are centered at different locations in CV space.

\subsubsection{Parallel-tempering metadynamics}
Metadynamics can be combined with parallel-tempering to help with sampling missing slow orthogonal degrees of freedom not included in the biased \hyperlink{ref:CV} {CV} set~\cite{Bussi-JACS-2006}. The exchange probability is obtained using
\begin{align}
\label{eq:exchange_prob_ptmetad}
\Delta_{i,j} = &
\left(\beta_{i} - \beta_{j}\right)
\left[U(\vx_{i}) - U(\vx_{j})\right]
\nonumber \\ & +
\beta_{i} \left[
U^{\mathrm{bias}}_{i}(\xi(\vx_{i})) - U^{\mathrm{bias}}_{i}(\xi(\vx_{j}))
\right]
\nonumber \\ & +
\beta_{j} \left[
U^{\mathrm{bias}}_{j}(\xi(\vx_{j})) - U^{\mathrm{bias}}_{j}(\xi(\vx_{i}))
\right].
\end{align}
The same idea can also be used for other ABP methods (e.g., variationally enhanced sampling). In a similar way, metadynamics (and other ABP methods) can be combined with other \hyperlink{ref:ReplEx} {replica exchange} methods such as replica exchange solute tempering~\cite{REST2_Wang_2011,HREX_Bussi_2013}, that have a better scaling in term number of replicas needed. Then the first term in Equation~\ref{eq:exchange_prob_ptmetad} would be adjusted while the last two terms would remain the same.

\subsubsection{Bias-exchange metadynamics}
\label{sec:be-metad}
Bias-exchange metadynamics~\cite{Piana2007_bemeta} allows for \hyperlink{ref:biasingE} {biasing} a large set of \hyperlink{ref:CV} {CV}s simultaneously by considering multiple replicas running at the same simulation temperature, but each \hyperlink{ref:biasingE} {biasing} a different set of \hyperlink{ref:CV} {CV}s using metadynamics. Generally, one considers one \hyperlink{ref:CV} {CV} per replica so the outcome are several one-dimensional free energy \hyperlink{ref:FES} {profile}s. By allowing for exchange of configurations between replicas, we can avoid the problem of missing slow orthogonal \hyperlink{ref:CV} {CV}s in each replica.
The exchange probability is obtained using
\begin{align}
\Delta_{i,j} = &
\beta \left[
U^{\mathrm{bias}}_{i}(\xi_{i}(\vx_{i})) - U^{\mathrm{bias}}_{i}(\xi_{i}(\vx_{j})) +
U^{\mathrm{bias}}_{j}(\xi_{i}(\vx_{j})) - U^{\mathrm{bias}}_{j}(\xi_{i}(\vx_{i}))
\right].
\end{align}
Traditionally, bias-exchange is used with conventional (non-well-tempered) metadynamics but it can also be used with well-tempered metadynamics. We can even imagine using bias-exchange with other ABP methods such as variationally enhanced sampling.

\subsubsection{Parallel-tempering in the well-tempered ensemble}
\label{sec:pt-wte}
Combining parallel-tempering with the well-tempered ensemble~\cite{Bonomi-PRL-2010} (i.e., well-tempered metadynamics \hyperlink{ref:biasingE} {biasing} the potential energy) allows to greatly reduce the number of replicas required for parallel-tempering simulations of solvated systems~\cite{Deighan2012_ptwte_efficient}. This comes from two effects. First, within each replica potential, energy fluctuations are enhanced by a factor of $\gamma$ while averages stay more or less the same, leading to a better potential energy overlap between replicas. Second, the $\Delta_{ij}$ factor used to calculate the exchange acceptance probability in Equation~\ref{eq:replexc_acceptance probability} is given as
\begin{align}
\Delta_{i,j} = &
\gamma^{-1}
\left(\beta_{i} - \beta_{j}\right)
\left[U(\vx_{i}) - U(\vx_{j})\right]
\end{align}
and thus reduced by a factor of $\gamma$ as compered to conventional \hyperlink{ref:ReplEx} {replica exchange} (Equation~\ref{eq:replexc_acceptance probability}),  which leads to higher exchange probability (if $\Delta_{i,j}<0$, otherwise if $\Delta_{i,j}>0$ the exchange acceptance probability is unity). Due to these two effect, one can use a larger temperature difference between replicas and thus require fewer replicas overall~\cite{Deighan2012_ptwte_efficient}. Parallel-tempering in the well-tempered ensemble can also be combined with metadynamics where other \hyperlink{ref:CV} {CV}s are biased separately~\cite{10.1073/pnas.1320077110}.

\subsubsection{Replica exchange with collective variable tempering}

The idea behind \hyperlink{ref:ReplEx} {replica exchange} with \hyperlink{ref:CV} {collective variable} tempering~\cite{Gil-Ley_JCTC-2015} is to reduce the number of replicas needed for replica-exchange simulations by focusing only on selected degrees of freedom. One considers $M$ replicas with the same temperature and within each replica, one performs concurrent well-tempered metadynamics simulations where one considers the same \hyperlink{ref:CV} {CV} set within replica and biases each \hyperlink{ref:CV} {CV} by a separate one-dimensional metadynamics potential. One can then bias a large number of degrees of freedom (e.g., all dihedral angles). The replicas are arranged in a ladder of increasing bias factor values where the lowest replica corresponds to the canonical \hyperlink{ref:Ensemble} {ensemble} ($\gamma=1$). Thus, by going up the replica ladder, the fluctuations of the biased degrees of freedom are enhanced. Though one considers a large number of degrees of freedom, this is considerably smaller than the total number of the system's degrees of freedom. Therefore, the number of replicas needed is considerably less than parallel-tempering, where fluctuations of all degrees of freedom are enhanced by heating the system.

\subsection{Combinations of metadynamics and other enhanced sampling methods}

Apart from the \hyperlink{ref:ReplEx} {replica exchange}-based hybrid methods discussed in the previous section, there are various other hybrid methods where metadynamics has been combined with other types of enhanced sampling methods. Furthermore, some of the variants of metadynamics listed in Section~\ref{sec:metad_variants} could be considered as hybrid methods, and vice versa, as the distinction between a variant and a hybrid method is not always so clear.

Metadynamics has been combined with methods that enhance the potential energy sampling (see Section~\ref{sec:abp_energy}) such as the multicanonical ensemble~\cite{Yonezawa_MulticanonicalMetaD_2011} and integrated tempering sampling~\cite{Yang_MetaD-ITS_1_2016,Yang_MetaD-ITS_2_2018}. Here the idea is similar as when metadynamics is combined with parallel tempering, this should help with sampling missing slow orthogonal degrees of freedom. In a similar spirit, variationally enhanced sampling has been combined with sampling in the multithermal-multibaric ensemble~\cite{Piaggi_MultiVES_2019,Piaggi_MultiVES+CV_2019}.

In driven metadynamics~\cite{Moradi_DrivenMetaD_JPCL2013}, metadynamics is combined with steered MD.
In orthogonal space random sampling (OSRW)~\cite{Zheng_OSRW-1_PNAS2008,Zheng_OSRW-2_JCP2009,Min_OSRW-3_JCTC2010}, metadynamics is combined with a procedure based on thermodynamic integration to facilitate sampling of orthogonal degrees of freedom. Metadynamics has been combined with umbrella sampling in various ways as we discuss in the following.

In Refs~\cite{Badin_MetaD-US-1_JCP2006,Autieri_MetaD-US-2_JCP2010}, metadynamics is used to generate a bias potential that leads to effective sampling and diffusion in the CV space. The bias potential is then used as a static bias potential in another simulations where the FES is calculated using umbrella sampling corrections (i.e., \hyperlink{ref:Reweighting} {reweighting} with a static bias potential).

In Ref~\cite{Zhang_MetaD-US-3_JCTC2013}, metadynamics is used to identify a pathway and then the free energy profile along the pathway is calculated using multiple window (localized) umbrella sampling.

In Refs~\cite{Johnston_MetaD-US-4_PLos2012,Awasthi_MetaD-US-5_JCC2016}, multiple windows (localized) umbrella sampling is used to bias some chosen CV while within each umbrella window, metadynamics is used to bias another set of CVs. The main idea behind this strategy is that umbrella sampling is more suited than metadynamics to bias CV whose \hyperlink{ref:FES} {free energy profile} is broad. Furthermore, the metadynamics bias potential within each umbrella window helps to sample degrees of freedom that are orthogonal to the CV biased in the umbrella sampling simulations and thus improve the convergence. This combined umbrella sampling and metadynamics strategy has been extended to incorporate temperature accelerated MD~\cite{Awasthi_TASS_JCP2017,Pal_TASS-2_JCC2021} or \hyperlink{ref:ReplEx} {replica exchange} with solute tempering (REST2)~\cite{Kapakayala_WS-MetaD+REST2_JCC2021} to further improve the sampling of orthogonal degrees of freedom.

The basic framework of metadynamics can also be used to update the weights in expanded ensemble simulations~\cite{Hsu_alchemical_metadynamics}. Since the $g_k$ weights can be updated by any methods the user might choose, one can simply use various metadynamics techniques to update them, but as a discrete variable rather than a continuous one. The GROMACS expanded ensemble implementation has been adjusted to allow the biasing functionality to be built using PLUMED (starting from version 2.8). This is not particularly better than existing expanded ensemble techniques for a single dimension, but becomes particularly useful in multiple dimensions, where one has one alchemical dimension, and one or more collective variable dimensions.  This allows one to perform binding free energy calculations using metadynamics and simultaneously accelerating transitions along slow degrees of freedom, such as accelerating crossing of a high intramolecular torsional barrier or overcoming a free energy barrier in the hydration of a host pocket by flattening the distribution of water molecules in that pocket~\cite{Hsu_alchemical_metadynamics}.

\subsection{Combinations of metadynamics and structural ensemble determination methods}

Structural ensemble determination methods~\cite{Bonomi_StructEnsembleDeterm_Review_COSB2017,Cesari_MaximumEntropyPrinciple_Review_Comp2018,Bottaro_Science2018}, for example based on maximum entropy principle~\cite{Pitera_MaxEnt_JCTC2012,Roux_MaxEnt_JCP2013}, are used to integrate experimental observations into molecular simulations and yield structural ensembles that are compatible with experiments. While such structural ensemble determination methods are not strictly enhanced sampling methods, they are somewhat related as they generally introduce external restrains in the form of bias potentials that can be fixed or \hyperlink{ref:Adaptive} {adaptively} updated. To accelerate the configurational sampling, structural ensemble determination methods are often combined with enhanced sampling methods such as metadynamics.

In Ref~\cite{Bonomi_MetadynamicMetainference_SciRep2016}, parallel-bias metadynamics is combined with metainference~\cite{Bonomi_Metainference_SciAdv2016} that is a structural ensemble determination methods that incorporates experimental errors via a Bayesian inference framework. In Ref~\cite{Amirkulova_PTWTE-EDS_JPCB2020}, parallel-tempering in the well-tempered ensemble (Section~\ref{sec:pt-wte}) is combined with experiment directed simulations~\cite{White_EDS_JCTC2014}.
An older work along this line is replica-average metadynamics~\cite{Camilloni_RAM_2013,Camilloni_RAM-2_JACS20214}, where metadynamics or bias-exchange metadynamics is combined with replica-averaging.

Related to the idea of structural ensemble determination methods are experiment directed metadynamics~\cite{White_EDM_2015}, ensemble-biased metadynamics~\cite{Marinelli_EnsembleBiased_2015}, and target metadynamics~\cite{GilLey_TargetMetaD_2016}. In these variants of metadynamics, the bias updating procedure is modified such that the biased CV distribution that the simulation converges to is some predefined \hyperlink{ref:targetdist}{target probability distribution}. Thus, by taking this \hyperlink{ref:targetdist}{target distribution} from experimental measurements, it is possible to obtain a structural ensemble that is compatible with the experimental results. In a similar spirit, the \hyperlink{ref:targetdist}{target distribution} in variationally enhanced sampling~\cite{Valsson_VES_PRL_2014,Valsson2020Handbook_VES} (Section~\ref{sec:ves}) can be taken from experimental measurements.

\subsection{Combinations of ABF and other enhanced sampling methods}
\label{sec:abf_hybrids}

The ABF method is applied within a defined region of collective variable space, where its application leads to improved sampling and yields an estimate of the free energy gradient. To reduce the time needed for diffusive sampling of a large volume of CV space, we can combine ABF with a stratification approach (Section~\ref{sec:localization}). In this case, ABF can be applied independently on several smaller, non-overlapping regions of the collective variable~\cite{Chipot2005} or several variables~\cite{Henin2010a}. Thanks to the local character of this gradient, the estimated gradient in all regions can be merged by simple concatenation, and then integrated in one piece~\cite{Henin2021integration}. Unlike energy-based methods like Umbrella Sampling, no particular precaution is necessary to match the data from different strata (windows).

Metadynamics has been combined with extended-system ABF~\cite{Fu_MetaD-eABF_JPCL2018,Fu_MetaD-eABF_JCIM2020} to improve the exploration properties of ABF.
For a review of these hybrid eABF methods, see~\cite{Fu2019}.
This combination has been further extended by incorporating Gaussian-accelerated MD~\cite{Chen_GAMD-MetaD-eABF_JCTC2021} to help sample orthogonal degrees of freedom not included in the biased CV set.

Orthogonal Space Tempering (OST) is a variant of OSRW based on an extended-system ABF method, using a finite sampling temperature in the orthogonal space~\cite{Zheng2012}.
In this method, the ABF-like sampling of an \hyperlink{ref:Alchemical} {alchemical} parameter $\lambda$ is completed by enhanced sampling of the force along $\lambda$, which correlates with slow orthogonal degrees of freedom, thereby accelerating orthogonal relaxation.

\section{Software implementations}
\label{sec:software}
The codes to run the various simulations outlined in the previous sections range from in-house scripts, to methods implemented natively in the largely used MD simulation packages GROMACS, AMBER,  NAMD or CP2K, and via the open-access libraries or plugins Colvars~\cite{Fiorin2013}, PLUMED~\cite{Bonomi-CPC-2009,Tribello2014,plumed-nest}, PMFlib~\cite{kulhanek2011pmflib}, and SSAGES~\cite{Sidky2018}. The most popular software options for each method have been mentioned in the dedicated sections and are summarized in Tables~\ref{Table:Codes} and~\ref{Table:Libraries}.

Several free and open source codes are gathered under a GitHub topic: \url{https://github.com/topics/enhanced-sampling}

\begin{table*}[!ht]
\caption {Built-in capabilities of widely used molecular dynamics simulations codes. Methods not discussed in the text may be included, see user guides for more discussion. Not an exhaustive list; enhanced sampling methods maybe be available in other MD codes.}
\label{Table:Codes}
\begin{tabularx}{0.95\textwidth}{
  || >{\raggedright\arraybackslash} l
  || >{\raggedright\arraybackslash}X
  | >{\raggedright\arraybackslash}l ||}
 \hline
  MD engine  & Main features                                 & Reference \\
\hline
\hline
CHARMM & Adaptively Biased Path Optimization (ABPO), Adaptive Umbrella Sampling, Constraints, Distributed CSA (Conformational Space Annealing), Dynamic Importance Sampling (DIMS), Enveloping Distribution Sampling Method, Replica Exchange, Free Energy Perturbation, Self-Guided Langevin Dynamics, String method, Targeted Molecular Dynamics, Transition Path Sampling, replica exchange& \cite{Brooks2009} \\
\hline
NAMD &  Simulated annealing, steered MD, (unconstrained variant of) targeted MD, replica exchange, accelerated and Gaussian-accelerated MD, custom algorithms via Tcl scripting, grid forces.        & \cite{Phillips2020} \\
\hline
GROMACS & Restraints (various potentials including harmonic potentials for umbrella sampling), simulated annealing, replica exchange, expanded ensemble (both as AWH and a separate implementation), \hyperlink{ref:OutOfEq} {non-equilibrium} pulling (steered MD), applying forces from three-dimensional densities.   &  \cite{lindahl_2021}\\
\hline
OpenMM & Simulated annealing, replica exchange (with OpenMMtools), applying external forces, versatile python framework to implement any scheme, expanded ensemble (as self-adjusted mixture sampling). & \cite{10.1371/journal.pcbi.1005659}\\
\hline
AMBER  &  Replica exchange, targeted MD, steered MD, accelerated and Gaussian-accelerated MD, Self-Guided Langevin Dynamics, external forces, umbrella sampling, string-of-swarms. & \cite{Case_2021} \\
\hline
CP2K & Constraints, harmonic restraints, targeted MD, steered MD, metadynamics. & \cite{CP2K_2020} \\
\hline
DESMOND & Umbrella sampling, Metadynamics, replica exchange & \cite{Desmond2006} \\
\hline
LAMMPS  & Harmonic restraints, applying external forces, replica exchange, parallel replica dynamics, temperature accelerated dynamics, original hyperdynamics, local hyperdynamics. &  \cite{LAMMPS_2022}\\
\hline
HOOMD-blue &  Restraints, applying external forces, versatile python framework to implement any scheme. & \cite{HOOMD-blue_2020} \\
\hline
Tinker-HP & Steered MD, Gaussian-accelerated MD, umbrella sampling & \cite{Celerse2019,Celerse2021} \\
\hline
GROMOS & Replica exchange, umbrella sampling, thermodynamic integration, enveloping distribution sampling.  & \cite{Gromos_2012} \\
\hline
GENESIS & Replica exchange, umbrella sampling, Gaussian-accelerated MD, restraints, targeted MD, steered MD  & \cite{GENESIS_MD_Code_2017} \\
\hline
SPONGE & Integrated tempering sampling, selective integrated tempering sampling, metadynamics & \cite{SPONGE_MD_Code_2022} \\
\hline
\end{tabularx}
\end{table*}

\begin{table*}[!ht]
\caption {Libraries and Modules for enhanced sampling}
\label{Table:Libraries}
\begin{tabularx}{0.95\textwidth}{
  || >{\raggedright\arraybackslash}l
  || >{\raggedright\arraybackslash}X
  | >{\raggedright\arraybackslash}l ||}
\hline
  Library name  & Main features                         & Reference \\
\hline
\hline
  Collective Variables Module (Colvars) & Definition and biasing of various \hyperlink{ref:CV} {CV}s.
  Multiple variants of Adaptive Biasing Force (ABF). and metadynamics. Support for multiple walker simulations. Scripted variables and biasing forces. Collective variables as custom functions.
  Built into in VMD for preparation and analysis of CVs. & \cite{Fiorin2013, Henin2022dashboard}\\
\hline
PLUMED        &  Definition of various \hyperlink{ref:CV} {CV}s that can be analyzed and biased. Various biasing methods (e.g., umbrella sampling, steered MD, metadynamics, parallel-bias metadynamics, bias-exchange metadynamics, extended-system adaptive biasing force, variationally enhanced sampling). Support for multiple walker simulations and replica exchange simulations. Methods for integrating experimental results (e.g., maximum entropy principle, metainference, experiment directed simulation). Modular design making it easy to add new features. Can be interfaced with a wide range of MD codes. Large number of tutorials are available~\cite{plumed_masterclass}. Large number of example input files are available in the PLUMED-NEST~\cite{plumed_nest_url}.         &  \cite{Bonomi-CPC-2009,Tribello2014,plumed-nest} \\
\hline
PMFlib        &  ABF, constrained dynamics, metadynamics, restraints, string method. &  \cite{kulhanek2011pmflib} \\
\hline
SSAGES & Definition of various \hyperlink{ref:CV} {CV}s that can be analyzed and biased. Various biasing methods (e.g., umbrella sampling, steered MD, metadynamics, adaptive biasing force, basis function sampling, artificial neural network sampling, combined force-frequency). Support for multiple walker simulations and replica exchange simulations. Other path-based methods such as nudged elastic band, finite temperature string, swarm of trajectories, forward flux sampling. & \cite{Sidky2018} \\
\hline
WESTPA        &   Weighted Ensemble     & \cite{Bogetti2019Suite, Russo_2022}  \\
\hline
Wepy          &  Weighted Ensemble  & \cite{samuel_d_lotz_2020_4270219} \\
\hline
\end{tabularx}
\end{table*}

\section*{Contributions}
J.H., O.V., T.L., M.R.S., and L.D.\ conceptualized the paper. All authors wrote initial drafts, in particular with L.D.\ and J.H.\ drafting Section~\ref{sec:Notion_Notation}, J.H.\ and T.L.\ drafting Sections~\ref{sec:Out-of-equilibrium_driven} and~\ref{sec:selective_accel}, J.H.\ drafting Sections~\ref{sec:localization} and \ref{sec:ABF}, O.V.\ drafting Sections~\ref{sec:ABP} and \ref{sec:hybrids}, M.R.S.\ drafting Sections~\ref{sec:fe_estimators} and~\ref{sec:generalized-ensemble}, L.D.\ drafting Section~\ref{sec:seeding}, and other sections jointly drafted by all authors. J.H., O.V., M.R.S., and L.D.\ significantly revised all sections of the paper. L.D.\ and J.H.\ provided additional management of the process.

\section*{Acknowledgments}
L.D.\ would like to thank the Science for Life Laboratory, the Göran Gustafsson Foundation, and the Swedish Research Council (Grant No.~VR-2018-04905) for support.
J.H.\ acknowledges support from the French National Research Agency under grant LABEX DYNAMO (ANR-11-LABX-0011).
M.R.S.\ acknowledges support from the National Science Foundation under grant numbers OAC-1835720 and OAC-2118174, and help from Arjan Kool for Figure~\ref{fig:EXEanalogy} and preliminary drafts of expanded ensemble analysis.
O.V.\ acknowledges support from the Deutsche Forschungsgemeinschaft (DFG, German Research Foundation) - Project number 233630050 - TRR 146 ``Multiscale Simulation Methods for Soft Matter Systems''.
O.V.\ thanks Benjamin Pampel for help with preparing Figure~\ref{fig:MetaD}.

We would also like to thank the people who have commented on GitHub to help make this document better, namely Soumendranath Bhakat, Giovanni Bussi, Ramon Crehuet, Michele Invernizzi, Yinglong Miao, and Adrian Roitberg.

\makeorcid


\appendix

\section{Abbreviations and acronyms}
\begin{itemize}
    \item ABF - Adaptive Biasing Force
    \item ABF-AR - Adaptive Biasing Force with Adiabatic Reweighting
    \item ABF-FUNN - ABF-Force Biasing using Neural Networks
    \item ABMD - Adaptive Biasing MD
    \item ABP - Adaptive Biasing Potential
    \item ABPO - Adaptively Biased Path Optimization
    \item AFED - Adiabatic Free Energy Dynamics
    \item ATLAS - Adaptive Topography of Landscapes for Accelerated Sampling
    \item BAR - Bennett’s Acceptance Ratio
    \item CAFES - Canonical Adiabatic Free Energy Sampling
    \item CSA - Conformational Space Annealing
    \item CVs - Collective Variables
    \item CZAR - Corrected Z-Averaged Restraint
    \item d-AFED - Driven Adiabatic Free Energy Dynamics
    \item DIMS - Dynamic Importance Sampling
    \item eABF - extended-system Adaptive Biasing Force
    \item FAST - Fluctuation Amplification of Specific Traits
    \item FEP - Free Energy Perturbation
    \item FES - Free Energy Surface
    \item GAMBES - Gaussian Mixture-Based Enhanced Samplin
    \item MBAR - Multistate Bennett’s Acceptance Ratio
    \item MC - Monte Carlo
    \item MD - Molecular Dynamics
    \item MetaD - Metadynamics
    \item MM - Molecular Mechanics
    \item MSM - Markov State Model
    \item mwABF - multiple-walker Adaptive Biasing Force
    \item NPT - Isothermal-Isobaric Ensemble
    \item NVE - Microcanonical Ensemble
    \item NVT - Canonical Ensemble
    \item OPES - On-the-fly Probability-Enhanced Sampling
    \item OSRW - Orthogonal Space Random Sampling
    \item OST - Orthogonal Space Tempering
    \item OTFP - On-the-fly Free Energy Parameterization
    \item PIGS - Progress Index Guided Sampling
    \item PMF - Potential of Mean Force
    \item QM - Quantum Mechanics
    \item QM/MM - Quantum Mechanics/Molecular Mechanics
    \item RAVE - Reweighted Autoencoded Variational Bayes for Enhanced Sampling
    \item REAP - Reinforcement Learning Based Adaptive Sampling
    \item REST - Replica Exchange Solute Tempering
    \item SAMS - Self-Adjusted Mixture Sampling
    \item SGMD - Self-Guided Molecular Dynamics 
    \item SGLD - Self-Guided Langevin Dynamics
    \item TALOS - Targeted Adversarial Learning Optimized Sampling
    \item TAMD - Temperature Accelerated Molecular dynamics
    \item TI - Thermodynamics Integration
    \item TICA - Time-Lagged Independent Component Analysis
    \item UFED - Unified Free Energy Dynamics
    \item VES - Variationally Enhanced Sampling
    \item WE - Weighted Ensemble
    \item WHAM - Weighted Histogram Analysis Method
\end{itemize}

\bibliography{refs}

%
%

\end{document}